\documentclass[usedcolumn,useAMS,usenatbib]{mn2e}
\usepackage{times}
\usepackage{graphicx}
\usepackage{aas_macros}
\usepackage{epstopdf}
\usepackage{color}
\usepackage{amssymb}
\usepackage[fleqn]{amsmath}
\usepackage[usenames,dvipsnames]{xcolor}
\usepackage{xspace}
\usepackage{lineno}

\newcommand{\myemail}{Boris.Haeussler@nottingham.ac.uk}
\newcommand{\deVa}{de Vaucouleurs\xspace}
\newcommand{\Haeussler}{{H\"au\ss ler}\xspace}
\newcommand{\galapagos}{{\scshape galapagos}\xspace}
\newcommand{\galfit}{{\scshape galfit}\xspace}
\newcommand{\galfitthree}{{\scshape galfit3}\xspace}
\newcommand{\galfitm}{{\scshape galfitm}\xspace}
\newcommand{\gimtwod}{{\scshape gim2d}\xspace}

\newcommand{\sersic}{S\'ersic\xspace}
\newcommand{\sigmakelvin}{{\scshape sigma}\xspace}
\newcommand{\sex}{{\scshape SExtractor}\xspace}

\newcommand{\sn}{S/N\xspace}

\newcommand{\HST}{{\slshape HST}\xspace}

\title[MegaMorph III]{MegaMorph -- multi-wavelength measurement of galaxy structure: complete \sersic profile information from modern surveys}
\author[\Haeussler et al.]{Boris~\Haeussler $^{1,2}$\thanks{E-mail: \myemail}, Steven~P.~Bamford$^1$, Marina~Vika$^2$, Alex~L.~Rojas$^2$, Marco~Barden$^3$, \newauthor 
Lee~S.~Kelvin$^{3,4,5}$, Mehmet~Alpaslan$^{4,5}$, Aaron~S.~G.~Robotham$^{4,5}$, Simon~P.~Driver$^{4,5}$, \newauthor 
I. K. Baldry$^6$, Sarah~Brough$^7$, Andrew~M.~Hopkins$^{7}$, Jochen~Liske$^{8}$, Robert~C.~Nichol$^9$, \newauthor Cristina.~C.~Popescu$^{10}$, Richard~J.~Tuffs$^{11}$
\smallskip\\
$^{1}$School of Physics and Astronomy, University of Nottingham, University Park, Nottingham, NG7 2RD, UK\\
$^{2}$Carnegie Mellon University in Qatar, Education City, PO Box 24866, Doha, Qatar\\
$^{3}$Institute of Astro- and Particle Physics, University of Innsbruck, Technikerstra\ss{}e 25, A-6020 Innsbruck, Austria\\
$^{4}$School of Physics and Astronomy, University of St Andrews, North Haugh, St. Andrews, Fife, KY16 9SS, UK\\
$^{5}$International Centre for Radio Astronomy Research, 7 Fairway, The University of Western Australia, Crawley, Perth, WA 6009, Australia\\
$^{6}$Astrophysics Research Institute, Liverpool John Moores University, Twelve Quays House, Egerton Wharf, Birkenhead CH41 1LD, UK\\
$^{7}$Australian Astronomical Observatory, PO Box 915, North Ryde, NSW 1670, Australia\\
$^{8}$European Southern Observatory, Karl-Schwarzschild-Str. 2, 85748 Garching, Germany\\
$^{9}$Institute of Cosmology and Gravitation (ICG), University of Portsmouth, Dennis Sciama Building, Burnaby Road Portsmouth, PO1 3FX, UK\\
$^{10}$Jeremiah Horrocks Institute, University of Central Lancashire, PR1 2HE, Preston, UK\\
$^{11}$Max Planck Institut fuer Kernphysik, Saupfercheckweg 1, 69117 Heidelberg, Germany
}

\begin{document}

%\linenumbers

\date{Accepted ... Received ...; in original form ...}

\pagerange{\pageref{firstpage}--\pageref{lastpage}} \pubyear{2012}

\maketitle

\label{firstpage}

\begin{abstract}
In this paper, we demonstrate a new method for fitting galaxy profiles which makes use of the full multi-wavelength data provided by modern large optical--near-infrared imaging surveys. We present a new version of \galapagos, which utilises a recently-developed multi-wavelength version of \galfit, and enables the automated measurement of wavelength-dependent \sersic profile parameters for very large samples of galaxies. Our new technique is extensively tested to assess the reliability of both pieces of software, \galfit and \galapagos\ on both real $ugrizYJHK$ imaging data from the GAMA survey and simulated data made to the same specifications. We find that fitting galaxy light profiles with multi-wavelength data increases the stability and accuracy of the measured parameters, and hence produces more complete and meaningful multi-wavelength photometry than has been available previously. The improvement is particularly significant for magnitudes in low \sn bands and for structural parameters like half-light radius $r_{\rmn{e}}$ and \sersic index $n$ for which a prior is used by constraining these parameters to a polynomial as a function of wavelength. This allows the fitting routines to push the magnitude of galaxies for which sensible values can be derived to fainter limits. The technique utilises a smooth transition of galaxy parameters with wavelength, creating more physically meaningful transitions than single-band fitting and allows accurate interpolation between passbands, perfect for derivation of rest-frame values.
\end{abstract}


\section{Introduction}
\label{sec_intro}
Studies of galaxy formation and evolution rely on accurate estimates of physical galaxy properties, such as luminosity, mass, star formation history (SFH), size and morphology.  Many of these properties are obtained from imaging data, via measurements of magnitude and profile shape.  Such measurements are nowadays relatively straightforward for individual objects.  However, many analyses benefit from being applied to as large a sample as possible.  Given the sizes of modern surveys, this typically means thousands, or even hundreds of thousands, of galaxies.  For such large numbers of measurements to be feasible, they must ideally be performed in a fully automated fashion, which significantly complicates matters.\footnote{Alternatively, for suitable tasks where automated tools are insufficient, one may instead resort to using large numbers of people, via citizen science methods, e.g., \citet{GZ1}, with their own set of complications.}

Most galaxy parameters may be obtained using a variety of methods. One common approach to measuring magnitudes is aperture photometry, in which one defines the extent of a galaxy in some manner and then sums all the flux within that area.  Ideally one would choose an aperture large enough to contain effectively all the galaxy flux.  However, one cannot simply use an arbitrarily large aperture, as that would introduce excessive noise from the sky and be more likely to be contaminated by flux from neighbouring objects.  Typical methods employed to define photometric apertures therefore seek a reasonable compromise.  As a result, aperture magnitudes necessarily miss a proportion of flux from the outer regions of each galaxy.  Furthermore, the aperture defined for a given galaxy, and hence the amount of missing flux will vary between images, depending on spatial resolution, signal-to-noise (\sn) and the exact shape of the galaxy light distribution.  Colour gradients within a galaxy also lead to the inferred extent of a galaxy varying with observed wavelength \citep[as recently reported by][hereafter K12, and others]{Kelvin2012}. 
Further refinements include applying convolutions to match the point spread functions (PSFs) of the images (e.g., \citealt{Hill}), and applying a minimal correction by estimating the flux that would be missed if the galaxy were a point source (e.g., \citealt{EDisCS, Graham}).

Varying definitions of photometric apertures can have a significant impact on the resulting science. 
Using a fixed surface-brightness threshold clearly misses more flux for objects that show less compact profiles.
More sophisticated methods still suffer from biases, for example Petrosian magnitudes recover essentially all the flux for exponential profiles, but miss $\sim\! 20$ to $\sim\! 70$ per cent of the flux \citep{Graham} for \deVa profiles \citep{deVauc}. The wavelength selected to define the aperture is also important.  For example, disk galaxies are typically redder in their centre, due to the presence of a bulge or dust.  Defining the aperture in a red photometric band will therefore result in total fluxes that are underestimated in bluer bands, and hence total colours that are systematically biased to redder values.

For measuring galaxy sizes, one could employ methods similar to those used to define photometric apertures. These approaches obviously suffer from many of the same issues described above, and generally do not provide a consistent, physically interpretable measure of galaxy size.  A more meaningful alternative is to determine the radius (or two axes of an ellipse) that contains a specified fraction of the total galaxy light.  Common examples include the half-light radius, $r_{50}$, and $r_{90}$, the radius which contains 90 per cent of the total galaxy light.  Of course, these measurements depend critically on a reliable measurement of the total magnitude.  
Sizes derived using aperture magnitudes will suffer from systematic biases with respect to galaxy profile shape, luminosity and distance.  There is again the issue of wavelength; sizes measured in blue bands will tend to reflect the extent of the young stellar populations, whereas red bands will more closely reflect the distribution of stellar mass.  Just as k-corrections are required to convert observed magnitudes to restframe values, similar corrections may be required to homogenise sizes when considering galaxies spanning a range of redshifts.  Finally, it is important to note that none of these size measurements are corrected for the effect of the PSF.  They will therefore be overestimated, particularly for intrinsically small or distant galaxies.

A variety of automated proxies for morphology have been proposed, the simplest of which focus on the shape of the azimuthally-averaged surface-brightness profile.  One widely used parameter is the concentration index, which is defined as the ratio of the radii containing two fractions of the total flux, e.g., $C = r_{90}/r_{50}$ \citep{Strateva} or $C = 5\cdot log(r_{80}/r_{20})$ \citep[as defined by CAS,][]{Conselice2003}.  All of the biases which affect these size estimates will therefore result in biases in the concentration index and similar non-parametric profile measurements.  In any case, even for well-resolved, bright galaxies, such simple proxies only give a very rough indication of true internal structure or morphology.

MegaMorph is a project aimed at improving our ability to measure and understand the structure of galaxies.  In particular, we endeavour to make optimal use of modern multi-wavelength imaging surveys.  Using data from multiple bands simultaneously in the fitting process increases the signal-to-noise, without greatly increasing the number of free parameters.  Importantly, combining multi-wavelength imaging provides information that is not available to techniques which operate on only a single band.  For example, this enables the fitting process to utilise the different wavelength dependence of each component to help separate their profiles, and produces a more physically consistent models.  We expect this to be particularly crucial when performing bulge-disk decompositions. However, in this work, we first consider only fits using single-\sersic profiles.  MegaMorph and the software developed and utilized is further discussed in Sections~\ref{sec_purpose} and \ref{sec_software}.

\subsection{Parametric methods}

To avoid many of the problems that empirical (aperture-based) methods suffer from, an increasingly popular approach to measuring galaxy properties involves fitting their surface brightness profiles with parametric models. This has a number of advantages: all the measurements are obtained in a consistent manner, varying PSFs can be easily accommodated, and the issue of missing flux is, at least partly, avoided.  The price is the assumption of a parametric form for the two-dimensional surface brightness distribution; typically exponential profiles for galaxy disks, \deVa profiles \citep{deVauc} for bulges and ellipticals, or more generally, \sersic profiles \citep{Sersic}.  A number of software packages have been produced to perform such fits, e.g., \galfit \citep{Peng2002, Peng2010}, \gimtwod \citep{Simard98, Simard2002}, {\sc BUDDA} \citep{budda}, {\sc 2DPHOT} \citep{2008PASP..120..681L} and {\sc GasPhot} \citep{gasphot}.

These tools can achieve good results, both when used manually to fit individual galaxies, or when applied to large surveys in a fully- or semi-automated fashion (at least when fitting single \sersic models to relatively bright galaxies, e.g., \citealt{Haeussler2007}). However, while a user of this technique can successfully employ a complex combination of profiles when fitting individual galaxies by hand \citep[e.g., nearby NGC galaxies, ][]{Vika}, automated model fitting in large surveys is considerably more challenging.  Many thousands of galaxies, each with their own individual complications, such as neighbouring objects and potentially varying sky level, PSF, imaging availability and profile complexity, must be dealt with in an automated fashion. Developing a fully automated code with sufficient complexity, accuracy, flexibility and speed to perform profile fitting in modern surveys is difficult.  Nevertheless, a number of studies \citep{schade97,Lilly1998,Allen2006,Simard2011,TascaWhite,Kelvin2012,LacknerGunn} have produced catalogues of galaxy profile parameters for large samples, with recent local studies often based on imaging provided by the Sloan Digital Sky Survey (SDSS, \citealt{York2000}). Some of these works focus on one-component \sersic models, although most also attempt bulge-disk decomposition.

Many existing approaches to galaxy profile fitting are primarily designed to work with only a single image, i.e. only one photometric band.  If applied to a multi-band dataset, each band must be fit separately.  One may choose to treat all bands equally, and allow the technique to fit a completely independent model in each band \citep[e.g.,][]{2010MNRAS.408.1313L,Kelvin2012}.  This enables one to study wavelength-dependent structural variations, e.g., due to colour gradients.  However, only a fraction of the data is used to constrain the profile in each fit.  Furthermore, the resulting colours may not be physically meaningful, particularly in the case of multiple components, due to unphysical variations in the structural parameters (as demonstrated later).  One might naively expect many model parameters to be identical in all bands, e.g., component centres, axial ratios and position angles, or vary smoothly with wavelength, e.g., \sersic index and size. Alternatively, therefore, one may select one dominant band in which to fit an initial profile, and then fit this profile to the other bands while holding various parameters fixed (e.g., \citealt{LacknerGunn}).  With the profile fixed across all bands the resulting component colours should be more meaningful, but again only one band has been used to determine that profile, wasting data. Also consider that the smooth variation of parameters with wavelength cannot be guaranteed which this method would assume to be the case.

One approach to using all the available data to constrain the profile would be to simply sum all the images together and fit a model to the resulting image.  Obviously, however, this does not allow colour information to be extracted. This profile could then be fit to the bands individually, with the structural parameters held fixed. Another solution is to fit a model to multiple images of the same object simultaneously. This is less common than fitting single images, but not a new idea.  \gimtwod \citep{Simard98} includes an option to fit two images with two bulge+disk models constrained to have the same structural parameters.  Only the flux of each component is allowed to vary independently between the two models.  This approach has been used to measure bulge and disk colours for over a million SDSS galaxies \citep{Simard2011}.  \gimtwod also provides an ability to fit a stack of images with identical profiles, with only the centre of the model allowed to vary between images.  However, we wish to (a) make use of an arbitrary number of multi-wavelength images, (b) constrain parameters to vary smoothly as a function of wavelength, being neither completely fixed or free, and (c) fit a variety of models, not just bulge+disk.  We would also prefer to fit neighbouring galaxies where appropriate, rather than relying on masking \citep[][hereafter H07]{Haeussler2007}.

To understand the desire for model parameters which vary smoothly with wavelength, consider the example of fitting a single-\sersic model to a normal disk galaxy, comprising a blue, exponential disk and a red, \deVa bulge.  In bluer bands, the disk will be dominant, and hence the profile is best represented by a low \sersic index, while in redder bands the bulge would become more dominant, resulting in a smoothly increasing \sersic index with observed wavelength (K12, Fig. 21). Additionally, gradients in the stellar populations within spheroids (e.g., \citealt{2009ApJ...699L..76L,2010ApJS..187..374S}) and discs (e.g., \citealt{2000MNRAS.312..497B,2004ApJS..152..175M,2010MNRAS.407..144T,2011MNRAS.411.1151G}) and centrally concentrated dust attenuation \citep{2007MNRAS.379.1022D,2010MNRAS.404..792M} can produce similar effects \citep{Moellenhoff,2012IAUS..284..306P}.
Fixing the profile shape as a function of wavelength would therefore give bad fits in some bands.  Allowing it to vary freely will often result in the parameters varying wildly with wavelength as the fit uses its increased freedom to fit the image noise.  This will also significantly increase the number of parameters to be fit.

We therefore propose that the preferred solution is to fit a full wavelength-dependent model to an arbitrary set of multi-band data, simultaneously.  This approach can use all the available data to define the profile, while enabling the measurement of physically meaningful component colours and colour gradients.  The form of the model parameters as a function of wavelength can be chosen to optimally balance consistency and flexibility.   This approach should also improve the number of galaxies for which a full set of robust photometry can be determined, as the bands can \lq help each other out\rq.  For example, in a simultaneous multi-band fit, low \sn bands would not contribute much to defining structural parameters, but would benefit from the constraints on these from higher \sn bands, resulting in robust measurements of the flux in each band.

\subsection{The purpose of this paper}
\label{sec_purpose}
In MegaMorph, we have developed a combination of tools in order to test our expectations regarding the benefits of multi-band parametric measurements.  This software is briefly described in Section \ref{sec_software}.  Details of the implementation, together with examples illustrating the advantage of this approach, appear in Bamford et al. (2012, in prep; hereafter Paper I).  Our technique is designed to be highly flexible, but for consistency we adopt a standard configuration for most of the work in this paper.  Our choices are explained in Section~\ref{sec_setup}.

This paper is accompanied by another paper (Vika et al., in prep; Paper II), which applies our technique to a sample of 168 nearby galaxies that have been artificially redshifted in order to assess its performance in fitting individual, realistic galaxies. The present paper complements that study by demonstrating the application of our technique to large surveys in an automated fashion, and with greater statistical power.  Using both real and artificial images, we will demonstrate how (and why) using multi-band fitting has advantages over single-band fitting, in terms of stability, improved accuracy and increased sample sizes, especially for the low \sn bands of a survey.
 
Most galaxies comprise multiple structural components, primarily a bulge and a disk. The most physically meaningful parameters should therefore be obtained by fitting multi-component models.  However, fitting such models is challenging, particularly on noisy, low-resolution, single-band imaging, as the parameters of the multiple components can be highly degenerate.  Using multi-band data to constrain the fit significantly alleviates this problem (see Paper I), as the different wavelength dependencies of the individual components (i.e. their colours) provides valuable information, which is not present in single-band fitting.

Ultimately we aim to decompose galaxies into physically meaningful structures, and measure reliable properties for each component.  However, simpler single-component fits still provide a great deal of useful information, and are less challenging to perform.  We will explore multi-component fits in future papers, but as a first step in demonstrating the advantages of multi-wavelength profile fitting, in our present work we focus on fitting single \sersic profiles to each object.

This paper is structured in the following way: In \S~\ref{sec_software} we introduce the idea of multi-band fitting, including a brief technical description on how this is carried out and what changes have been applied to both \galfit (see \S~\ref{sec_galfitm}) and \galapagos (see \S~\ref{sec_galapagos}). 
\S~\ref{sec_setup} explains the setup of both codes used throughout this paper. 
In \S~\ref{sec_real} we show tests from applying this software to real GAMA data (Galaxy And Mass Assembly) and compare the values to show in how much multi-band fitting improves the fitting results both on individual galaxies and on the galaxy population as a whole.
\S~\ref{sec_sims} carries out similar tests, but uses simulated images, e.g. galaxies whose true intrinsic values are known. This comparison, while not containing any real physical meaning about galaxy populations, allows to show the improvement by using multi-band fitting in more detail.
\S~\ref{sec_other} takes other considerations than fitting accuracy into account, e.g. fitting time and disk-space required.
Finally, \S~\ref{sec_cmd}, both as a sanity check and to further show improvements enabled by the new technique presented in this paper, discusses the colour-magnitude diagram of galaxies. This chapter is aimed to be a motivation for users to apply the software developed, tested and presented in MegaMorph papers in order to improve their scientific results.

\section{Multi-wavelength profile fitting}
\label{sec_software}
In order to evaluate the advantages of fitting wavelength-dependent models to multi-band data, we have implemented software to perform such fits.  For the sake of efficiency and reliability, we chose not to re-implement all the functions required for a profile fitting code ourselves.  Instead we elected to build upon existing, well-tested software and make only those changes necessary to enable multi-wavelength fitting.  However, in the course of modifying the software, we have also added additional features where required or deemed convenient, and generally improved the efficiency of the code where possible.

We selected \galapagos \citep{galapagos} and \galfitthree \citep{Peng2010} as the starting point for our development, due to their reputation for reliability, flexibility and speed, as well as the extensive experience of members of our team in using these software tools (H07).  \galfit performs the fit for each target image while taking the image PSF into account; \galapagos, after initial preparation of the data, takes care of everything else required to run \galfit in an automated manner on a large survey, including book-keeping, object detection (using \sex; \citealt{Bertin}), cutting images of each target, masking, determination of the sky level, estimation of initial parameters, writing setup files and load-balancing.  In this section we briefly describe our choices in implementing multi-wavelength fitting, and the modifications we have made to the standard versions of these codes.

\subsection{\galfitm}
\label{sec_galfitm}
We have adapted \galfitthree \citep{Peng2010} for the requirements of this project, with permission of the original developer, C.~Peng.  To differentiate our modified version from the standard release we refer to it as \galfitm.  For reference, all the work in this paper uses \galfitm version 0.1.2.1.  The code will be publicly released in the near future.  \galfitm-0.1.2.1 is based on \galfit version 3.0.2, although the additions in \galfit-3.0.4 (the latest standard version) will be incorporated in \galfitm before public release.  Development is continuing, primarily to improve ease-of-use and incorporate the additional features mentioned above. However, the general performance of the technique is expected to remain as presented in this paper.

\galfit constructs model images by summing one or more components, which potentially include a sky background (with optional gradient), elliptical \sersic functions, point sources and a variety of other profiles.  \galfit fits the parameters of its model to the input data (weighted by an error map, which may be provided or internally created) by employing the widely-used Levenberg-Marquardt (LM) algorithm to minimise the weighted sum of the square residuals ($\chi^2$).  In addition to the model image itself, \galfit calculates the derivative of the model image with respect to each free parameter, as required for the LM algorithm.  The model and all of its derivatives are convolved with the provided PSF for comparison with the input image.  The reader is advised to consult \citet{Peng2002,Peng2010} for a detailed description of \galfit.

The standard version of \galfitthree accepts only a single input image with which to constrain the model fit.  It was therefore necessary to make fairly substantial modifications to enable the use of multi-band data.  However, most of the original code and its structure is maintained, and we intend our modified version to be backward compatible when used with single-band data. In this subsection, we briefly describe the significant changes. For full details we refer the reader to Paper I.

\subsubsection{Wavelength-dependent model parameters}
\label{sec_cheb}

In order for \galfitm to be able to fit multi-band data, we replaced every galaxy model parameter with a wavelength-dependent function,
\begin{align}
M(x, y;\,p_1, \dots, p_n) \rightarrow M[x, y;\, &\tilde{p}_1(\lambda;\, q_{1,1}, \dots, q_{1,m_1}), \dots,\\
\notag\ & \tilde{p}_n(\lambda;\, q_{n,1}, \dots, q_{n,m_n})]\,,
\end{align}
where $M(x, y;\, \cdot)$ is the model (describing the surface brightness as a function of pixel coordinate, before PSF convolution), the $p_i$ are the $n$ original parameters of the \galfitthree model, and each $\tilde{p}_i$ is some function, with $m_i$ parameters $q_{i,j}$, which describes the variation of the model parameter $i$ versus wavelength, $\lambda$.  Whereas standard \galfit fits the $p_i$, in \galfitm the parameters of the fit are the set of $q_{i,j}$.

In the case of a standard \sersic profile used in this paper, these parameters are position, magnitude, half-light radius, \sersic index, axis ratio and position angle. However, the approach is implemented in a general fashion and works for any of the model functions provided by \galfitthree.  The choice of function is somewhat arbitrary, although various properties are clearly desirable, including a straightforward way of selecting the function's flexibility and independence of the function parameters.  We chose to use a series of Chebyshev polynomials \citep[of the first kind;][]{cheb_poly}, $T_j(z)$, for all of the functions:
\begin{equation}
\tilde{p}_i(\lambda;\, \{q_{i,j}\}) = \sum_{j=0}^{m_i} q_{i,j}\, T_j[z(\lambda)]\,.
\end{equation}
The Chebyshev polynomials are restricted to the domain $z \in [-1, +1]$, and hence the wavelength range of the input bands is linearly mapped on to that interval.
The Chebyshev polynomial $T_j(z)$ is of order $j$, i.e. $T_2(z)$ is a quadratic function of $z$.  Cheybshev polynomials of the first kind are defined by the recurrence relation
\begin{align}
T_0(z) &= 1 \notag\\
T_1(z) &= z \notag\\
T_{n+1}(z) &= 2z\, T_n (z) - T_{n-1}(z)\,.
\end{align}
The fit parameters, $q_{i,j}$, are therefore the Chebyshev coefficients of the series for parameter $i$.

The flexibility of this function may be varied by selecting the maximum order of each series, $m_i$, i.e. limiting to zeroth-order implies that a parameter must be constant with wavelength, second-order allows quadratic dependence with wavelength, while choosing the order as one less than the number of bands gives the function freedom to interpolate the data precisely.

Chebyshev polynomials form an orthogonal basis set, but full orthogonality only occurs when the function is constrained at the corresponding set of Chebyshev nodes.  This is possible when approximating a smooth function, but in our case we are not free to choose the position of the constraints, they are set by the locations of the available set of photometric bands on the selected wavelength scale.  Nevertheless, the independence of individual Chebyshev polynomials, even if only partial, is expected to limit the degeneracies between parameters, and hence aid the stability of the fitting process.

We stress that the purpose of the functions, $p_i(\lambda)$, is to connect the parameter values in the different bands with a user-specified degree of smoothness.  For example, we might choose the \sersic index to vary quadratically, position angle constant, and magnitude to be completely free as a function of wavelength\footnote{To be exact, we use 3543\AA, 4770\AA, 6231\AA, 7625\AA, 9134\AA, 10305\AA, 12483\AA, 16313\AA, 22010\AA\ for $ugrizYJHK$-band, respectively.}.  The functions themselves are not intended to be physically meaningful, although they may be used to approximate parameter values at wavelengths between the observed bands, e.g. to determine restframe parameters.  In this paper we use the Chebyshev polynomials as a function of wavelength.  However, the variable used in the polynomials need not be true wavelength.  Frequency, the logarithm of wavelength, or a variety of other variables may be suitable (see Paper I for a more thorough discussion). For the purpose of this paper, we have chosen to use linear scaling with wavelength.

All of the free parameters of the model, the set of $q_{i,j}$, are fit to all the multi-band data simultaneously by minimising a single quantity, defined as:
\begin{equation}
\label{eqn:chi2}
\chi^2 = \sum_{u,v,w}{\frac {[d_{u,v,w} - M(x_u, y_v; \lambda_w, {q_{i,j}})]^2}{\sigma_{u,v,w}^2}}\,,
\end{equation}
where $u$ and $v$ index the pixels, at positions $x_u$ and $y_u$, in image $d_w$, with wavelength $\lambda_w$, and uncertainty image $\sigma_w$.
We write Eqn.~\ref{eqn:chi2} in this way to emphasize that the data comprises a set of discrete measurements, while the model $M$ is, in principle, a continuous function, evaluated at the position and wavelength of the data in order to compute $\chi^2$.
Further technical details will be presented in Paper I.

\subsubsection{Parameter constraints}

One side effect of our multi-wavelength modifications is that the approach taken to constrain model parameters in \galfitthree required revision.  These constraints take two forms: hardcoded limits (such as ensuring that sizes cannot become negative) and user specifiable limits, but both are treated similarly.  Constraints are useful to guide the fitting process, by eliminating regions of parameter space which are ruled out by other considerations.  They can therefore improve the efficiency of the early stages of the fitting process.  However, if the fitting process repeatedly encounters constraints, this is an indication that a good model fit to the data cannot be achieved.

Appropriate handling of constraints is particularly important when using \galfitm to fit multiple objects simultaneously (as is common with \galapagos).  When considering a single galaxy (possibly with multiple components), if the fitting process ends with a parameter very close to a constraint boundary, it is reasonable to discard the resulting fit from subsequent analysis. However, in the case of a target galaxy with one or more neighbours, we would not want difficulties encountered in obtaining an unconstrained fit for a neighbouring object to negatively impact the fit to the primary target, or result in a potentially good fit to the primary target being discarded.
 
In \galfitthree, the physical parameters, $p_i$, feature directly in the fitting algorithm.  Constraining these physical parameters to lie on specified intervals can therefore be achieved in a straightforward manner.  At each iteration of the fitting process, the LM algorithm proposes a step for each parameter. If that step would violate a constraint, \galfitthree typically resolves the conflict by simply setting the offending parameter to the value at the constraint boundary.  For multi-band fits, however, constraints on the physical parameters may be violated in some bands but not others. There is also a complicated relationship between the physical parameter at a given wavelength, $\tilde{p}_i(\lambda)$, and the fit parameters, $q_{i,j}$.  An alternative approach is therefore required.  We briefly outline this here.  For further details and discussion see Paper I. 

The LM algorithm interpolates between the Gauss-Newton (GN) algorithm and the method of gradient descent (GD), with the degree of interpolation controlled by a damping parameter, $\Lambda$.  The GN algorithm will generally attempt to make relatively large steps, whereas GD is more conservative.  Increasing $\Lambda$ leads to dominance of GD over GN, and increasingly smaller steps.  The LM algorithm includes a prescription for varying $\Lambda$ to appropriately balance GN and GD as the fit progresses: if a proposed set of parameter steps successfully improves $\chi^2$, then $\Lambda$ is decreased by a factor, and the steps are accepted; otherwise it is increased by the same factor, and the steps are rejected.  (This factor is 10 in the case of \galfit.)

In \galfitm, if a proposed set of steps in the fitting parameters $q_{i,j}$ would violate a constraint on the standard parameter $p_i$ in any of the wavelength bands, then the steps are not performed for parameters $q_{i,\cdot}$.  All other (unoffending) parameters are stepped as usual and a trial value of $\chi^2$ generated.  This approach avoids the difficulty of determining how to limit the parameters $q_{i,\cdot}$ so as to avoid the resulting $\tilde{p}_i(\lambda;\, \{q_{i,\cdot}\})$ from violating any constraints at the wavelengths of the input bands.

For a moment assume that this is the only change. In that case a failure to improve $\chi^2$ would result in a decrease in $\Lambda$, and rejection of the entire proposed set of parameter steps.  The steps in the next proposed set would be smaller, and more likely, though far from guaranteed, to avoid violating constraints.  However, often the proposed steps (without those which would cause violated constraints) will result in an improved $\chi^2$, an acceptance of those steps and an increase in $\Lambda$.  The next iteration is therefore likely to propose a step in the offending standard parameter that is similar to, or larger than, the previous.  Tests have shown that this often results in that parameter remaining fixed for the entire duration of the fit, even though a small movement toward the constraint may lower $\chi^2$ and result in a more acceptable model.

To mitigate this issue, and encourage movement of constraint-violating parameters towards (but not beyond) constraint limits, we impose a schedule for $\Lambda$.  This is designed to substantially increase $\Lambda$ occasionally in the case of violated constraints, resulting in the next set of proposed steps being much smaller than the previous proposal, and so more likely to avoid overstepping the constraint boundary.  This adopted schedule was developed through trial and error, but appears to do a reasonable job of meeting our requirements.

The result is that constraints can be specified in \galfitm, on both individual fitting parameters, $q_{i,j}$, and more usefully on the standard parameters, $p_i$, at any, or all, of the input band wavelengths.  For example, in the fits in this paper the \sersic index, $n$, is constrained to lie on the interval $0.2 \leq n \leq 8$ at all input wavelengths.  Details of all the constraints applied in the present work are given in Section \ref{sec_setup}.

\subsubsection{Other modifications}

While making the changes described above, we have attempted to retain backward compatibility as far as possible. While the setup is, of course, slightly more complicated for multi-band data, we ensure for any original \galfitthree start file (for single-band data) to work unaltered with \galfitm. The additional multi-wavelength features are simple to enable, and any user already familiar with \galfit should have no difficulties using \galfitm.

The output format of \galfitm is also slightly modified. In addition to the image, model and residual (which of course \galfitm provides for each band), for convenience it also stores the PSFs used and provides all the fitting information, including setup details and the full results, in FITS tables within the output file.  Information is also still provided via header keywords for backward compatibility.

Several minor fixes and efficiency improvements have also been made.  For example, all variables are now stored as double precision. Besides being more accurate, this also provides a modest speed improvement on modern 64-bit machines.  Again, we refer interested readers to Paper I for full details of our \galfit modifications.

As part of our MegaMorph project we are investigating several other modifications to \galfit, including the incorporation of non-parametric components, and alternatives to the LM algorithm.  The results of these investigations will be described in future papers.

Throughout the development of \galfitm, we have compared its output on single-band data to that of \galfitthree, generally finding very close agreement.  For the vast majority of galaxies, the results of \galfitm and \galfitthree are identical.  The greatest differences relate to our modified implementation of constraints.  In cases where constraints are encountered during the fit, this can result in somewhat slower convergence, but this is often accompanied by a formally better fit, in terms of a slightly reduced $\chi^2$ compared with \galfitthree.  

\subsection{\galapagos}
\label{sec_galapagos}
We have adapted the current public version of the IDL script \galapagos (version 1.0) to support \galfitm, and hence utilise multi-wavelength data, in close collaboration with its original developer, and coauthor of this paper, M.~Barden.  In the process, we have implemented a number of improvements and additional features, some for efficiency or convenience, and others that were required by the nature of our chosen dataset.  We refer to our new version as \galapagos-2.  Specifically, version 2.0.2 was used to produce most of the results shown in this paper. Section \ref{sec_cmd} uses version 2.0.3, but the two versions only differ in very minor details and should produce nearly identical fitting results. Changes became necessary in order to be able to target specific objects instead of every object that was detected, hence speeding up the analysis for Section 7, where only a subset of the detected objects are considered. In all versions of the code, we have attempted to preserve backward compatibility in the case of single-band data.  The code will be publicly released in the near future. In this subsection, we will briefly explain our modifications.

Prior to this work, \galapagos was designed for use with space-based imaging, specifically for surveys performed by the \textsl{Hubble Space Telescope (HST)}.  The \HST PSF is very stable on the multi-drizzled images that are usually used for fitting purposes, both temporally and across the field of view.  \galapagos-1 therefore only used one PSF for the entire survey.  For the MegaMorph project, we wished to apply \galapagos to ground-based imaging.  While it is possible to approximately homogenise the PSF across an entire survey, by applying appropriate convolutions, the resulting resolution must necessarily correspond to the worst-case and thus a great deal of spatial information would be discarded. Instead, we adapted \galapagos to work with a spatially variable PSF.

In principle, it would be desirable for \galapagos to construct an estimated PSF for each target galaxy (as was done in K12).  However, this would require providing \galapagos with knowledge of the survey strategy.  As each survey will typically adopt a different tiling strategy, PSF creation is not trivial to generalise, particularly given the importance of the PSF in correctly modelling galaxy profiles. Sophisticated software already exists to take a set of point sources and combine them to produce an accurate PSF (e.g., PSFEx\footnote{\texttt{http://www.astromatic.net/software/psfex}}) and a user of our code should use those to pre-determine a set of suitable PSFs. For the GAMA survey, PSFs had already been determined by K12 and were used throughout this analysis.

For generality, we implemented a selection of the PSF from a provided list of filenames and sky coordinates.  \galapagos simply selects the closest PSF to each target position.  As K12 generate PSFs at the position of each galaxy with $m_r < 19.8$, the PSFs will correspond exactly for these galaxies, and the sampling is sufficiently dense that, in the vast majority of cases, fainter galaxies will be well represented by their nearest PSF.  The PSF selection is performed for each band individually.

We ultimately aim for our technique to be applicable to the largest surveys available.  During this proof-of-concept stage we are content to restrict ourselves to more modest datasets, but still wished to work with a single GAMA II region, of area $\sim 60$ deg$^2$.  We therefore improved the efficiency of the code in several places where it became apparent that, for large datasets such as ours, a major speedup was possible.  Some of these changes may alter the outcome of the code very slightly (e.g., sub-scripts of \galapagos now know only about the neighbouring frames at times, instead of the entire survey, potentially changing deblending decisions).  However, we think that in practice this will not produce any significant differences in the results, as we were always very conservative in our modifications.  Overall, we were able to speed up the code by approximately a factor of four in terms of CPU time.  A further simple optimisation was made in the loop over all objects in the survey.  \galapagos-1 determines whether the next object in queue is close enough to be influenced by an object currently being processed, and, if so, waits for that object to finish before starting the next. Our new code simply continues with a different object further away, thus keeping more CPUs busy at any given time.

Most importantly, of course, the code can now handle multi-wavelength datasets. In addition to defining all the bands, each comprising a set of images, which are to be used for fitting (using a setup similar to \galapagos-1), the user must define an additional set of images on which \sex is run for object detection.  Of course, the detection images could simply correspond to one of the fitting bands.  In our case, we chose to use a co-added image of all bands for detection, in order to detect sources with extreme colours that would be missed if only one band were used for detection.  In this way, we use the whole dataset in order to derive an object list. Even so, the fitting bands are not all quite equal: one must be defined as the primary band, on which all deblending and masking decisions are made. We chose the $r$-band in our dataset, because it typically has the highest \sn.  Our construction of the multi-band detection image is very simple, and could admittedly be improved upon by weighting the input images more carefully (e.g., \citealt{1999AJ....117...68S}).  Nevertheless, it suffices for the purposes of this paper.

The output returned by \galapagos-2 provides much more information than previously; namely most of the details from the FITS tables in the \galfitm output file.  However, \galapagos-2 may continue to be used with \galfitthree, for single-band data, in which case it provides the same information as the previous version.

Further planned changes include the implementation of bulge-disk decomposition, ideally together with model selection, so that \galapagos itself is able to decide whether a single-profile fit or bulge-disk decomposition provides the better representation of the imaging for each object and hence the more useful set of parameters.  We also plan to adapt \galapagos to run in supercomputing environments, in order to achieve the speed necessary for larger samples and/or surveys.  However, the general performance of single-\sersic profile fitting is expected to remain as presented in this paper.

\section{Choice of multi-wavelength model, initial parameter values, and constraints}
\label{sec_setup}

\galapagos requires various choices to be made regarding its operation and the setup information it provides to \galfit.  Our generalisation to multi-band data adds a number of additional options.  This section describes the choices we have made for the analysis described in this paper.

If \galfitm is used to fit a single-component profile with structural parameters (i.e., all except magnitude) that are constant with wavelength, this is mostly equivalent to using \galfitthree to first fit a single co-added image to obtain these parameters, and then measuring each magnitude by fitting each band with this fixed profile. When using multiple-component profiles, this is no longer true, and retaining the multi-band information throughout the fit leads to more accurate and reliable measurements (see Paper I).  In the case of single-component fits, the greatest advantage of our new fitting technique comes from allowing profile structural parameters to vary systematically with wavelength; the use of co-added images would lose this information. We will quantify the benefit of our multi-band fitting approach in Section \ref{sec_sims}.  An important set of choices are therefore the degree of wavelength dependence we allow for each parameter.

%magnitude
It is critical that we obtain an accurate magnitude for each band, and hence colours.  The magnitudes in each band are clearly correlated, and the variety of possible galaxy spectral energy distributions (SED) are very well known.  These cannot be reproduced with a low-order polynomial, and so we must ensure that sufficient freedom is given to the magnitudes such that they are accurately recovered.  Full freedom is implied by using a polynomial with as many coefficients as data points, in which case the function is capable of perfectly interpolating the data.  To describe the wavelength dependence of magnitude, we therefore use an 8th-order polynomial (with 9 free coefficients, equal to the number of bands in our dataset).  Note that the use of high-order interpolating polynomials is afflicted by
Runge's phenomenon, whereby the function oscillates excessively between data points, particularly at the edges of the considered interval.  The entire function itself therefore does not well-represent the galaxy SED and so cannot be used to estimate magnitudes at wavelengths other than those for which there is constraining data (see Section \ref{sec_sim_results}).  However, the magnitudes obtained for each band in the dataset remain reliable.  If required, e.g., for k-correcting magnitudes, the resulting magnitudes may be interpolated using different codes, based on realistic SED templates (e.g., \textsc{kcorrect}; \citealt{kcorrect}).  Finally, note that while we allow full freedom for magnitudes in this paper, in further work it may be appropriate to consider using slightly lower order polynomials to reduce this issue, while still retaining sufficient flexibility to recover accurate values (see Paper I).

% sizes
Equally important to accurate magnitudes is the determination of physically meaningful structural parameters of the galaxies, e.g. half-light radius and \sersic index. We chose to allow profile half-light radius and \sersic index to vary with wavelength quadratically (i.e., second-order polynomials, with three coefficients). This was decided after examining the wavelength dependence of these quantities from single-band fits to bright galaxies.  For most of these a linear function was sufficient to model the trend, but in some cases a mild curvature was seen. We therefore elected to fit a polynomial one order higher than linear, in order to examine this effect.

In our simulations (described in Section \ref{sec_sims}), we create simulated galaxies for which half-light radius and \sersic index vary according to known second-order functions. When fitting these simulations, we allow these parameters to vary with third-order, in order to investigate our ability to recover the correct higher-order coefficients.

As the images for each wavelength band are accurately registered (although read \citealt{Kelvin2012} and \S 4.1), the centre of the profile is constant with wavelength (although the centre is allowed to vary during the fit; we do not constrain the position to that given by \sex).  Similarly, for position angle and axis ratio we also choose to fit constant values, with no wavelength dependence.  While this ignores variations that might be expected for typical spiral galaxies (red bulges are round, blue disks appear elongated), this seemed to be a reasonable approximation for our purposes in this paper.

The sky values for each band are pre-determined by \galapagos and held fixed during the fit\footnote{At this point, it should be mentioned, that the GAMA data shows imperfect flat-fielding of the provided Swarped images due to the use of large filters. These are chosen to avoid removing too much real local structure but lead to sky backgrounds not being accurately measured around small objects. As this effect would be present in both single and multi-band fitting, we ignore this effect in this paper.}. 
We have shown in H07 that this is the most reliable approach for single-band fits, and there is no obvious reason why this should not also be the case when using multi-band data.

Largely following H07, we adopt the following constraints on the parameters in \galfitm. In the case of multi-band fitting, these constraints apply to the parameter values for all bands (but not the entire polynomial).
\begin{description}
\item[Position ($x$, $y$):] These are simply constrained to lie within the image cut-out for this object. In practice, this constraint is rarely encountered during the fit, but is retained to prevent the centre running out of the image in the case of a nearby bright source.
\item[Magnitude ($m$):] $-5 \leq m_{\rmn{fit}}-m_{\rmn{input}}\leq 5$, where $m_{\rmn{input}}$ is derived by adding an empirically estimated offset to the \verb+Mag_Best+ derived by \sex during object detection. Additionally, we use $0\leq m\leq 40$, to ensure sensible values; in practice this constraint is rarely hit.
\item[Size (half-light radius; $r_{\rmn{e}}$):] $0.3 \leq r_{\rmn{e}}\leq 400$ pixels. This maintains values in a physically meaningful range and prevents the code from fitting very small sizes, where, due to oversampling issues, the fitting iterations become very slow. Pixel sizes in the data used are 0.339 arcsec/pixel, hence we constrain the half-light radii to be $r_{\rmn{e}}\gtrsim 0.1$ arcsec. For reference, it should be noted that $r_{\rmn{e}}\gtrsim 0.1$ arcsec corresponds to $r_{\rmn{e}}\sim 0.3$ kpc at a typical objects redshift of $z\sim 0.2$ in the GAMA survey.
\item[\sersic index ($n$):] $0.2 \leq n \leq 8$. Fits with values outside these ranges rarely represent good models of a target galaxy. The upper value of 8 is a conservative choice as objects with higher \sersic indices are rarely seen and, from earlier visual inspection, are usually associated with spurious galaxy fits or cases where the target object is a star. It should be stated that some luminous elliptical galaxies with $n>8$ do exist \citep[e,g,][]{Graham}, hence this constraint will be removed and loosened to higher values in future works.
\item[Axis ratio ($q$):] $0.0001\leq q \leq 1$. Again, this ensures the fit value is physically meaningful, but is mostly superfluous, as \galfit includes a hardcoded constraint on $0 \leq q \leq 1$. The main reason for this constraint being applied is that \galfit, when very small values are reached becomes very slow due to oversampling issues.
\item[Position angle ($\theta$):] $-180\deg<\theta<180\deg$. This constraint is hardcoded in \galfit. Following the same definition as \galfit, position angle is defined as a major axis positioned vertically is $0\deg$ (nominally north if rotated to the standard orientation) and increases counterclockwise (nominally toward the east).
\end{description}

These constraints are implemented in order to improve and speed up the fitting process when fitting galaxies. For stars, the situation is slightly different. An unsaturated star should technically return a point source when PSF-correction is used during the fit, i.e., $r_{\rmn{e}}=0$. However, for practical reasons, enforced by the constraints specified above, the fit is not allowed to do this. Instead, the fit results usually end up on one of these fitting constraints (typically $r_{\rmn{e}}=0.3 [pix]$ and $n=8$) and thus removes the star from the image in a slightly non-optimal way. However, when this model is subtracted from the image after PSF convolution, we generally find these constraints produce a reasonable residual image, thus not significantly influencing the galaxy fits. For saturated stars, fitting a \sersic\ fitting, while masking out the saturated part of the profile, is more suitable to remove the wings of the profiles, more important in order to improve the fit of the neighbouring galaxies.

We use these constraints to remove both stars and galaxies with bad fits, from our catalogue, by identifying objects with fit results lying on one or more constraint boundaries.  As just explained, the vast majority of stars result in values on one of the above constraints.  This also occurs for galaxies when the object in question cannot sensibly be fit with a \sersic profile, and hence any returned values should not be used in a scientific context. Please read Section~\ref{sec_cleaning} for more details about the cleaning of the fitting results catalogue.

One obvious potential improvement would be to fit stars with PSF profiles (as has been done in K12) instead of \sersic profiles. \galfit generally does allow this, but there are two main reasons why we have chosen not to do this. Firstly, such a procedure would require a reliable galaxy/star classifier, to make the decision of which profile to use for each object. While this feature is desirable, it is not straightforward to implement, and was not deemed to be high-priority, given the low impact we expect it to have on our results.  Secondly, there are bright stars present in the images, which possess significant flux at radii beyond the size of the PSF images we use. This is especially true for very bright, highly-saturated stars, for which the profile core strongly deviates from the PSF due to saturation effects.  Furthermore, the wings of bright stars often vary, and are therefore not well represented by an averaged PSF.  In these worst cases, it was found that fitting a \sersic profile, rather than a PSF, results in much better residual images, particularly at large radii, as it has more flexibility to mimic and remove the outer wings of the PSF profile.  In our data setup, we additionally use a masking scheme to identify saturated areas in the image. These areas are consequently (a) smoothed in the \sex detection image, such that star images with the internal structure typical of saturated sources in SDSS are detected as one object only; and (b) masked in the images used by \galfit, hence sufficiently removing the wings of the stars and masking out the innermost areas, so a fit to a neighbouring source should not be significantly influenced by either of these areas.

\section{Application to real imaging}
\label{sec_real}

We evaluate our multi-band galaxy profile fitting technique on both real and simulated datasets. Our real data obviously have the advantage of showing actual galaxies, and we can compare to results from other studies.  However, there is no definitive \lq truth\rq\ to which we can compare.  Simulated data, on the other hand, are idealised and do not capture all the subtleties of real data, and of course cannot be used to study the real universe.  However, they do provide us with a way of testing the results of our method against known values.  We first present our work using the real dataset.

In this section, we describe the data, and show the results of extensive tests and comparisons between single-band and multi-band fitting techniques, using an otherwise identical code.

\subsection{Data}
\label{sec_data}
For the purpose of these tests, we have chosen to use imaging data provided by the GAMA survey \citep{Driver2011}, as it comprises one of the largest multi-wavelength datasets currently available, in terms of both area and wavelength.  GAMA is focussed around a redshift survey, but, crucially, this is supplemented by a highly homogeneous and complete set of multi-wavelength data, spanning from the far-UV to radio, making it a superb tool for studying the local universe.  We plan to use the galaxy profile fits from this paper and future work to perform a variety of scientific studies for which the wealth of the GAMA dataset is extremely valuable.

The present GAMA survey (phase I) covers an area of $\sim\! 144$ deg$^{2}$, of which half is in three $\sim\! 48$ deg$^{2}$ equatorial fields, with high spectroscopic completeness to a depth of $r < 19.4$. Current efforts (GAMA II) are focussed on two additional, more southerly, fields as well as expanding the equatorial fields to $\sim\! 60$ deg$^{2}$, increasing the depth of all fields to $r < 19.8$ and the total survey area to $\sim\! 290$ deg$^{2}$.  As we only require a relatively modest sample size for the purposes of this paper ($\sim 10^4$ galaxies), and because of the computing time required to fit large samples, in this paper we limit ourselves to one region of GAMA, the equatorial field at $9$h R.A., and often to only a subregion of that.  In future work we will expand our analysis to the full GAMA II survey.

GAMA has prepared its imaging data in a very convenient form for our purposes \citep{Hill}. These data include five-band optical ($ugriz$) imaging from SDSS plus four-band near-infrared ($YJHK$) imaging from the Large Area Survey (LAS) component of the UKIRT Infrared Deep Sky Survey (UKIDSS; \citealt{ukidss}).  All of these bands have a depth and resolution amenable to \sersic-profile fitting.  Importantly, the images for all nine bands have been \lq micro-registered\rq\ onto the same pixel grid, using SWarp \citep{terapix}\footnote{Note that small systematic sub-pixel offsets between the bands have been reported in K12, but we ignore these here.}. This procedure also homogenises the photometric zeropoints and sky background in order to produce an artefact-free mosaic.

\citet{Kelvin2012} has presented the results of single-band \sersic-profile fitting for all GAMA spectroscopic targets ($r < 19.8$) in the GAMA I fields. This catalogue has provided a very useful comparison dataset, enabling us to carry out initial tests and aiding the early development of our code.  We have used our own software to perform single-band fits, to ensure a fair comparison with our multi-band results, and so fit parameters from K12 are not shown in this paper other than in Fig. \ref{fig0}. A more complete comparison between the two different single-band codes used here and in K12 is beyond the scope of this paper and requires a detailed discussion about minor details in the two codes. However, we have made detailed comparisons between our single-band results and those of K12, finding generally excellent agreement. The PSFs used in this paper are those obtained by K12 with \galapagos' choosing the closest PSF to the targeted object for the fit.

The imaging data used in this analysis will be made public in GAMA data release 2.  Compared to the already public DR1 images, these images cover a larger area with a slightly different pixel scale to better match forthcoming GAMA datasets. The only modifications we make to these imaging data is the addition of a background pedestal so that \galfit can construct correct sigma images. For each band, we simply use a typical background value from the original SDSS imaging\footnote{For reference, the values added for $ugrizYJHK$-band were 6646, 2833, 3951, 5610, 25338, 5315, 13620, 59374 and 65015 counts, respectively.}. We also cut the images into overlapping tiles, which makes the data easier to handle in \galapagos.

A second set of images is used within GAMA to carry out the aperture photometry used in Section \ref{sec_cmd}. In order to derive aperture matched photometry, all images have been blurred to a common PSF size of 2 arcseconds FWHM, creating a homogeneous dataset and avoiding artificial colour biases that would be introduced by the individual seeings in the 9 observed bands.

\subsection{Profile fitting}
\label{sec_cleaning}

We used our multi-band version of \galapagos (itself utilising \galfitm), to create several catalogues for all objects detected in a small subregion ($\sim1.3\degr \times 1.3\degr$) of the 9-hour GAMA field (G09) using the same versions of our codes to ensure maximum compatibility of the results. In all runs on the data, we have used the same setup as closely as possible, including \sex setup. However, the imaging data used for object detection varies. For each single-band fitting run, we have used the same single-band image for object detection, using an identical \sex setup for each (Mode\_S1 in Table \ref{tab_modes} and throughout this paper). For our multi-band fitting run (Mode\_M), we used a co-added image that was created by simply adding all individual images without further normalization\footnote{All GAMA data provided is normalized in the AB system, co-adding the images is the correct way of combining the different images in an energy driven fashion.}. Given the background level and noise of the data, this creates a slightly biased image towards red bands (e.g. faint, $u$-band-only detections may vanish in the noise from the $K$-band image), but by adding up the images, we increase the \sn for each pixel in the image, allowing for a very deep and detailed object detection. Most very blue objects, that would not be detected in red bands, are bright enough to still be detected in the co-added image.  Again, the same \sex setup is used for this co-added image.  However, the number of detected objects increases dramatically compared to all individual bands (see numbers in Table~\ref{tab_1}).
\begin{table}
\centering
\caption{Summary of the 3 different fitting modes used in this paper.}
\begin{tabular}{@{}llrr@{}}
\hline
Mode    &    \#\galapagos runs & Method of detection & Method of fitting\\
\hline
\hline
Mode\_S1 & 9 (1 run each band) & individual bands & single-band \\
Mode\_S2 & 9 (1 run each band) & co-added image & single-band \\
Mode\_M & 1 & co-added image & multi-band\\
\hline
\label{tab_modes}
\end{tabular}
\end{table}

Using this arrangement, we run the new codes on each single-band ($ugrizYJHK$) individually (Mode\_S1) and one multi-band dataset (Mode\_M). Bear in mind that, in the multi-band approach, only the detection uses a co-added image, the fitting process utilises all nine bands individually, though simultaneously. The resulting ten catalogues (nine single-band, one multi-band) from these runs are then matched using a simple RA/Dec.\ source correlation and all parameters are copied into the one single catalogue that is used throughout this section. For consistency checks and to check whether pre-existing knowledge about additional neighbouring objects gives Mode\_M fitting an unfair advantage over Mode\_S1 fitting, we have also run all nine single-band fits using detection on co-added images (Mode\_S2). The results of this test are discussed in \S~\ref{sec_real_results} and \S~\ref{sec_sim_results}.

Along with additional information, the resulting catalogue contains, for each mode:
\begin{itemize}
\item all \sex output
\item all setup values for \galfit
\item all fitting values, including $\chi^2$ values and fitting times
\item all uncertainties for the fitting values
\item file names and folders
\item flags for neighbours, fitting status, constraints hit during the fitting process
\item software versions
\end{itemize}

Please note that all magnitudes in the catalogue and used throughout this paper are total-\sersic magnitudes as directly returned by \galfit, i.e., integrating the light profiles to infinity. It is perhaps more physically meaningful to consider magnitudes based on integrating the \sersic profile to some finite radius, e.g. $10 r_{\rmn{e}}$, but this is not done in this work.  We refer the reader to K12 for further discussion of this issue.

As our sample contains all detected objects instead of objects with $r < 19.8$ for which spectra have been obtained, redshifts are only known for the minority of objects in our sample. Requiring these would reduce our sample dramatically and restrict the analysis to a much smaller number of bright objects. Instead, for the majority of this paper, we work with apparent magnitudes and sizes, giving $r_{\rmn{e}}$ and other sizes in pixels.  This allows us to consider all objects measured by \galapagos.  Physical values are only used in the last sections of this paper where we restrict our consideration to those objects with known spectroscopic redshifts, but use a much larger area to create a catalogue for a sufficient number of galaxies.

Before examining the parameters from our \galapagos catalogue, it must first be cleaned in order to select only the objects that have been successfully fit by \galfitm. In particular, we wish to identify and discard fits with one or more parameters lying on (or very close to) a fitting constraint, as described in Section~\ref{sec_setup}.  Such a fit is unlikely to have found a true minimum in $\chi^2$ space and is indicative of a serious mismatch between the model profile and the object in question.  This also serves to remove stars from the catalogue, as explained in Section~\ref{sec_setup}. We keep all objects which meet the following criteria:
\begin{itemize}
\item $m_{\rmn{input}}-5<m<m_{\rmn{input}}+5$,
\item $0<m<40$,
\item $0.201<n<7.99$,
\item $0.301\ [pix]<r_{\rmn{e}}<399.0\ [pix]$.
\item $0.001 < q \leq 1.0$,
\item $95 - 5*mag\_best < fwhm\_image$ and $fwhm\_image<1\ [pix]$. These relations were found in the mag-size diagram to well separate saturated stars (unsaturated stars) from galaxies and is used within \galapagos\ for this purpose.
\item flag $= 2$.
\end{itemize}
The magnitude input, $m_{\rmn{input}}$, is the \sex MAG\_BEST for each object, with offsets where the multi-band detection image is used. This offset was empirically determined using previous results, to adjust (on average) magnitudes measured on the multi-band detection image to those for individual bands. The third to fifth criteria are slightly more restrictive versions of the fitting constraints used, the next criterion is aimed at separating stars from galaxies in the magnitude--size plane, and the last uses a flag that is returned by \galapagos.  This flag is used to keep track of the fitting status of objects and is initially set $0$ if the fit has not started/tried, $1$ if fit has been started -- e.g. it stays at 1 if the fit failed for some reason-- and $2$ if \galfit finished the fit and returned a result.

This cleaning is done on a band-by-band basis, i.e., the decision for each band is entirely independent of the others. For the multi-band catalogue, however, we apply these criteria to all bands simultaneously, i.e., if values of the fit fail to meet the above criteria for any band, the entire fit is considered unsuccessful (but not the entire polynomial is checked, e.g. interpolated values could in places violate these criteria). Hereafter, we refer to the objects in this cleaned catalogue as \lq successfully fit\rq\ or as having a \lq valid fit result\rq, to distinguish them from objects that were detected by \sex, but for which \galfitm \lq failed\rq\ to find a valid fit ($flag = 1$ and the objects violating the above criteria), and, in the case of simulated data, galaxies that were too faint to be detected at all. Tables~\ref{tab_1} and \ref{tab_2} summarize the numbers of objects and success rates. The reader should be advised here to be careful in interpreting this success rate as a true success rate (especially when comparing to success rates in K12). On the order of half the detected objects in SDSS/GAMA imaging are stars, e.g. we would both want and expect those to be filtered out by our catalogue cleaning, e.g. lowering the success rate to 50\%, although every single galaxy could have valid fitting results. Whereas K12 applies techniques to recover fitting values for objects that initially failed (e.g. by using different settings and re-running the fit), such a scheme is not present in our software. Although this could be introduced, it creates a less homogenous dataset and for the purpose of the single to multi-band comparison in the context of MegaMorph, we avoided this by simply ignoring these failed fits in our analysis.

\subsection{Results}
\label{sec_real_results}
After running \galapagos in both Mode\_S1 and Mode\_M, cleaning the catalogues and correlating the objects, using the same codes, same procedures and with a setup as similar as possible, we are now in a position to compare the single and multi-band techniques. Some example images, fits and fitting residuals for all bands are shown in the Appendix.

\begin{figure*}
\begin{center}
\includegraphics[width=0.98\textwidth, trim=20 60 50 0, clip]{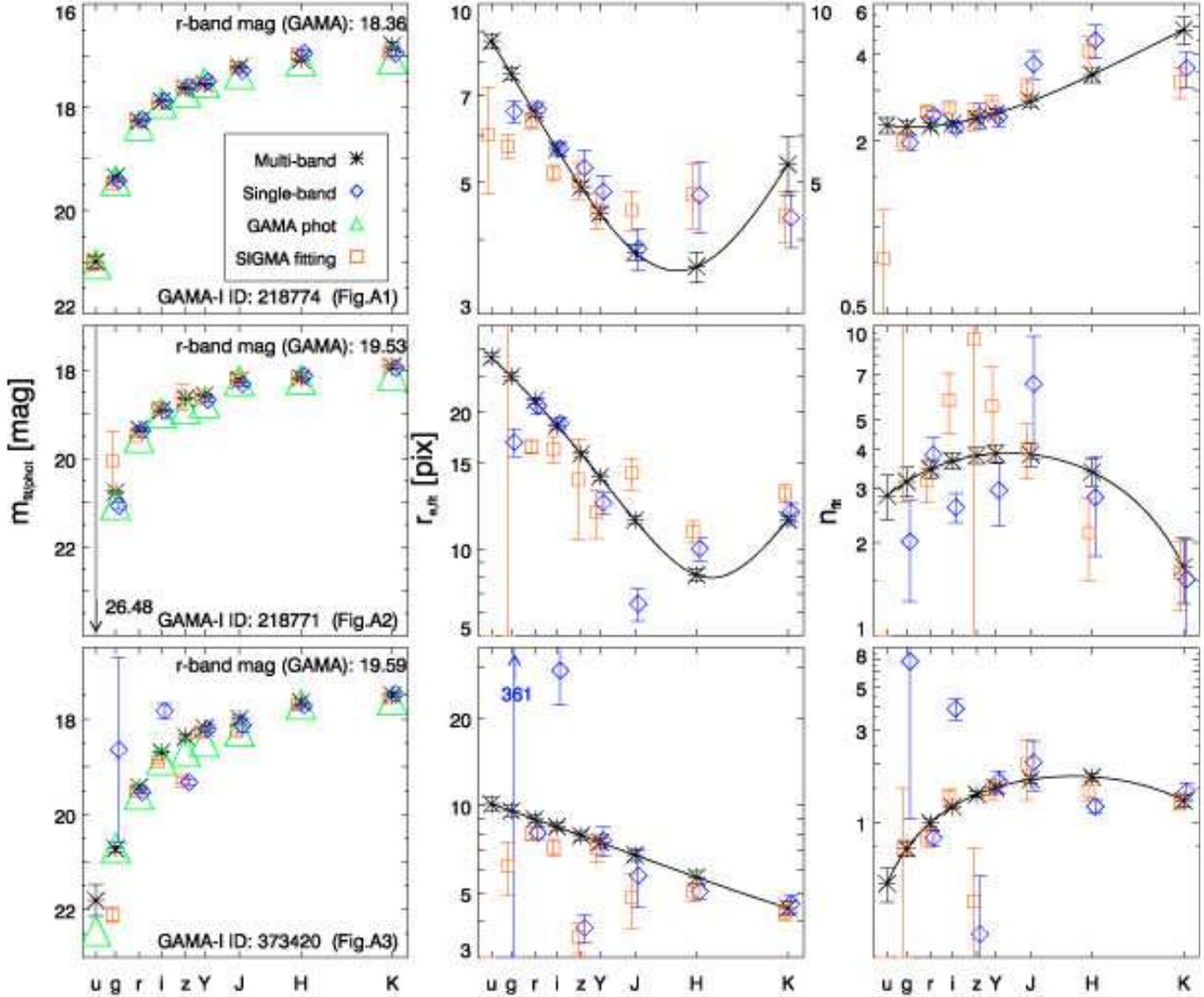}
\caption{Example fitting results for three individual galaxies as a function of wavelength. The images, models and fitting residuals of both single-band (Mode\_S1) and multi-band fitting (Mode\_M) are shown in Figs. \ref{figA1} to \ref{figA3} in the Appendix.  
We show recovered magnitudes on the left, where we compare to both GAMA (\sex MAG\_AUTO from catalogue ApMatchedCatv03; green triangles) values and single-band fits by (K12, orange boxes).
Black asterisks show multi-band results, blue diamonds show results from single-band fits (please note that single-band values have been slightly offset to the right, values from K12 slightly to the left, in order to make the error bars more visible). 
Error bars shown in all panels are parameter uncertainties as returned by \galfit. 
It is clear that single band fitting introduces larger scatter than multi-band fitting, particularly for fainter objects. 
In the middle column we show the same comparison for galaxy sizes. For comparison, we show the $r_{\rmn{e}}$ from the \sersic fits of K12. 
Here, by design, the multi-band fitting returns sizes which vary as a polynomial of second order with wavelength, indicated by the black line. 
In the right column, we show similar figures for \sersic index. 
Again, by design, multi-band fitting returns a smooth dependence over wavelengths. 
Multi-band fitting generally follows similar trends, but largely reduces the scatter and the error bars.
In all panels, please note that missing single-band points (both in our own analysis and in the values by K12) indicate that these fits were unsuccessful, and so no sensible value is available, e.g. all three of these objects have no single-band fit for the $u$-band in our own analysis and only the first object has a value by K12. Please see the text for more details.}
\label{fig0}
\end{center}
\end{figure*}

Figure~\ref{fig0} shows some fitting results as a function of wavelength for three of the objects in our real GAMA sample (top to bottom panels). 
The left column shows recovered magnitudes, as a function of wavelength, from both single and multi-band fitting, in comparison to GAMA photometric data and single-band fits performed by K12. 
The middle column shows the sizes recovered for the same galaxies, the right column shows \sersic indices. 
Please note that the x-axis in this figure -- and all figures throughout this paper -- shows linear scaling with wavelength. 
Although $log(\lambda)$ might be physically more meaningful, the scaling parameter in the fitting process was chosen to be linear with wavelength in this work and this should be resembled in the figures, e.g. a linear polynomial would only then appear linear in the figures. 
Similarly, the slightly distorted shape in the plots for size and \sersic results from the logarithmic scaling of the $y$-axis. 

One visible effect is that the magnitudes recovered by both fitting techniques are nearly always brighter than those from aperture photometry.
This offset is expected, as aperture photometry always misses some fraction of the light, whereas the magnitudes from \galfit integrate the profile out to infinity\footnote{For a more detailed discussion of this effect of \sersic\ profiles, please see external literature, e.g. \citet{GrahamDriver}}. 

It also becomes clear that even in case of bright galaxies, some of the single-band fits (e.g., $u$- and $z$-band in the second example), fail to return a valid result (with \lq valid\rq\ being defined in Section~\ref{sec_cleaning}). 
For fainter galaxies (e.g. the lowermost example), the success rate for single-band fitting decreases and the scatter increases with respect to multi-band and aperture-based results. 
Bear in mind that for magnitudes, we do not constrain the fitting values directly; the smoothness of the recovered SED is an indirect result of constraining the profile structural parameters, and not forced by direct constraints on the magnitudes themselves. 

For multi-band size measurements, by design, the multi-band fitting results lie on smooth curves, which greatly reduces the scatter in this parameter. 
Especially in the lowermost example, the single-band values vary strongly (and arguably un-physically, with a size difference between $g$- and $z$-band of nearly a factor of 100). 
Generally, for these relatively bright galaxies (chosen to be GAMA spectroscopic targets, with $r < 19.8$), the sizes from multi-band fitting follow the trends of the single-band results, but with a more physically realistic smoothness. 
Even for relatively bright galaxies, single-band sizes vary greatly from one band to the next, often by a factor of a few. 
Please keep in mind that the error bars shown in all these plots are parameter uncertainties as returned by \galfit, which have been shown to underestimate the true values (H07) and should be interpreted as a lower limit of the true uncertainty. 
While we do not believe these error bars to be realistic, they do allow a comparison between single and multi-band fitting. 
However, we would possibly not consider the upturn towards K-band sizes real. 
Especially these galaxies show only individual examples and do not represent the population as a whole. 
An upturn in the K-band sizes in the entire population is not found, see e.g. Fig.~\ref{fig4}.

A comparison for \sersic indices is shown in the right column of this figure, including a comparison to values of K12 where they exist. 
A comparison to other values from the literature is difficult because no such values exist for most of our objects. 
Generally, a trend from lower $n$ in blue bands to higher $n$ in red bands is visible for most objects in our sample.
We will investigate the recoverability of \sersic indices more in Section \ref{sec_sims}, where a true value is known and an analysis is both easier and more thorough.

In the figures of size and \sersic index $n$ versus wavelength (middle and right column of Fig.~\ref{fig0}), we not only show the individual band sizes for the \galfitm multi-band fit results, but we also show the full polynomial function ($\tilde{p}_i(\lambda;\, \{q_{i,\cdot}\})$, c.f. Section~\ref{sec_cheb}) as a black line. An elegant side-effect of fitting these polynomials, rather than values specific to the wavelength of each band, is that it allows easy estimation of sizes (as well as other parameters) at intermediate wavelengths.  For scientific analyses, one often wishes to compare restframe parameter values.  The polynomial parameter functions, inherent to our multi-band approach, provide a simple way to determine these, with greater accuracy than could be obtained by interpolating between single-band values.
However, as discussed in Section~\ref{sec_setup}, we cannot take the same approach with our magnitudes, because the high-order polynomials used for these suffer from Runge's phenomenon.  This issue, and ways around it, are discussed in Paper I.  In this work, we take the conventional approach and treat the magnitudes as discrete values, and determine restframe values via SED template fitting using \textsc{kcorrect} \citep{kcorrect}.

\begin{figure}
\begin{center}
\includegraphics[width=0.50\textwidth, trim=55 30 10 30, clip]{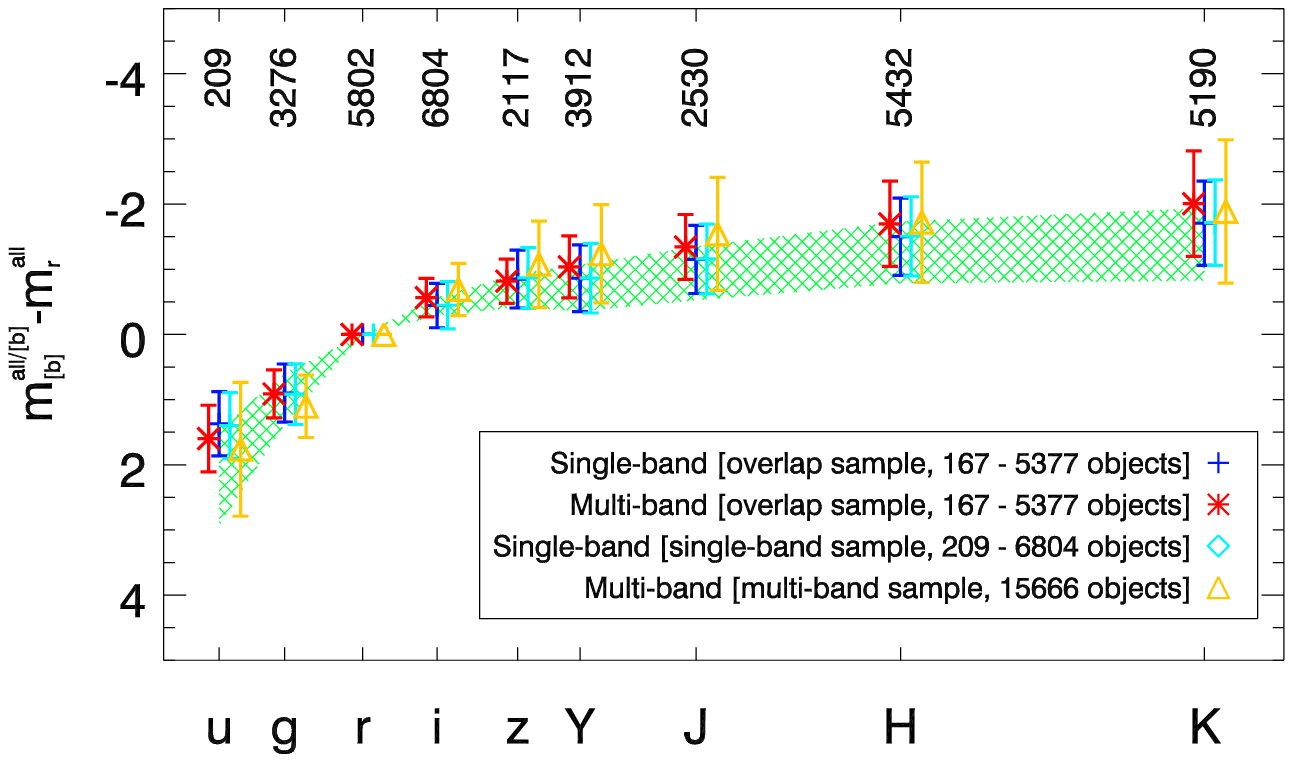}
\includegraphics[width=0.50\textwidth, trim=55 30 10 30, clip]{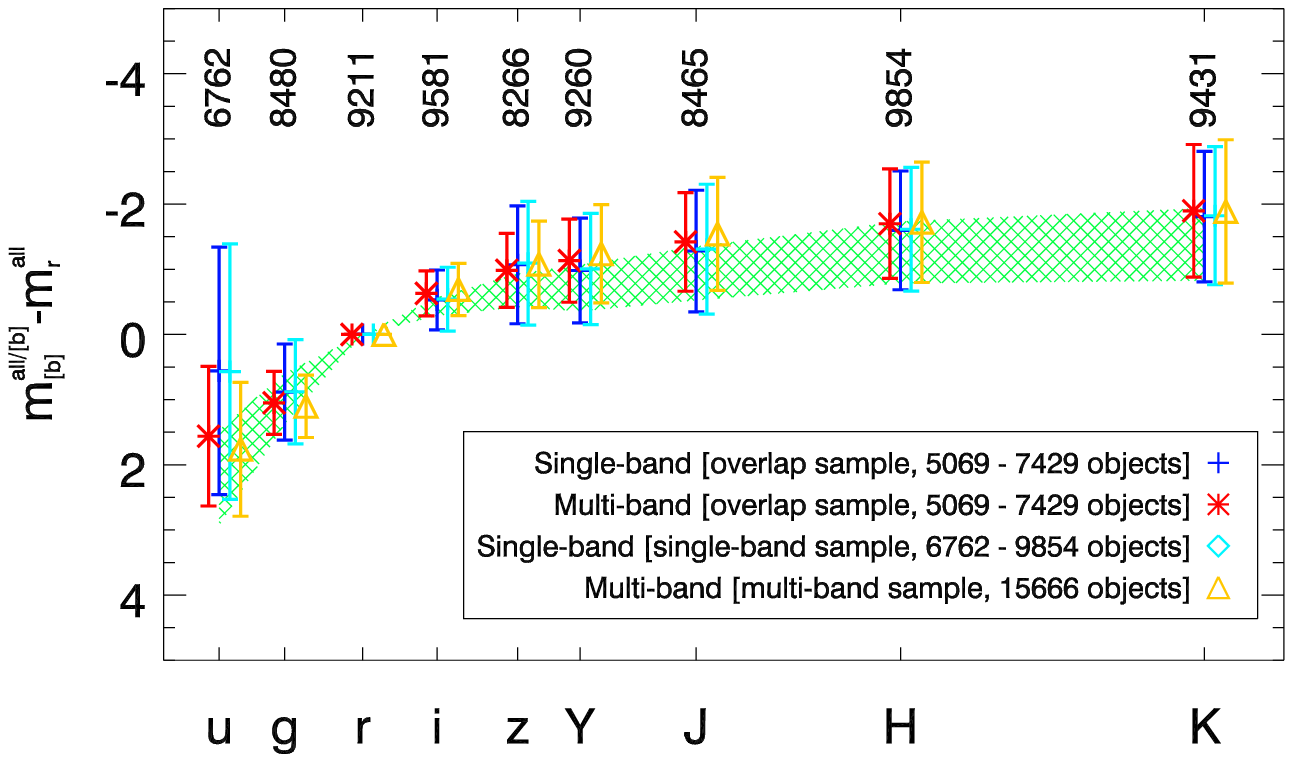}
\caption{The average normalised SED for various methods and samples. 
The upper figure shows Mode\_S1 as blue data points, the lower figure uses Mode\_S2 (increasing sample sizes, but introducing additional scatter), please see text for discussion. 
All input SEDs are normalised to zero magnitude in the $r$-band. 
Dark-blue crosses and red asterisks indicate the average SED from single and multi-band fits to an overlapping sample, in which each galaxy has valid fitting results from both techniques. 
Light-blue diamonds display the same trend for all single-band fits, irrespective of whether the multi-band fitting result and especially the other single-band fits in the other bands were valid. 
Conversely, orange triangles show the SED for all multi-band fits, irrespective of whether the single-band fit returned a valid result. 
Points other than dark-blue are slightly offset in wavelength for readability of the figure. 
The error bars on each point show the rms scatter of the normalised SEDs (resistant mean, iteratively clipped at 3 sigma).  
The numbers above the points for each band refer to the number of valid single-band fits for that band (light blue sample).  
The numbers in the legend give the ranges of the other samples.  
Note that multi-band fits are either valid for all bands or for none at all, so the multi-band sample contains the same number of objects in each band. 
For comparison, the green shaded area shows the normalised SED, and its rms scatter, for GAMA aperture photometry of 972 GAMA objects that were identified in the same survey area, e.g. they would be a subset of galaxies in our total sample. 
As our samples do contain fainter objects, not targeted by GAMA, no perfect match is expected.
In the lower panel, we can see that the improvement of Mode\_M versus Mode\_S2 appears very large. 
However, please see the text for further discussion.}
\label{fig1}
\end{center}
\end{figure}

In a similar manner to the individual examples shown in Fig.~\ref{fig0}, in Fig.~\ref{fig1} we show trends for magnitudes recovered using both single (both Mode\_S1 and Mode\_S2) and multi-band fitting for our entire sample of real galaxies. Lacking \lq true\rq\ values, we cannot show the offset and scatter of the two methods, instead we consider the average SED. All the individual galaxy SEDs were normalized to an $r$-band magnitude of zero before averaging, to minimize the scatter due to different galaxy brightnesses. As a comparison, we show the average SED for a bright galaxy sample based on aperture photometry from the GAMA survey, normalized in the same way. This comparison, while not being perfect due to differences in the samples shown, gives an indication of the intrinsic scatter in galaxy SEDs. 

In the upper panel of Fig.~\ref{fig1}, we show the comparison between Mode\_S1 fitting (e.g. only using single-band data for the entire process, including object detection) and multi-band fitting. In the lower figure, we show what happens when \emph{multi-}band detection is used for \emph{single}-band fitting (Mode\_S2). 

First, we will discuss the upper figure here. Overall, both single and multi-band fitting show the same trend. Both results show slight offsets with respect to the general GAMA SED (as determined by aperture photometry, see discussion below), as derived from 972 GAMA objects identified in the region.
Most of the scatter in the normalised SEDs is due to intrinsic variation between the galaxies. Lacking a \lq true\rq\  comparison value makes it difficult to make more stringent tests, but there are hints that the scatter (for the same sample please compare dark-blue and red data points in Fig.~\ref{fig1}) is slightly reduced in most bands when multi-band fitting is used (especially in low \sn bands, $u$ and $z$).
The normalised SEDs for the entire multi-band sample (orange) shows larger scatter and offsets even when compared to the full single-band sample (light blue). This is a result of the multi-band sample containing fainter galaxies than the others. Using real data, it is not clear whether the increased scatter compared to the general GAMA SED is a result of worse fitting results, or real variations that are not reflected in aperture photometry. However, this effect will be examined using simulated data without such scatter in Section~\ref{sec_sims}.

The most dramatic advantage of multi-band fitting becomes apparent when comparing the sample sizes of galaxies for which parameter values can be derived. Whereas the single-band fits return valid results for between 209 ($u$-band) and 6804 ($i$-band) galaxies, the multi-band fitting returns valid values in \emph{all} bands for 15666 objects. However, there are two effects at work here; the number of valid fits in each band depends upon both the number of objects detected and the fitting success rate. The former is a result of the chosen detection image and \sex setup. The Mode\_S1 results in this figure use single-band detections, while the multi-band results are based on detections on a co-added (and hence deeper) image. Table~\ref{tab_1} gives the number of objects detected in the imaging for each band, and the number of those objects which are successfully fit by single-band \galapagos.  It also shows the resulting numbers of objects with valid single-band measurements for every band ($ugrizYJHK$), or just the six highest \sn bands ($griYHK$).  Finally, Table~\ref{tab_1} gives the number of objects detected in the co-added multi-band detection image, and the number of these with valid multi-band fits (and hence meaningful measurements in all $ugrizYJHK$ bands).
While the number of detections plays an important role (e.g, the number of $u$-band detected sources is 20 per cent of that for $r$-band), the fit success rate is significantly higher for multi-band than for any of the individual single-band fits, and much greater when one requires complete multi-band data.
While of course benefiting from detecting more objects, this shows that the multi-band approach is more stable and thus more often returns valid measurements that are more likely to resemble the true parameters of the galaxy.

The substantial difference in the number of objects returned by Mode\_S1 and Mode\_M methods is partly due to the initial object detection.  The multi-band fits are based on a multi-band detection image, and hence fits are attempted for many objects which are undetected in some of the single-band images, especially $u$-band.  It is possible that single-band fitting may be able to return meaningful values for these objects, if it were to be aware of their presence.
To investigate this, we have repeated the above analysis, but using the multi-band detection image even when fitting single-band data (Mode\_S2). 
In addition to making the single-band method attempt to fit more targets, the additional objects will result in differences in deblending, masking and starting parameters. 
This potentially gives single-band fitting a better chance of measuring reliable galaxy parameters.

We will discuss this test in more detail in Section~\ref{sec_sims} where performance comparison is easier.  
Here we only show an example for magnitude, in the lower panel of Fig. \ref{fig1}. 
Two effects are strikingly evident. 
Firstly, Mode\_S2 fitting now indeed does return valid fitting results for many more objects (as apparent from the numbers at the top of the figure). 
The samples are still smaller than for multi-band fitting, indicating that \galfit still fails to return a valid fit more often than for multi-band fitting. 
Secondly, those objects that the code does succeed on have a much larger scatter than the multi-band results (compare dark blue to red error bars, which show the scatter for an identical sample of objects). 
While the use of multi-band detections does greatly increase the sample sizes, the scatter in the single-band fitting results has increased dramatically as well, especially in the $u$- and $z$-bands. 
This confirms that the improvement in fitting quality and sample size when using multi-band fitting does indeed result from the strength of the multi-band fitting approach presented in this paper, and not simply due to the advantage of using a multi-band detection image. 
Our multi-band fitting technique can recover reliable measurements in bands where the objects are too faint to be reliably fit (using a single-band method) or even detected.

The relatively poor behaviour of the single-band fits to multi-band detections can be attributed to a number of potential causes.  
The simplest explanation is that without the constraints on the size and shape of the profile, which naturally come from higher \sn bands in multi-band fitting, single-band fits produce more uncertain results in the low-\sn bands. 
However, there are two additional effects that may be at work. 
Firstly, given that the fitting position of the profiles is only constrained to be within the postage stamp, potentially one or more neighbouring (secondary) objects may \lq wander off\rq, away from their intended targets (which may be undetectable in the single band) and on to the (primary) target object. 
Such behaviour would lead to a fainter magnitude being returned for the primary, as its flux is distributed between multiple profiles. 
Secondly, the opposite is possible. 
When the primary source is much fainter than any of the secondaries, or even invisible in the single-band image that is being fit, the primary profile may \lq wander off\rq\ to settle on a secondary object. 
While the stacked profiles would split the secondary flux, the resulting magnitude might be brighter than that of the true primary source. 
We did not further investigate which of these effects dominates, as \galapagos does not return the fitting values of secondaries and this investigation is beyond the scope of this paper. 
In the case of multi-band fitting, these issues becomes much less significant as the profile position is effectively constrained using information from all the bands.

Our Mode\_S2 fitting results could hence potentially be improved by either imposing tighter constraints on the positions, thus preventing profiles from \lq wandering off\rq\ their intended target, or by using profile information from one fit (e.g., on the $r$-band) to constrain or fix parameters in a subsequent fit on a lower-\sn band. 
However, this is not as natural nor effective a solution as the multi-band approach we advocate in this paper.

The last thing to note from our detection-image test is that some of the objects that are missing from our Mode\_S1 fit results, but which are recovered in Mode\_S2 fits, have relatively bright magnitudes. 
These may be objects with low surface brightness, which are undetected in some of the single-band images, particularly $u$-band, despite their integrated brightness.  
Furthermore, even when single-band fits are aware of these objects in Mode\_S2, these fits frequently fail to extract meaningful information, in contrast to multi-band fitting.  
The implication of this is that, without taking the multi-band approach, we preferentially lose information about galaxies with less-peaky profiles, i.e., disks.

\begin{figure}
\begin{center}
\includegraphics[width=0.50\textwidth, trim=55 30 10 30, clip]{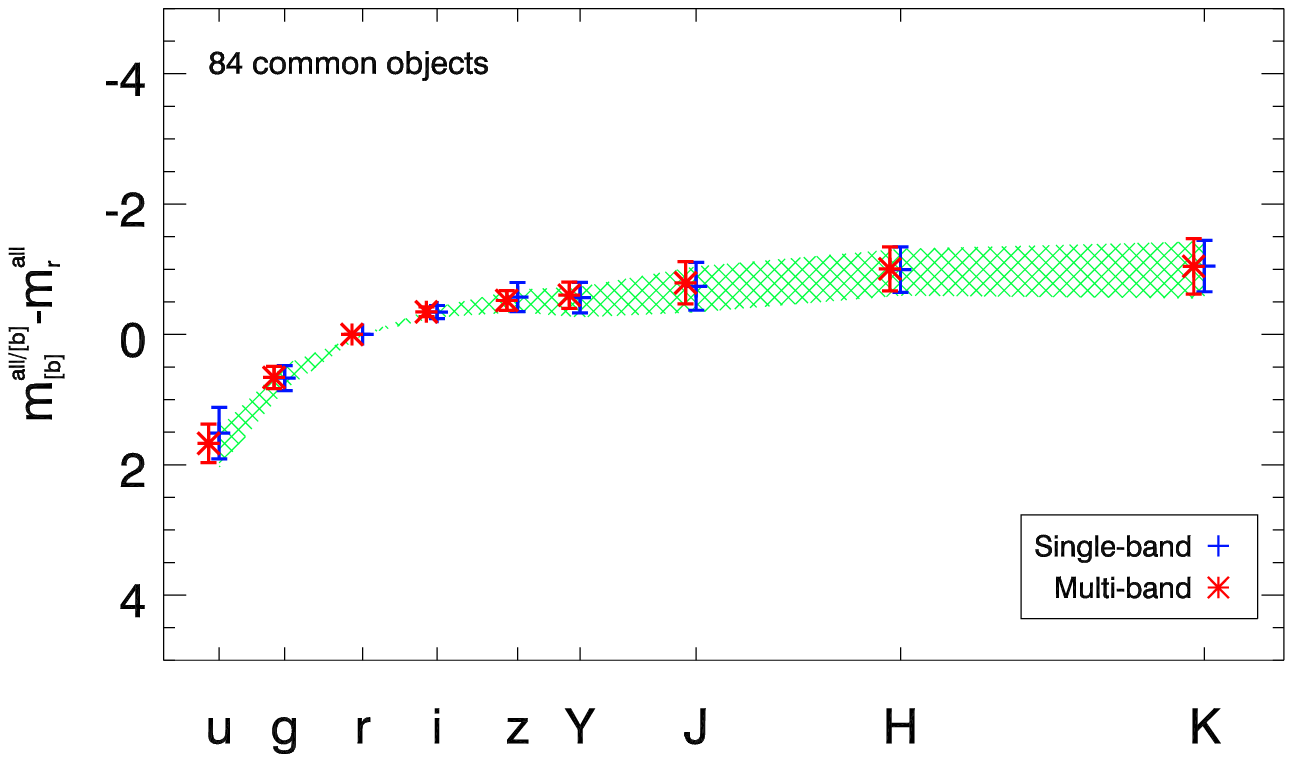}\\
\includegraphics[width=0.50\textwidth, trim=55 30 10 30, clip]{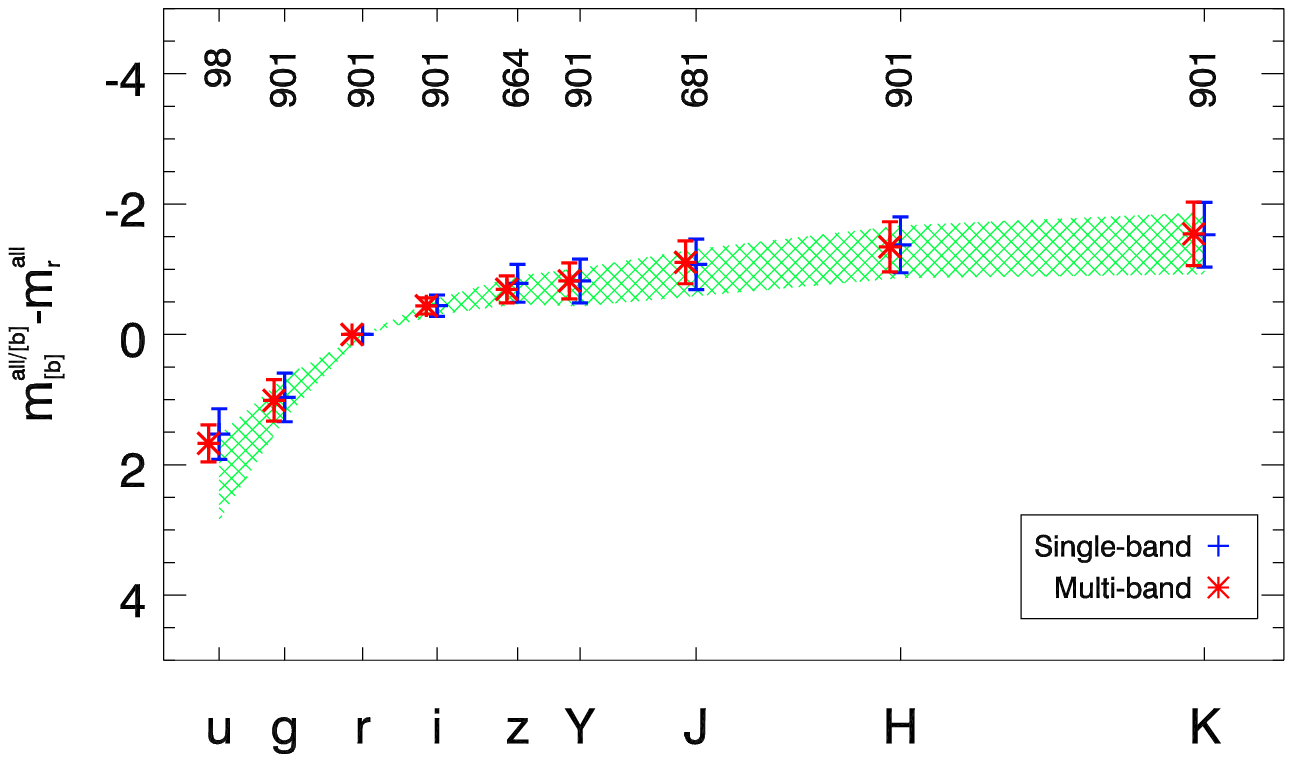}
\caption{Similar to Fig.~\ref{fig1}, these figures show the average normalised-SED and the scatter around that average. Both figures use Mode\_S1 data, but now for a small sample of 84 galaxies for which \emph{all} single-band and multi-band fits return a valid results (top panel) and a sample of 901 galaxies for which the $griYHK$ bands return valid results (bottom panel). All points within each figure therefore incorporate the same objects (unless numbers indicate differently in the lower panel). The shaded green areas in each figure show normalised-SEDs based on GAMA aperture photometry, using the galaxies in our samples for which these values exist. This sample hence forms a very similar sample to the ones used in the plotting symbols, which explains the much better agreement between points and shaded green area compared to Fig~\ref{fig1}.}
\label{fig2}
\end{center}
\end{figure}

Figure~\ref{fig2} is similar to Fig.~\ref{fig1}, but only considers two stringently-defined subsets: 84 galaxies for which \emph{all} single band fits returned a valid result, and 901 galaxies for which we obtained valid values in the six highest \sn ($griYHK$) bands. This makes the comparison between the codes much easier and cleaner, as the same sample of galaxies is compared at all times. Both the Mode\_S1 and Mode\_M fitting results closely agree with one another and the GAMA aperture photometry.  However, while subtle, it is also apparent that multi-band fitting slightly reduces the scatter for most bands.  This is shown more clearly using simulations in Section~\ref{sec_sim_results}.

We do not explicitly show a figure similar to Fig.~\ref{fig2} for Mode\_S2 results. However, we can show from our analysis discussed above, that, while again the sample size increases when multi-band detection is used, the scatter of the single-band fitting results increases dramatically.

\begin{table*}
\centering
%\begin{minipage}{140mm}
\caption{The number of objects in the real dataset that are detected and successfully fitted, and the corresponding success rate, for each single-band image (only Mode\_S1), for the Mode\_S1 results combined, and for the multi-band (Mode\_M) technique. \lq Combined\rq\ in this context does not mean detection on the co-added image (Mode\_S2), but combination of all the single-band detections and fit results to create a sample of all detections and successful fits when using $ugrizYJHK$-bands or $griYHK$-bands, respectively.
The reader should take care in interpreting this success rate as a true success rate. On the order of $50$ per cent detected objects in SDSS/GAMA imaging are stars, that are thrown out during the cleaning of the catalogue. This lowers the success rate dramatically, even if all galaxies had valid fitting results. An interpretation of these rates in an absolute sense should take this effect into account, but is not given in this analysis. The fraction of stars in the detections is possibly higher in the shallower bands due to their higher detectability, which would partly explain the lower success rates in these band. As a reliable star/galaxy classification is difficult in order to correct these numbers, we instead show the same values on the right side, but restrict the sample to targeted GAMA galaxies, e.g. no stars and only galaxies with $r<19.8$. Success rates are indeed much higher when only this bright galaxy sample is considered. Comparing the numbers at the bottom of the table, we can successfully derive all fitting values for 885 GAMA objects out of the 1251 GAMA objects that were detected when using multi-band fitting. Requiring all $ugrizYJHK$ fitting results from single-band fitting reduces this sample to 75 objects.}
\begin{tabular}{@{}l|rrr|rrr@{}}
\hline
   &  \multicolumn{3}{c|}{all objects} & \multicolumn{3}{c}{GAMA objects}\\
Band    &    \#objects fit        & \#objects detected & Success rate & \# fit & \# detected & Success rate\\
\hline
\hline
u & 209 & 3773 & 5.5\% & 108 & 281 & 38.4\%\\
g & 3276 &13547 & 24.2\% & 827 & 1229 & 67.3\%\\
r & 5802 & 19169 & 30.3\% & 896 & 1237 & 72.4\%\\
i & 6804 & 22059 & 30.8\% & 907 & 1235 & 73.4\%\\
z & 2117 & 12496 & 16.9\% & 725 & 1178 & 61.5\%\\
Y & 3912 & 16200 & 24.1\% & 859 & 1221 & 70.4\%\\
J & 2530 & 13289 & 19.0\% & 686 & 1150 & 59.7\%\\
H & 5432 & 17750 & 30.6\% & 885 & 1212 & 73.0\%\\
K & 5190 & 16117 & 32.2\% & 770 & 1171 & 65.8\%\\
 & & & & & & \\
\textbf{combined single-band ($ugrizYJHK$)} & \textbf{87} & \textbf{24995} & \textbf{0.35\%} & \textbf{75} & \textbf{1251} & \textbf{6.0}\%\\
\textbf{combined single-band ($griYHK$)} & \textbf{992} & \textbf{24866} & \textbf{4.0\%} & \textbf{544} & \textbf{1251} & \textbf{43.5}\%\\
\textbf{multi-band} & \textbf{15666} & \textbf{29205} & \textbf{53.6\%} & \textbf{885} & \textbf{1251} & \textbf{70.7}\%\\
\hline
\end{tabular}
\label{tab_1}
%\end{minipage}
\end{table*}

The sample sizes in Fig.~\ref{fig2} highlight an important problem with single-band fitting, at least in the simplest case of treating each band completely independently, as we do here.  As detailed in Table \ref{tab_1}, we only obtain meaningful parameter values in \emph{all} bands for 87 objects (992 objects in the case that only $griYHK$ values are needed).  With multi-band fitting, we derive valid parameters in all bands for 15666 objects, an increase in sample size of a factor of $\sim180$ ($\sim16$), and obviously a valuable advantage.  As mentioned above, this is partly due to the use of a multi-band detection image (which can only be reliably used with multi-band fitting), and partly due to an increase in the fitting success rate, thanks to the stability achieved by constraining the profile shape across multiple bands. The advantage of the multi-band approach becomes especially evident in science cases where parameter values are required in several bands, which, in case of single-band fitting, decreases the sample size dramatically. While multi-band fits derive valid values for $\sim54$ per cent of detected objects ($\sim71$ per cent of galaxies considering only a sample of bright GAMA galaxies), a combined single-band approach only recovers valid parameters for $<0.4$ per cent of these objects (6.0\% of bright galaxies), and $\sim4$ ($\sim 44$) per cent when only values in the higher \sn bands ($griYHK$) are needed. Such an science case is discussed in Section~\ref{sec_cmd}.

\begin{figure}
\begin{center}
\includegraphics[width=0.50\textwidth, trim=45 30 10 30, clip]{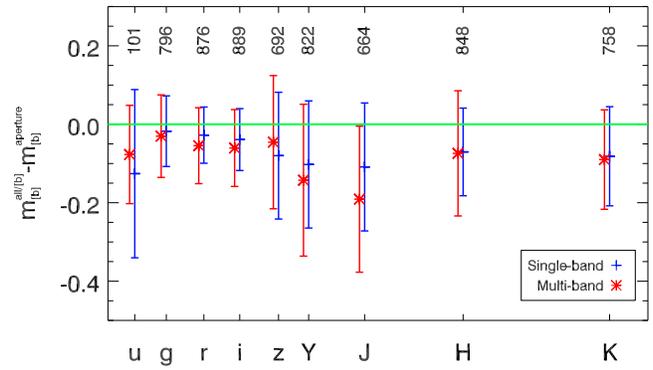}
\caption{The difference between magnitudes obtained from \galfit (blue crosses: Mode\_S1, red asterisks: Mode\_M) and aperture photometry measured by the GAMA survey, using \sex. We show the figure for a sample of galaxies with $r < 19.8$ for which, for each filter band, all of these measurements exist. We recover fainter magnitudes below the green line, and brighter magnitudes above.
At first glance, the multi-band fitting results seem slightly worse, returning significantly brighter magnitudes. We will demonstrate, however, in Section~\ref{sec_sims}, specifically Fig.~\ref{fig7.2}, that \sex magnitudes are bad proxies for this kind of test and that the multi-band fits are better at recovering true total magnitudes.}
\label{fig3}
\end{center}
\end{figure}

The normalised SEDs, given in Figs. \ref{fig1} and \ref{fig2} allow us to evaluate the band-to-band variation of our measurements, and compare these to aperture photometry.  However, the scatter in these SEDs are dominated by intrinsic variations in the galaxy population.  We therefore perform a further comparison of our \galfit magnitudes directly against GAMA aperture photometry, a test that would be carried out by most authors as performance tests of profile fitting codes. The results of this test are shown in Fig. \ref{fig3}\footnote{This figure looks very similar if we use magnitudes from the individual \sex runs performed for object detection in \galapagos.}.

This comparison in Fig. \ref{fig3} is only possible for a subset of galaxies in each band, those which have valid fitting results for the method in question.
As previously mentioned, some offset is expected as aperture photometry will always miss some fraction of the light from the outskirts of the galaxies. On the other hand integrating galaxy light profiles out to infinity can add fictional flux particularly for high-$n$ systems, making the definition of a true value difficult. The \galfit magnitudes are indeed systematically brighter than aperture photometry by $0.1$--$0.2$ mag. The scatter between \galfit and aperture magnitudes is also $0.1$--$0.2$ mag, depending on typical \sn of the band, suggesting that much of this scatter could result from variations in the flux missed by aperture photometry for galaxies of different profile shapes.
The single and multi-band fitting results appear quite similar, although the multi-band magnitudes generally display slightly greater offsets. However, from this figure, it is difficult to determine which magnitudes best represent the total light and colours of the target galaxies.  We explore this in a more satisfactory manner in Section~\ref{sec_sims}, with the use of simulations.

\begin{figure}
\begin{center}
\includegraphics[width=0.50\textwidth, trim=45 30 10 30, clip]{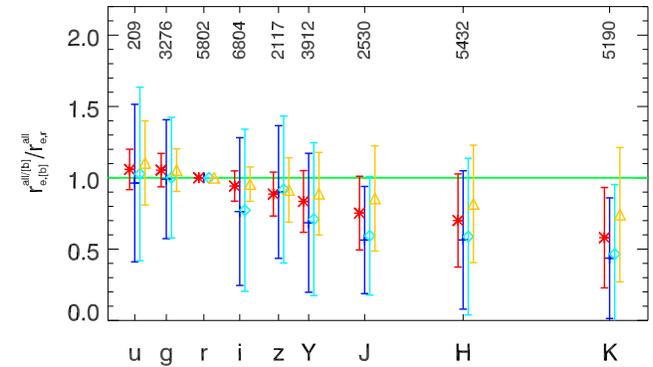}
\caption{A comparison of the trend in measured half-light radius versus wavelength for single and multi-band fitting. Symbols and colours indicate the same samples as in Fig.~\ref{fig1}. The half-light radii for each galaxy are normalized using the $r$-band, and the average variation in relative size versus wavelength is shown by the points. We recover larger half-light radii above the line, and smaller below. The rms scatter around this trend is indicated by the error-bars. Both single and multi-band approaches reproduce no significant average trends, but clearly the multi-band technique reduces the scatter substantially.}
\label{fig4}
\end{center}
\end{figure}

Magnitudes are integrated quantities, and are therefore generally the easiest galaxy properties to extract from imaging.
Another important galaxy property is size.  Size is a natural, direct product of \sersic profile fitting.  It may be defined in many ways, but a convenient and commonly used value is the half-light radius. This is the size measure returned by \galfit.

\begin{figure}
\begin{center}
\includegraphics[width=0.50\textwidth, trim=45 30 10 30, clip]{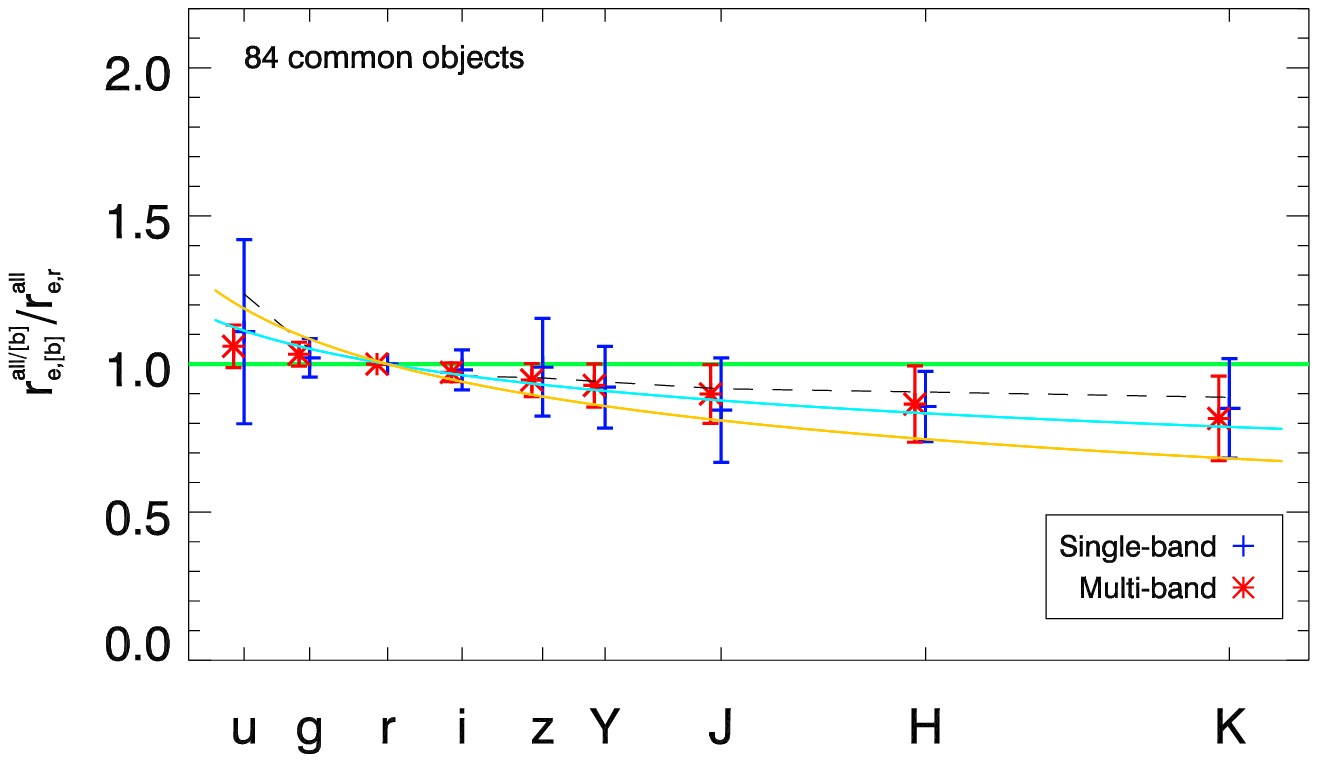}
\includegraphics[width=0.50\textwidth, trim=45 30 10 30, clip]{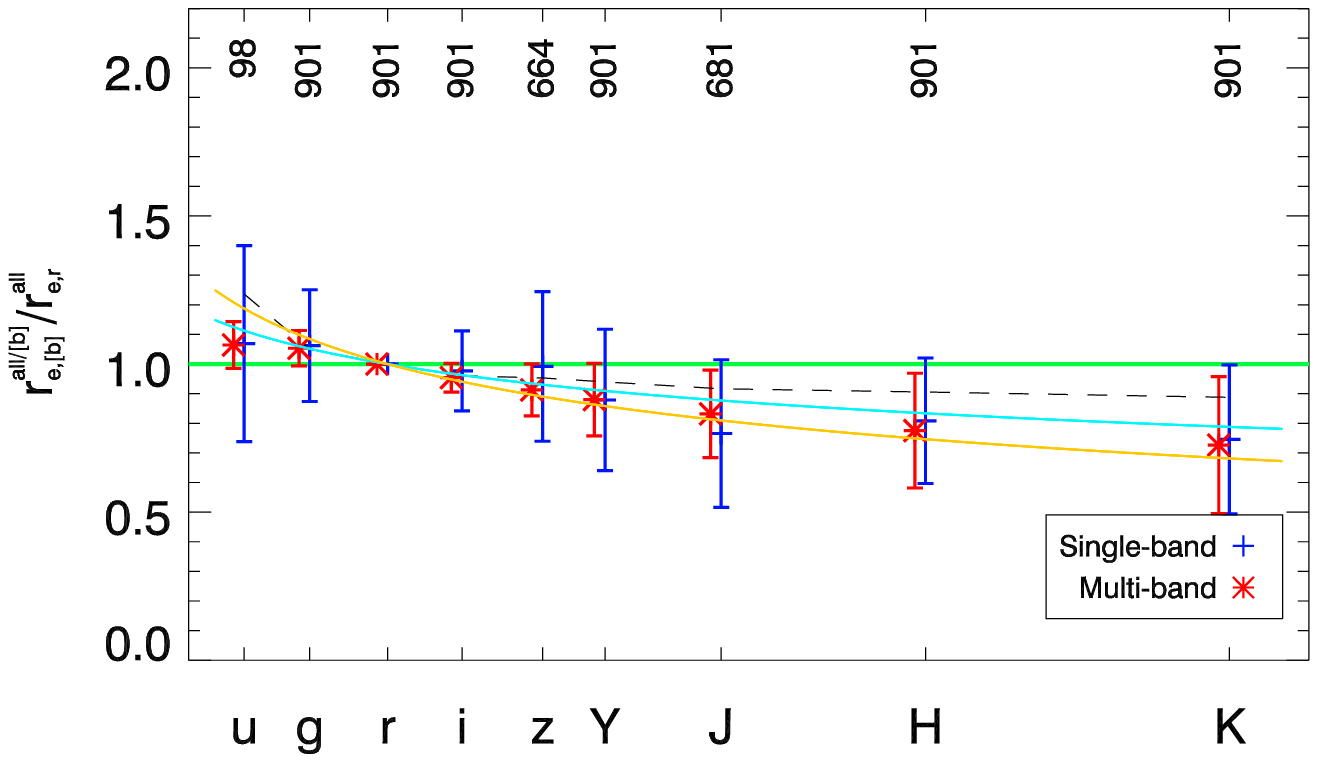}
\caption{Similar to Fig.~\ref{fig4}, these figures compare trends of half-light radius versus wavelength, but now for (top panel) a small sample of 87 galaxies for which all Mode\_S1 and Mode\_S1 fits return a valid results and (bottom panel) a sample of 901 galaxies for which the $griYHK$ bands return valid results in Mode\_S1. All the points within each figure therefore correspond to the same objects. The improvement in scatter when using multi-band fitting is especially evident in this figure. For comparison we show the relations described in K12 for spheroid-dominated galaxies (orange) and disk-dominated galaxies (light blue) which describe the same trends found in our data. The black dashed line shows the effect of dust content in the galaxy on the size measurements (while ignoring intrinsic stellar population gradients) from Pastrav et al (in prep), based on the radiation transfer model of \citet{Popescu}.}
\label{fig5}
\end{center}
\end{figure}

We follow the same approach as for magnitudes (c.f. Figs.~\ref{fig1} and \ref{fig2}).  Figures~\ref{fig4} and \ref{fig5} show the trend in galaxy half-light radii, normalised to the $r$-band, as a function of wavelength \citep[similar trends have been reported by][see Fig.~\ref{fig5}]{Kelvin2012}.  
Error bars indicate the scatter in these normalised sizes. Figure~\ref{fig4} compares the Mode\_S1 and Mode\_M results using samples of valid objects for each technique, and an overlap sample, containing only objects with valid results in both techniques (dark blue and red datapoints). 
Figure~\ref{fig5} limits the comparison sample further to those for which single-band fits for \emph{all} bands, or the six highest \sn ($griYHK$) bands, returned a valid result.
In contrast to magnitudes, for which the scatter is only slightly improved, the scatter on the recovered sizes are hugely improved by multi-band fitting.  
These figures also display an average trend, recovered by both techniques, such that galaxies appear smaller in redder bands.  
This can be explained by two processes. 
Firstly, blue bands galaxy light is typically dominated by galaxy disks, whereas in red bands bulges, which are usually smaller than disks, contribute a greater fraction of the light (see magenta and orange lines in Fig.~\ref{fig5}). 
Secondly, a combination of dust and stellar populations in pure disk galaxies can create a similar effect (\citealt{Moellenhoff,2012IAUS..284..306P}; Pastrav et al., in prep., see dashed line in Fig.~\ref{fig5} for the effects of dust on the measured half-light radius).

The reduction in scatter we see in Figs.~\ref{fig4} and \ref{fig5} is at least partly by design, as the half-light radii are constrained to lie on a second order polynomial as a function of wavelength.  Such a reduction in scatter would therefore be seen even if the real sizes of galaxies did vary wildly with wavelength.  However, from inspection of the single-band fits, i.e. figures like Fig.~\ref{fig0} for many more examples, we see that sizes do show a smooth, and nearly linear, trend with wavelength.  Consideration of physically realistic stellar populations also supports smooth variations in half-light radius with wavelength.  Such was our justification for using second order polynomials in the first place.  As we lack an independent estimate for the \lq true\rq\ sizes of the objects, we can not investigate this further with real data.  However, in the following section we use simulations to demonstrate that the multi-band approach results in significantly improved size measurements, suggesting that the reduced scatter in Figs.~\ref{fig4} and \ref{fig5} corresponds more closely to the true variation in relative size with wavelength.

For comparison, we illustrate the trends described in K12 for both spheroid-dominated and disk-dominated galaxies in Fig.~\ref{fig5}. Given the clean sample in this figure, which mostly consists of bright galaxies, that would also be targeted by K12, we can confirm to find very similar trends of $r_{\rmn{e}}$ with wavelength.

Determining galaxy size may be thought of as finding a radius for which the integrated flux within and without corresponds to a given ratio.  The behaviour of the profile within each of those two regions does not matter, and there is no ambiguity in the definition for any arbitrary monotonic profile. Determining \sersic index requires further information, particularly regarding the behaviour of the profile at its peak and in its tail.  Higher \sersic indices imply a greater shift of flux from around the half-light radius and into both the centre and outskirts of the profile. However, these pieces of information need not be consistent for an arbitrary monotonic profile. The \sersic index is therefore more difficult to measure accurately and consistently.

\begin{figure}
\begin{center}
\includegraphics[width=0.50\textwidth, trim=45 30 10 30, clip]{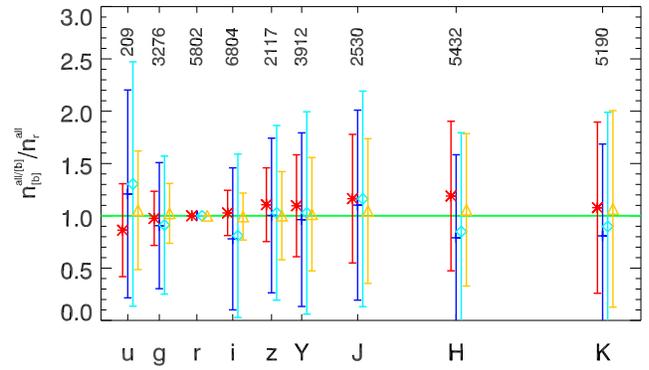}
\caption{A comparison of the trend in \sersic index versus wavelength for Mode\_S1 and Mode\_M fitting. This figure includes all galaxies, e.g. no \lq morphology\rq\ selection has been done. Symbols and colours indicate the same samples as in Fig.~\ref{fig1}.  The \sersic indices for each galaxy are normalized using the $r$-band, and the average variation in $n$ versus wavelength is shown by the points. We recover larger \sersic indices above the line, and smaller below. The rms scatter around the observed trend is indicated by the error-bars.  Single and multi-band approaches reproduce similar average trends, but the multi-band technique reduces the scatter substantially.}
\label{fig4.2}
\end{center}
\end{figure}
\begin{figure}
\begin{center}
\includegraphics[width=0.50\textwidth, trim=45 30 10 30, clip]{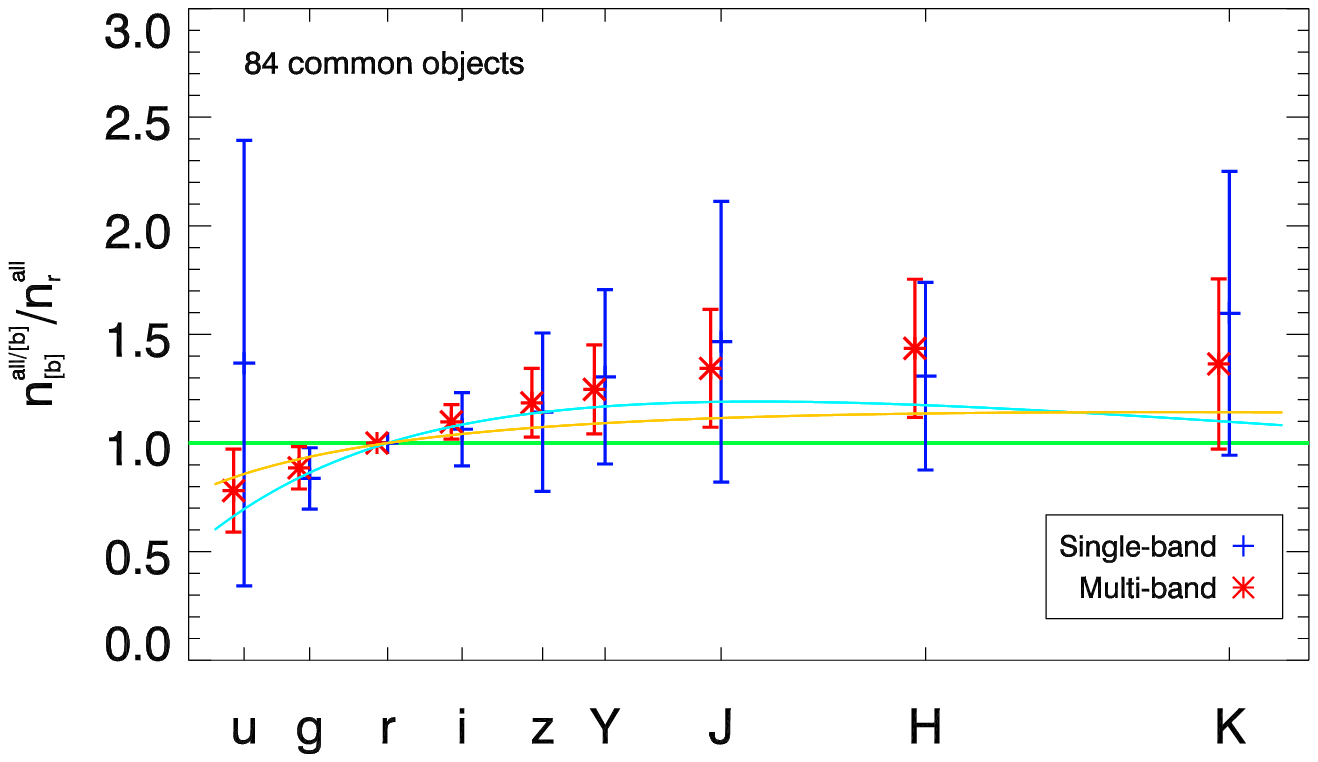}
\includegraphics[width=0.50\textwidth, trim=45 30 10 30, clip]{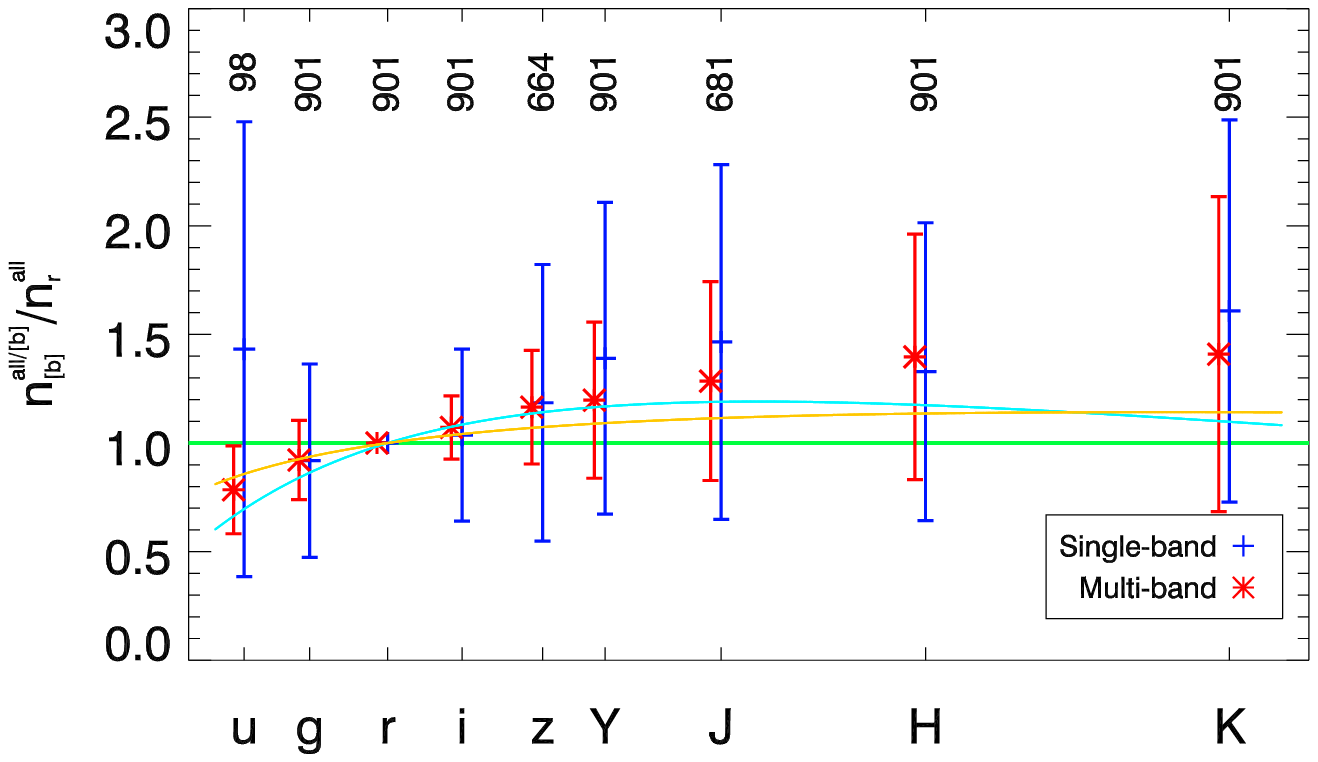}
\caption{Similar to Fig.~\ref{fig2}, these figures compare trends of \sersic index versus wavelength, but for a small sample of 87 galaxies for which all single-band and multi-band fits return a valid results (top panel) and a sample of 901 galaxies for which the $griYHK$ bands return valid results (bottom panel). All the points within each figure therefore correspond to the same objects. The improvement in scatter when using multi-band fitting is very evident in this figure.}
\label{fig5.2}
\end{center}
\end{figure}

In Fig.~\ref{fig4.2} and~\ref{fig5.2}, we present a comparison of Mode\_S1 single band and Mode\_M multi-band fitting results for \sersic index along the same lines as for half-light radii, again, slightly arbitrarily, normalizing all objects to the $r$-band \sersic index. Again, a general trend in $n$ with wavelength, similar to the ones reported in K12, is recovered by both single and multi-band fitting, although, given the scatter in single-band, it is much more visible in multi-band results. This trend confirms our argument that in blue bands one fits a light profile dominated by a galaxy disk (low $n$) whereas in red bands, the bulge dominates the light profile (high $n$). The same effects regarding the scatter of the fitting values are seen, multi-band fitting hugely reduces the scatter in $n$ and increases fitting quality. As expected, the scatter in both single-and multi-band fitting is larger for $n$ than for $r_{\rmn{e}}$, strengthening the argument that \sersic indices are indeed the hardest values to measure.

We can compare our data to the trends described in K12, which we show as lines in Fig.~\ref{fig5.2}. We generally find a qualitatively similar trend, but find it to be stronger than discovered in K12. The reason for this is that the sample in K12 has been split up into disk-dominated and spheroid-dominated samples using a cut in the (U-V)-$n_k$ plane. This split and the normalization to the $r$-band \sersic index in Fig.~\ref{fig5.2} suppress the trend found in our work and lead to shallower trends. This effect can be confirmed in Fig.~\ref{fig5.2}. As splitting up the sample into disk and spheroid dominated sample is beyond the scope of this paper, this effect is expected in this comparison. 

In a similar manner to half-light radius, the reduction in scatter for \sersic index is partly by design, as we constrain the values to lie on a second order polynomial as a function of wavelength with the caveats discussed above. Again, from inspection of the single-band result, we decided that fitting a second order polynomial to the \sersic index as a function of wavelength generally provides sufficient flexibility to fit most real objects. Again, we refer to \ref{sec_sim_results} for a more detailed and normalization-independent evaluation of the fitting accuracy of the \sersic parameter.

\section{Application to simulated imaging}
\label{sec_sims}
Unlike observed data, which has the advantage that the galaxies in the data truly are real and include internal galaxy structures, such as spiral arms, star-forming regions and asymmetries, simulated data has the advantage that one knows the input galaxy parameters. This allows for a direct comparison of recovered magnitudes, half-light radii and \sersic indices to their true values, instead of being limited to comparing to alternative estimators, such as aperture photometry or Petrosian radii, or similar \sersic profile measurements from other studies. The results are therefore much cleaner and free of the intrinsic scatter which dominates most figures in the previous section. Additionally, one can examine in more detail the ability of multi-band fitting to recover the variation of galaxy sizes and \sersic indices with wavelength.

On the other hand, simulations have the disadvantage that the objects now all possess the smooth, symmetric profiles for which we know that \sersic profiles (with some polynomial wavelength dependence of their parameters, and convolved by a known PSF) will precisely fit the data. Our simulation can only test the fitting routines under the assumption that galaxies are intrinsically representable by perfect \sersic profiles. We know that for some galaxies, and for all galaxies observed in sufficient detail, this is not true.  In this sense, simulations are an idealised case and, while affording a good comparison of single-band to multi-band fitting, they can only give a lower limit on the error bars for real data.

\subsection{Creating the simulations}
\label{sec_sim_sims}
To create our simulated data, we follow the methodology of H07, making adaptations to the scripts as necessary to enable the simulation of a multi-wavelength dataset.

As in H07, we started from a catalogue of results from fitting profiles to real data, in order to get the distribution of galaxy parameters in the simulated images as close to that of a real sample as possible.  We decided to start from the sample used in Section \ref{sec_real} and \lq cleaned\rq\ as described above. We first divide the $r$-band $m$--$\log(r_{\rmn{e}})$ plane into bins of width $0.4$ in $m$ and $0.1$ in $\log(r_{\rmn{e}})$. Following H07, to create the parameters for each simulated galaxy we select an $r$-band $m$--$r_{\rmn{e}}$ bin, and use the centre of this bin as our starting value for $r$-band magnitude.  We then pick a real galaxy at random from that bin, from which we obtain realistic values for $r$-band effective radius, \sersic index and axis ratio, as well as the wavelength-variation for each parameter, including colours.  The centre position and position angle were chosen randomly for each object.

To produce a simulated dataset with the same magnitude and effective radius distribution as the real data, we selected $r$-band $m$--$r_{\rmn{e}}$ bins with a probability weighted by the distribution of real galaxies in this plane.  However, in order to test the code at the detection limits we extend the faint side of the magnitude distribution.  This was achieved by finding the peak in the number density of galaxies versus magnitude, for each $r_{\rmn{e}}$ bin, and extending this peak level two magnitudes fainter (as was done in H07).

As in H07, we slightly varied the galaxy parameters produced by the above procedure, in order to create a smooth distribution, instead of simply re-simulating the same few galaxies over and over.  These applied variations are as follows:

Position: The $x$ and $y$ centre coordinates in the simulated field are randomly chosen (although of course neighbouring tiles are consistent in their overlap regions), but are identical for all bands (no statistical or systematic offsets are applied).

Magnitude: Starting from the initially chosen $r$-band magnitude, we first apply an offset, drawn from a Gaussian distribution with standard deviation of $0.4$ mag.  To define the magnitudes in the other bands, colours (as differences of the other bands to $r$-band) are directly taken from a real galaxy within the corresponding $m$--$r_{\rmn{e}}$ bin and added to the $r$-band magnitude.  The magnitudes in the individual bands are finally modified by adding Gaussian noise with a $0.1$ standard deviation. The magnitudes were allowed to vary freely as function of wavelength in the fitting process applied to the real galaxies, so there are no smoothness constraints on magnitude in the simulated dataset.

Half-light radius: The fits to the real data were carried out using a polynomial of second order to describe the wavelength dependence of $r_{\rmn{e}}$. We start with the function taken from a real galaxy in the selected $m$--$r_{\rmn{e}}$ bin. We then add an additional linear slope, drawn from a Gaussian distribution with standard deviation corresponding to a 10 per cent change in $r_{\rmn{e}}$ over the full wavelength range. No noise is added to the individual bands, so that simulated values follow a second order polynomial exactly.

\sersic indices: Similar to $r_{\rmn{e}}$, in the fits to the real data the wavelength dependence of $n$ was described by a quadratic. For most of the simulated galaxies, starting values are taken from the same real galaxy as before, but in this case we simply modify the values in all bands by a factor drawn from a Gaussian with unit mean and standard deviation of 0.3. This offsets the values, but preserves the linear and quadratic coefficients. The simulated values therefore follow a second order polynomial exactly.  However, for 5 per cent of objects we randomly chose $n$ uniformly from the range $0.2$--$8$, with a linear slope as a function of wavelength.  This was done in order to cover areas of parameter space that are not covered by the cleaned catalogue of real galaxies, either because no such galaxies exist or such galaxies defy reliable detection or modelling. As these areas are of particular interest to us, we simulate some galaxies to cover them.

Axis ratio: This is simply taken from the chosen real galaxy and offset by a Gaussian random variable with standard deviation $0.1$ (subject to the allowed $0$--$1$ range of this variable). The same value is adopted by all bands, and no additional noise is added.

Position angle: This is randomly chosen, with the same value applied to all bands, with no further noise added.

Following H07, we added Poisson noise to the simulated images and then add them into an image that is made up of empty sky patches from the real GAMA imaging. This means that sky noise is present in the simulated images with the same properties as in the real data.

The result of this procedure is a realistic-looking set of images with thousands of galaxies that mostly show the same parameter distributions as the real galaxies, and for which we know the true parameter values.  It should be noted that these images do not contain any stars, due to the cleaning of the catalogue prior to parameter selection.  A fraction of real galaxy images are contaminated by stars, which must be masked or deblended by \galapagos, potentially leading to additional uncertainties which are neglected by our simulations.  However, blended galaxies are naturally included by our simulation method, and so the effects of blending are included in our results.

\subsection{Results}
\label{sec_sim_results}
In this section, we examine the single and multi-band fitting results by comparing the recovered parameter values to the true, simulated values directly. Where possible, we recreate figures similar to those in Section \ref{sec_real}, but we are able to examine issues in more detail and add a set of figures which could not be created using real data.

When running the codes we use exactly the same setup and procedures as were used for real data in the previous section. In particular, we use the same versions of our codes (\galfitm version 0.1.2.1, \galapagos version 2.0.2). We run the same three scenarios, i.e., Mode\_S1, Mode\_S2 and Mode\_M, and carry out the same procedure to combine the individual catalogues (9 single-band, 1 multi-band) in order to create one single catalogue which is used throughout this section. This catalogue is additionally correlated to the simulated catalogue using RA and Dec., so that true values for all objects are known (disregarding the potential for rare mis-identifications).

\begin{figure*}
\begin{center}
\includegraphics[width=0.98\textwidth, trim=0 0 0 0, clip]{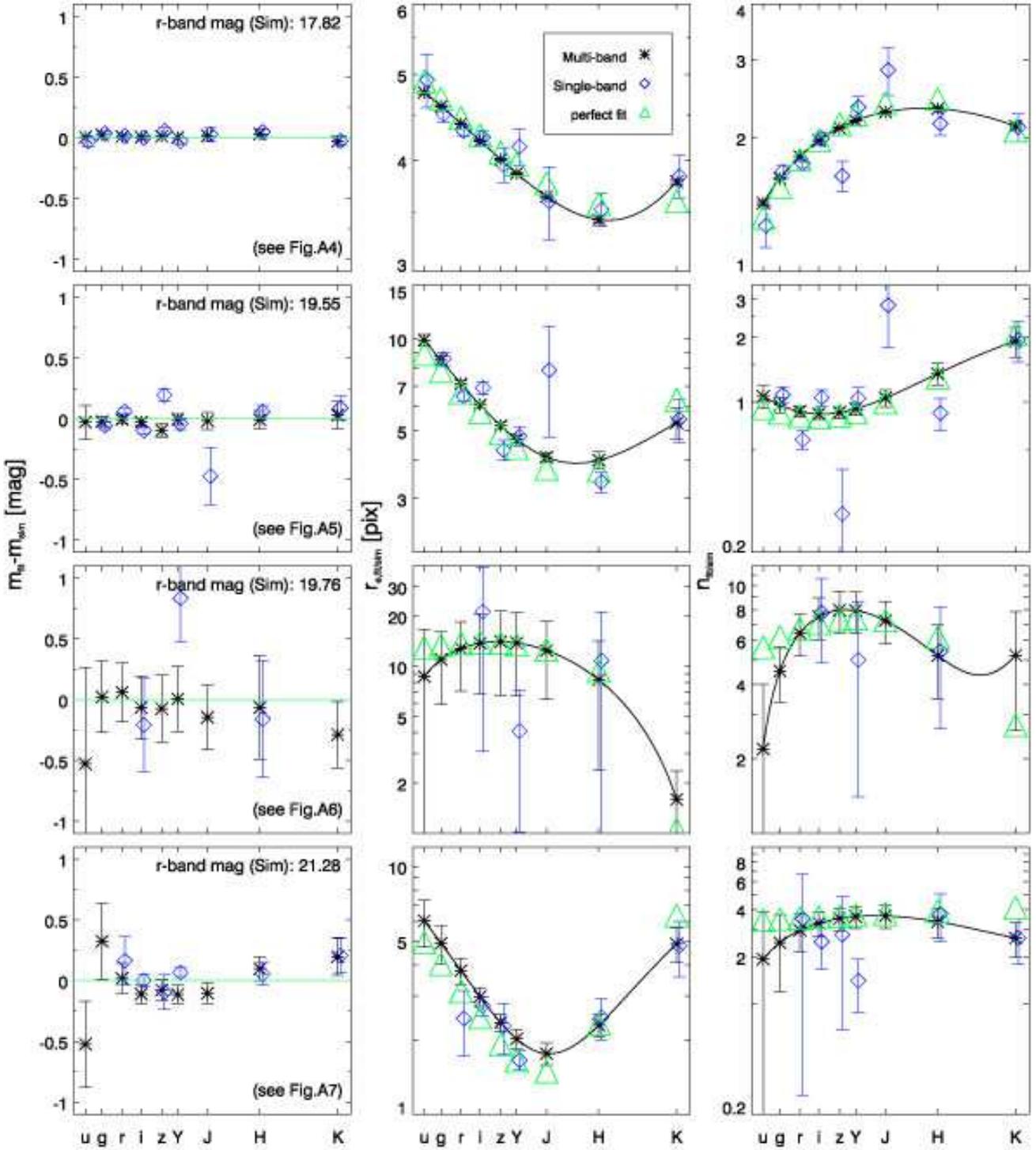}\\
\caption{Fits for individual simulated galaxies, similar to Fig.~\ref{fig0} in case of real galaxies. The images of these objects, together with models and fitting residuals from both Mode\_S1 and Mode\_M fitting, are shown in Figs.~\ref{figA4} to \ref{figA7} in the Appendix. The left column shows the magnitude difference between fitted and simulated values. The green line/values indicates an ideal fit (no offset between fit and simulated values), blue diamonds show Mode\_S1 fitting results, black asterisks show Mode\_M results. The middle column shows both simulated and recovered half-light radii in all bands. The black line shows the entire polynomial that is fit to the data for Mode\_M; in contrast to single-band fitting, where only some of the individual values exist, we derive the entire function in the case of multi-band fitting. The right column shows the simulated and fitting values for \sersic indices. The $r$-band magnitude of the objects is indicated in the leftmost figure. See further discussion in the text.}
\label{fig6}
\end{center}
\end{figure*}

We show some fits of individual simulated galaxies in Fig.~\ref{fig6}. As for real galaxies, it is apparent that multi-band fitting returns values for more bands, making studies using multiple sets of values (e.g., colours) possible for a larger sample of galaxies. It also reduces the scatter on all the values that are measured. Whereas the single-band magnitudes (and even sizes in the brighter galaxies) generally return sensible values, the $r_{\rmn{e}}$ and \sersic indices recovered are typically much noisier in single-band fitting compared to multi-band fitting. Multi-band fitting generally recovers the shape of the wavelength-dependence well, but this is of course a function of magnitude, where fainter galaxies are generally less well recovered. 

In addition to returning more accurate values at the specific wavelengths of the input images, multi-band fitting also provides the entire polynomial over the wavelength region.  Estimations of values at intermediate wavelengths, e.g. restframe values, therefore become trivial.  As long as they are within the wavelength range covered by the input images, they can simply be obtained from the polynomial. As mentioned in Section~\ref{sec_setup}, this interpolation is not sensible for magnitudes, for which we use a higher-order polynomial, due to Runge's phenomenon.  In any case, we have a better understanding of galaxy SEDs, and so it is wise to use this knowledge, e.g., via template spectra, to estimate accurate magnitude k-corrections.

After visually inspecting a large fraction of individual fits, we found that multi-band fitting was able to successfully outperform single-band fitting in all cases when the multi-band fit was not rejected due to proximity to a fitting constraint. The opposite case -- objects where multi-band fails but single-band fits return results for at least some bands -- does occur, but is less common. Numbers to illustrate this are included in Fig.~\ref{fig8}.
As expected we also find that for very faint galaxies ($r$-band magnitude $\sim23$, close to the SDSS detection limit), the wavelength dependence of the sizes and \sersic indices do not match that simulated, and the magnitudes are very noisy. On most of these objects single-band fitting is not able to recover many, if any, values. However, parameter estimates from the Mode\_M fits, though noisy, were generally still in agreement with the simulated values, given the big error bars.

\begin{figure}
\begin{center}
\includegraphics[width=0.50\textwidth, trim=43 30 10 30, clip]{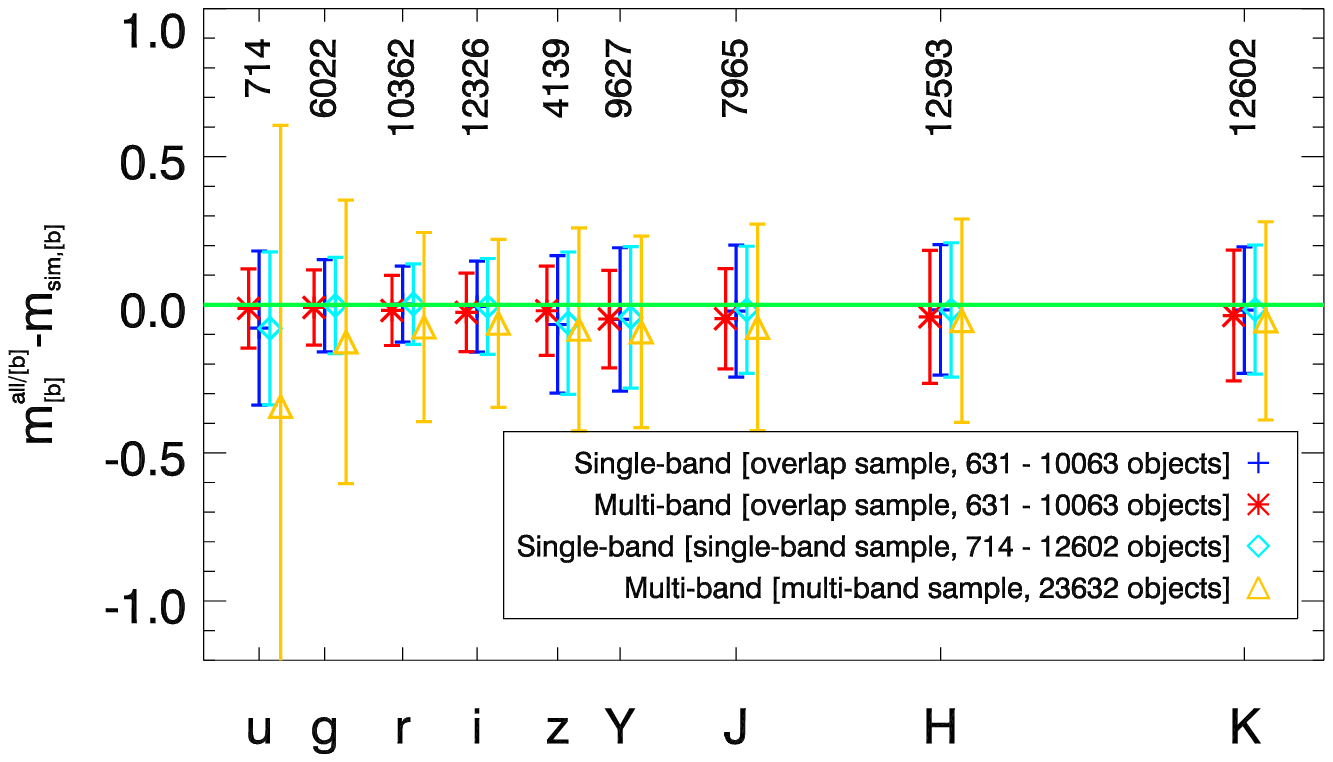}
\includegraphics[width=0.50\textwidth, trim=43 30 10 30, clip]{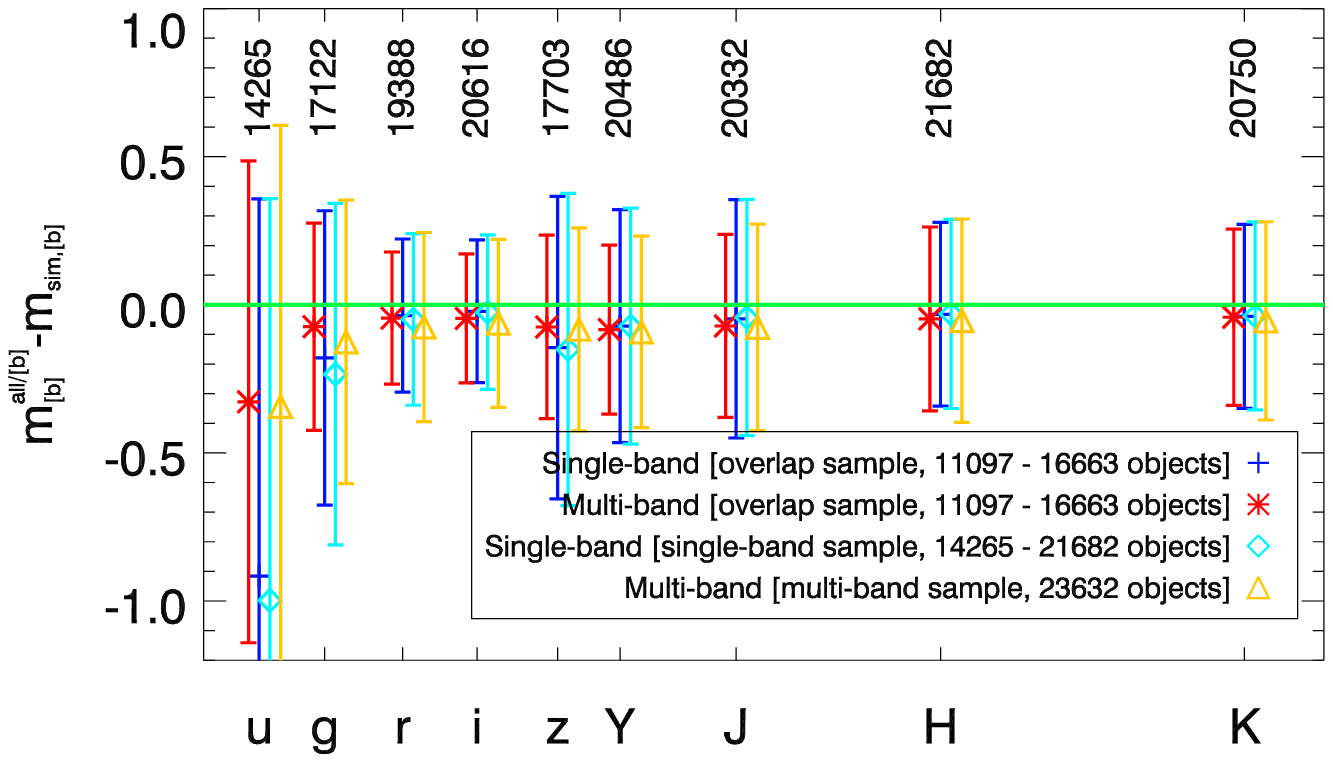}
\caption{Magnitude recoverability for simulated galaxies (Mode\_M and Mode\_S1 in the top panel, Mode\_M and Mode\_S2 in the bottom panel). Note especially the change of quantity on the y-axis compared to Fig.~\ref{fig1}; for simulated galaxies we are able to demonstrate the difference between simulated and measured values.  This means that the measurement uncertainties are no longer convolved with the intrinsic scatter on the colours, permitting a more stringent comparison between techniques. For example, while on real galaxies the scatter in $u$-band is around $0.5$ mag, it is only about $0.2$ mag here. Multi-band fitting produces a smaller scatter in comparison to single-band results. Systematics are very small, although both techniques do display a bias to recover brighter magnitudes by $< 0.05$ mag. The sample size is enlarged by the multi-band approach from 714 ($u$-band) or 12602 ($K$-band) in single-band fitting to 23632 objects in multi-band fitting, nearly a factor of 2 more objects even for the best single bands.  These numbers are compared in more detail in table \ref{tab_2}). In the bottom panel, the bias introduced by combining multi-band detection with single-band fitting (Mode\_S2) becomes very apparent.  This is simply a result of attempting to model objects that are below the noise limit.}
\label{fig6.2}
\end{center}
\end{figure}

In Fig. \ref{fig6.2}, we illustrate the performance of single- and multi-band techniques for recovering the magnitudes of our simulated galaxies.  This is an equivalent figure to Fig.~\ref{fig1}, but compares directly to the simulated parameter values, rather than arbitrarily normalizing the SED to the $r$-band.  The intrinsic scatter in galaxy colours, which obscured improvements in scatter in Fig.~\ref{fig1}, is not present in this figure.  The offset and scatter in this figure represent the true bias and uncertainty in our ability to recover object parameters.   As in Fig.~\ref{fig1}, one can see that the scatter is significantly reduced in most bands when using multi-band fitting. Multi-band fitting also somewhat reduces the small offset that is apparent in low-\sn bands (mainly $u$- and $z$-band).

The scatter and bias for the complete multi-band sample (shown in yellow) are significantly larger. This is largely due to the inclusion of large numbers of faint objects in this sample and will be examined in detail later (in Fig.~\ref{fig8}).

The lower panel in Fig.~\ref{fig6.2} shows the same figure as the upper panel, but for Mode\_S2 results. Similar to what was seen using real galaxies, the improvement from Mode\_S2 to Mode\_M fitting is more dramatic than from Mode\_S1 to Mode\_M. Our arguments for this are similar to the ones presented in Section~\ref{sec_real_results}. The results of Mode\_S2 fitting are worse largely because the effect of neighbouring galaxies. Either the primary object wanders off to fit a secondary (in which case the fit would likely appear to be too bright), or a secondary can centre on the primary (in which case the fit would appear to be too faint).  Multi-band fitting is largely immune to such issues, as the higher \sn images naturally constrain the position of the target in the lower \sn bands.

\begin{table*}
\centering
\begin{minipage}{140mm}
\caption{Object numbers in the simulated dataset and fraction with successful fits for Mode\_S1 and Mode\_M. Please note that the success rate in this simulated sample resembles an actual success rate and does not include stars that are purposefully excluded from the final analysis, but which suppress the apparent success rate in Table~\ref{tab_1}. In that respect, the success rates in this table are more comparable to those presented for the GAMA galaxy sample in Table~\ref{tab_1}.}
\begin{tabular}{@{}l|rrr@{}}
\hline
Band    &    \#objects fitted        & \#objects detected & Success rate\\
\hline
\hline
u & 714 & 1213 & 58.9\%\\
g & 6022 & 8839 & 68.1\%\\
r & 10362 & 13739 & 75.4\%\\
i & 12326 & 16490 & 74.7\%\\
z & 4139 & 6196 & 66.8\%\\
Y & 9627 & 13123 & 73.4\%\\
J & 7965 & 11068 & 72.0\%\\
H & 12593 & 16992 & 74.1\%\\
K & 12602 & 17755 & 71.0\%\\
 & & & \\
\textbf{combined single band ($ugrizYJHK$)} & \textbf{305} & \textbf{26308} & \textbf{1.2\%}\\
\textbf{combined single band ($griYHK$)} & \textbf{2560} & \textbf{25631} & \textbf{10.0\%}\\
\textbf{mwl} & \textbf{23632} & \textbf{31798} & \textbf{74.3\%}\\
\hline
\end{tabular}
\label{tab_2}
\end{minipage}
\end{table*}
Table \ref{tab_2} presents the fitting success rates for the simulated galaxy sample, in a similar manner to those shown in Table~\ref{tab_1} for the real data. The fitting success rate here is much higher than in case of real galaxies. There are two distinct effects responsible for this. Firstly, the simulated images do not contain stars, due to the cleaning of the fitting catalogue prior to parameter selection. As stars typically result in \lq failed\rq\ fits, by design, their absence from the simulations naturally means that the resulting success rate is boosted. The remaining improvement is largely due to the fact that all the simulated galaxies actually resemble smooth \sersic profiles, which makes the fitting process much more stable, even in single-band fitting and in low S/N bands.

The multi-band technique is not the one with the highest success rate on simulated data. Due to a more sensitive detection in Mode\_M compared to Mode\_S1, the sample includes many more faint galaxies which are harder to fit and decrease the success rate. The overall number of objects for which parameters can be derived is still higher. When comparing the overlap of the samples of single-band with each other and the sample in multi-band fitting, the increase is still substantial, effectively providing fits for seven times more galaxies in a sample that is only twice the size to start with. This emphasises how dramatically the size of scientific samples can be increased by using multi-band fitting.

\begin{figure}
\begin{center}
\includegraphics[width=0.50\textwidth, trim=43 30 10 30, clip]{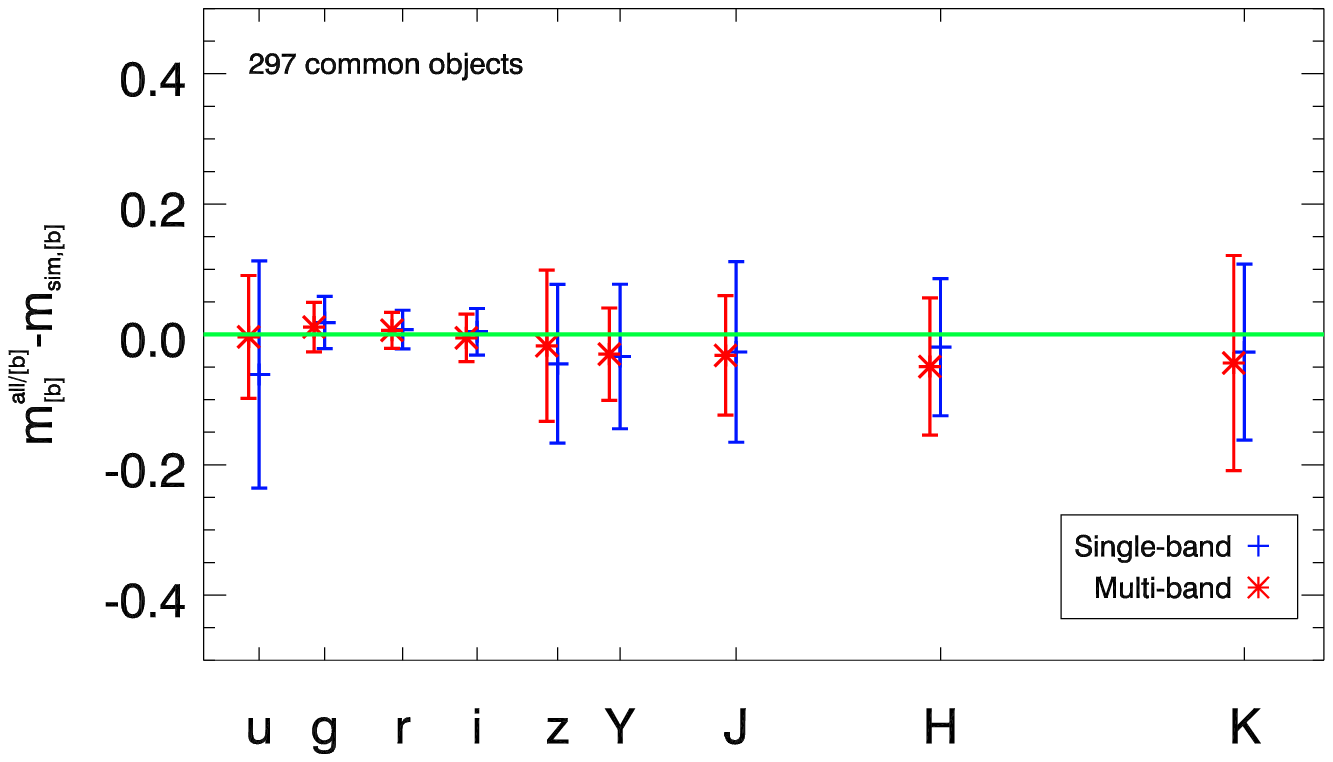}\\
\includegraphics[width=0.50\textwidth, trim=43 30 10 30, clip]{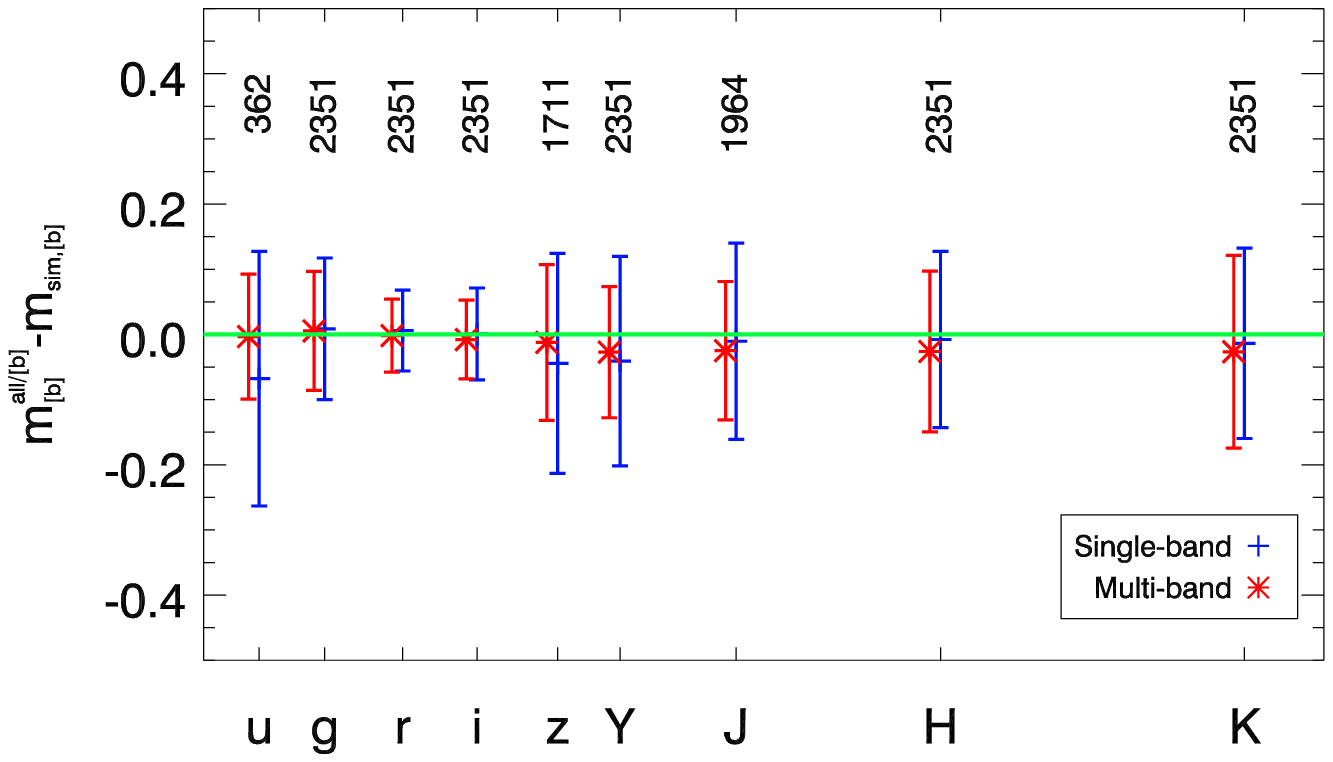}
\caption{Magnitude recoverability in simulated data for galaxies with complete sets of single-band fits. These figures are the same as the top panel of Fig.\ref{fig6.2}, but show only a common sample of 297 galaxies for which all Mode\_S1 fits return a result (top) or 2351 galaxies that have fits in $griYHK$ (bottom). For these common samples, multi-band fitting clearly increases the accuracy and precision for magnitude measurements. For reference, the values shown in the top panel are given in Table~\ref{table_mag_error}.}
\label{fig7}
\end{center}
\end{figure}
To emphasise the effect on sample sizes, and ensure a consistent comparison. Fig.~\ref{fig7} again shows the recoverability of magnitude by both- single and multi-band fitting, but for matched subsamples for which all, or most, Mode\_S1 results return valid fitting values. In both cases, the reductions in both scatter and offsets are clearly visible for most bands. Low S/N bands particularly benefit from the multi-band approach. The values from the top panel in Fig.~\ref{fig7} are also given in Table~\ref{table_mag_error}. Biases are somewhat reduced at most wavelengths, while scatter is reduced significantly in the low \sn bands and slightly in the other bands.

\begin{table}
\centering
%\begin{minipage}{140mm}
\caption{Mean and scatter of the magnitude offsets for Mode\_S1 and Mode\_M for a sample of objects with complete Mode\_S1 measurements, as shown in the top panel of Figure~\ref{fig7}.}
\begin{tabular}{@{}lcc@{}}
\hline
Band    &   offset $\pm \sigma$ Mode\_S1 &   offset $\pm \sigma$ Mode\_M \\
\hline
\hline
u & -0.06$\pm$0.17 & 0.00$\pm$0.09\\
g & 0.02$\pm$0.04 & 0.01$\pm$0.04\\
r & 0.01$\pm$0.03 & 0.01$\pm$0.03\\
i & 0.00$\pm$0.04 & -0.01$\pm$0.04\\
z & -0.05$\pm$0.12 & -0.02$\pm$0.12\\
Y & -0.03$\pm$0.11 & -0.03$\pm$0.07\\
J & -0.03$\pm$0.14 & -0.03$\pm$0.09\\
H & -0.02$\pm$0.11 & -0.05$\pm$0.11\\
K & -0.03$\pm$0.13 & -0.04$\pm$0.17\\
\hline
\label{table_mag_error}
\end{tabular}
%\end{minipage}
\end{table}

\begin{figure}
\begin{center}
\includegraphics[width=0.50\textwidth, trim=45 30 10 30, clip]{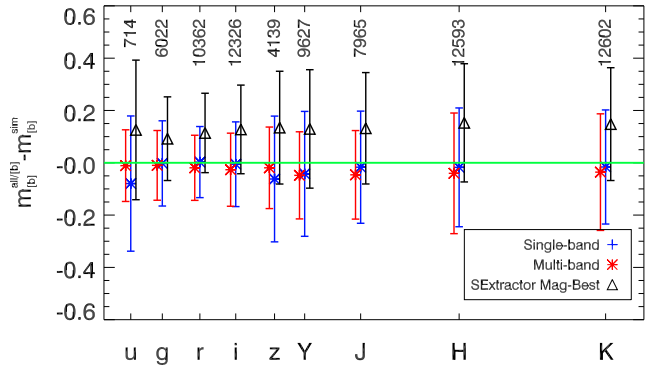}
\caption{Equivalent to Fig. \ref{fig3} we show here the deviations of recovered magnitudes for both Mode\_M (red) and Mode\_S1 fits (blue) compared to \sex\ \lq Mag-Best\rq. For comparison, we show the deviations of the \sex magnitudes from the simulated values as black data points. Clearly visible, \sex magnitudes produce a much larger offset to simulated magnitudes and are not a good proxy for \lq real\rq\ magnitude. Either multi- or single-band fitting recover the simulated magnitudes better. This puts Fig.~\ref{fig3} and the offsets of Mode\_S1 and Mode\_M values in this figure in perspective.}
\label{fig7.2}
\end{center}
\end{figure}
In  Fig. \ref{fig3} we compared magnitudes derived by \galfit with those obtained using aperture photometry, finding a consistent offset of $\sim 0.1$ mag, which appeared greater when using the multi-band technique.
In Fig.~\ref{fig7.2} we follow up on this by comparing magnitudes from Mode\_S1 and Mode\_M fits, as well as  \sex\ \lq Mag-Best\rq\ aperture magnitudes, to their simulated values.
The \galfit results are the same as those shown in the top panel of Fig.~\ref{fig6.2}. It is obvious that the offset in Fig.~\ref{fig3} is a result of \sex underestimating the total flux. Both single and multi-band fitting provide better estimates of total magnitude.  Fig.~\ref{fig3} therefore does not show which method recovers the better values. However, as Fig.~\ref{fig3}, together with Fig.~\ref{fig6.2}, serves as a simple illustration of the issues in comparing aperture and model photometry, we retain it.   Note that our simulated magnitudes are based on \sersic models integrated to infinity, as are the magnitudes recovered by \galfit. Aperture magnitudes necessarily do not measure the flux in the distant tails of the profile, and it is arguable whether this flux actually exists, as much of it is at surface brightnesses fainter than can possibly be measured.  This is discussed in much more detail by K12, who advocate using magnitudes obtained from truncated \sersic profiles for all science analyses.

\begin{figure*}
\begin{center}
\includegraphics[width=0.88\textwidth, trim=0 0 0 0, clip]{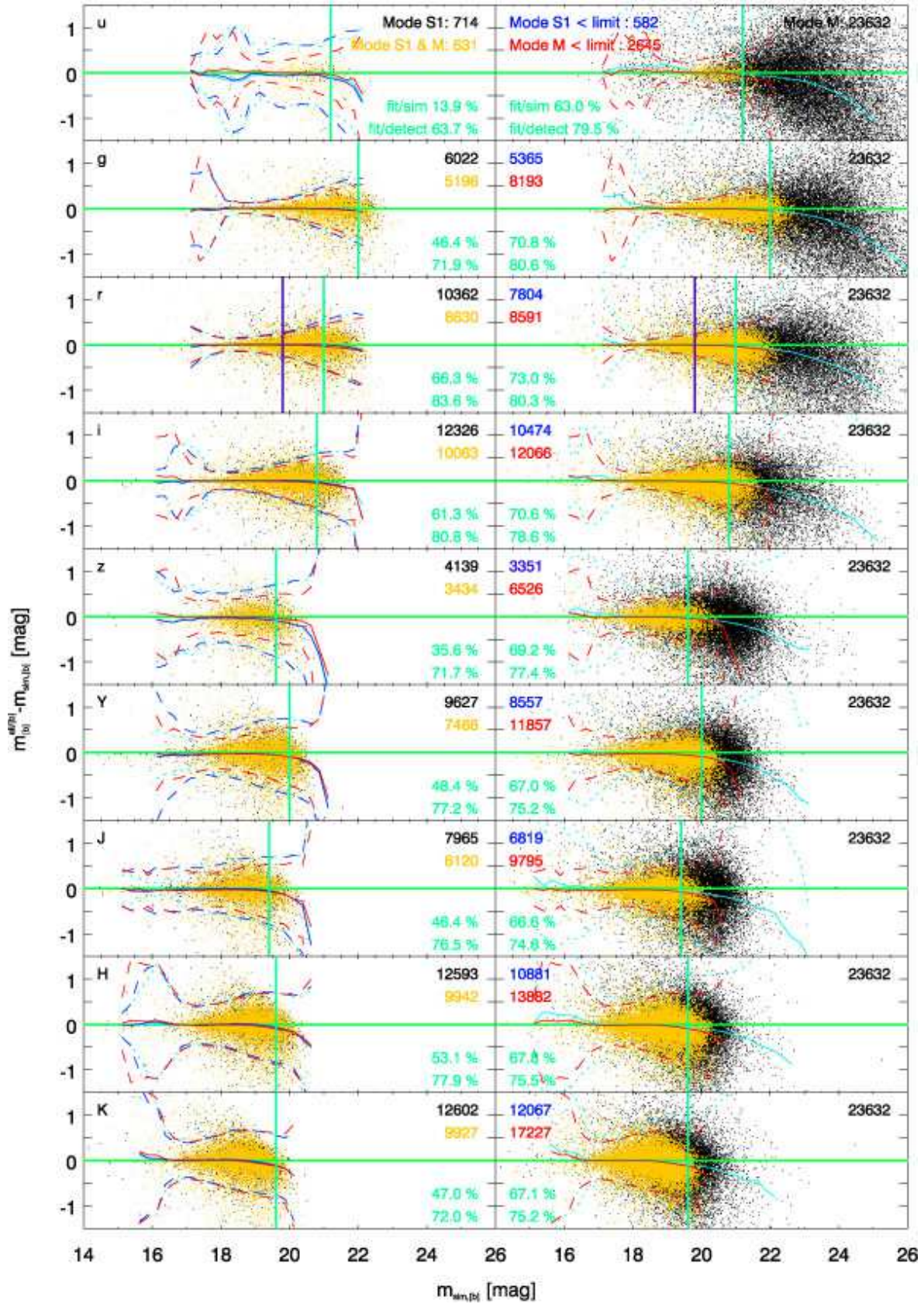}
\caption{Magnitude fitting accuracy as a function of simulated magnitude. 
The blue and red dashed lines show the median scatter of the distribution as a function of magnitude for single- and multi-band fits for a common sample of objects for direct comparison.
Vertical lines show magnitude limits used in the analysis and to derive the objects numbers shown in the different panels. 
These object counts show the sizes of different samples for which useful values can be derived. 
Detailed description and discussion of these numbers and this figure can be found in the text in Section \ref{sec_sim_results}, starting on page \pageref{Fig13_discussion}.}
\label{fig8}
\end{center}
\end{figure*}

\label{Fig13_discussion}
A more complete presentation of the results from fitting simulated data is presented in Fig.~\ref{fig8}. For each band ($u,g,r,i,z,Y,J,H,K$ from top to bottom) we show the magnitude offset from the fitting values, but instead of putting everything into one bin as in the previous figures, we show them as a function of simulated (true) galaxy magnitude\footnote{We note that when using observed surface brightness on the x-axis, the figures look qualitatively very similar}. The left column shows results from Mode\_S1 fits with the mean offset and rms scatter overlaid in light blue (dotted lines indicate the $\pm 1 \sigma$ scatter); the right column shows the same for Mode\_M fits, with the mean offset and rms scatter for the whole object sample in light blue. Black numbers in the top right corner of each panel show the number of objects appearing in the figure, i.e., simulated objects which have been detected and successfully fit by the method in question.  Additionally, in both columns, we overplot the sample of objects for which both Mode\_S1 and Mode\_M fits return a valid result in yellow.  Yellow numbers in the top right corner of the left-column panels give the number of objects in these samples. One can see that not all objects with single-band fits have successful multi-band fits, but the fraction is much higher than the other way around. We show the mean trends and rms scatter for this overlap sample for Mode\_S1 fits in blue in the left column and, to aid comparison, for Mode\_M fits in red in both columns.

For an ideal fitting code, the mean and scatter lines should be close to zero at all input magnitudes, indicating that no systematic or statistical biases exist in the fitting routines. As one can see, both single-band and multi-band fitting values are not significantly biased for most galaxies, but the multi-wavelength fitting reduces the scatter somewhat, particularly in lower \sn bands. The improvement is not significant for the overlap sample of galaxies in some higher \sn bands, but even in these cases the multi-band approach significantly improves the completeness of the sample.

At this point, we would like to remind the reader that before creating these figures we cleaned our catalogue of objects that ended up close to fitting constraints. As mentioned above, the total number of successfully fit objects is indicated by the black number in the corner of each panel. As one can see, the numbers of objects for which we can get valid results is much larger in the case of multi-band fitting, confirming our findings from Section~\ref{sec_real}. Whereas in $r$- and $i$-band the sample size only doubles, we can obtain fitting parameters for $30$ times more galaxies in the $u$-band. In most bands, we can also recover values for fainter galaxies. As these are harder to fit and result in larger error bars, they also lead to a large \lq scatter\rq\ in Figs.~\ref{fig6.2} to \ref{fig7.2} (orange symbols). This is expected and does not imply that Mode\_M fits are less accurate than Mode\_S1 fits, as could be wrongly interpreted from the error bars in those figures.

For reference, we show, as a vertical purple line in the $r$-band panels in Fig.~\ref{fig8}, the $m_r=19.8$ magnitude limit of the spectroscopic survey carried out as part of the GAMA project. For galaxies brighter than this limit, both single-band fitting and multi-band fitting do a similar job of returning good fitting values.  However, the accuracy and precision of the recovered parameters are clearly improved by the multi-band method, particularly for lower \sn bands and, as we shall see, size and \sersic index.  One distinct advantage of the multi-band technique is its ability to obtain valid results for fainter objects.  However, if one is only interested in bright galaxies, e.g. for which spectroscopy has been obtained, then the difference in terms of sample size is less clear.  To examine this we estimate fitting limits from the single-band data, indicated by the vertical green lines in each panel of Fig.~\ref{fig8}.
These lines are defined as being 0.5 magnitude fainter than the peak in the magnitude histogram for Mode\_S1 detections. The blue number in the right column shows the number of galaxies brighter than this limit that returned valid results in Mode\_S1 fitting, the red number shows the corresponding number using Mode\_M fitting. As one can see, the multi-band fitting returns useful results for appreciably more galaxies, even in this bright sample of objects. This illustrates that the increase in overall sample size is not only due to adding fainter galaxies, but also due to deriving valid fitting results more often for bright galaxies.

To investigate this topic further, we print two fractions in each panel in green. The top number shows the fraction of galaxies simulated brighter than the single-band fitting limit (vertical green lines) for which we obtained valid fits. However, this fraction includes the issue of single-band versus multi-band detection completeness and is therefore not a direct comparison of the robustness of the fitting techniques.  In order to take out the detection issue, we determine the fraction of objects detected in each band (and above the fitting limits) for which we obtained valid fitting results. These are given by the lower green percentage in each panel.  These fractions are higher and correspond more closely to the numbers given in Table~\ref{tab_2} for the full sample.  They are also similar between single- and multi-band techniques, indicating that for bright samples, at least in the case of single-\sersic fitting, the multi-band technique does not make large differences to the number of objects which can be successfully fit.  In some bands this success fraction is actually higher for single-band fitting than for multi-band fitting.  This probably reflects the fact that the cleaning process for multi-band is more stringent than that for single band: proximity to a single constraint in any band results in rejection of that multi-band fit.  Note that even though the individual band success rates are comparable between single and multi-band methods, when one desires a sample with a complete set of measurements across all bands, the single-band method suffers dramatically.  This is clearly demonstrated in the final three rows of Table~\ref{tab_2}.  It is also important to remember, as shown throughout this paper, that the accuracy of the recovered parameters, particularly in lower \sn bands, is significantly improved by adopting the multi-band technique.

One effective approach to improving the fractions of detections and valid fits is repeat attempts.  In the case of detection, if one knows an object is present in one band, one can attempt to detect it in other bands by varying detection thresholds and deblending parameters.  For fitting, whenever a fit ends up on a constraint, one can restart the fit with slightly different initial conditions or masks, hoping that one of these repeats will produce acceptable values.  Both of these approaches are adopted by K12, with the result that their fractions of valid fits are significantly higher than ours.  However, the fact that restarting \galfit with slightly different parameters can produce different final results is concerning, and it is uncertain whether parameter values obtained after several attempts are reliable.  In our case we take constraint violations as a signal that reliable results cannot be obtained for the object in question.  It is not yet clear which approach is most appropriate, although we aim to investigate this question further in future work.

\begin{figure}
\begin{center}
\includegraphics[width=0.50\textwidth, trim=45 30 10 30, clip]{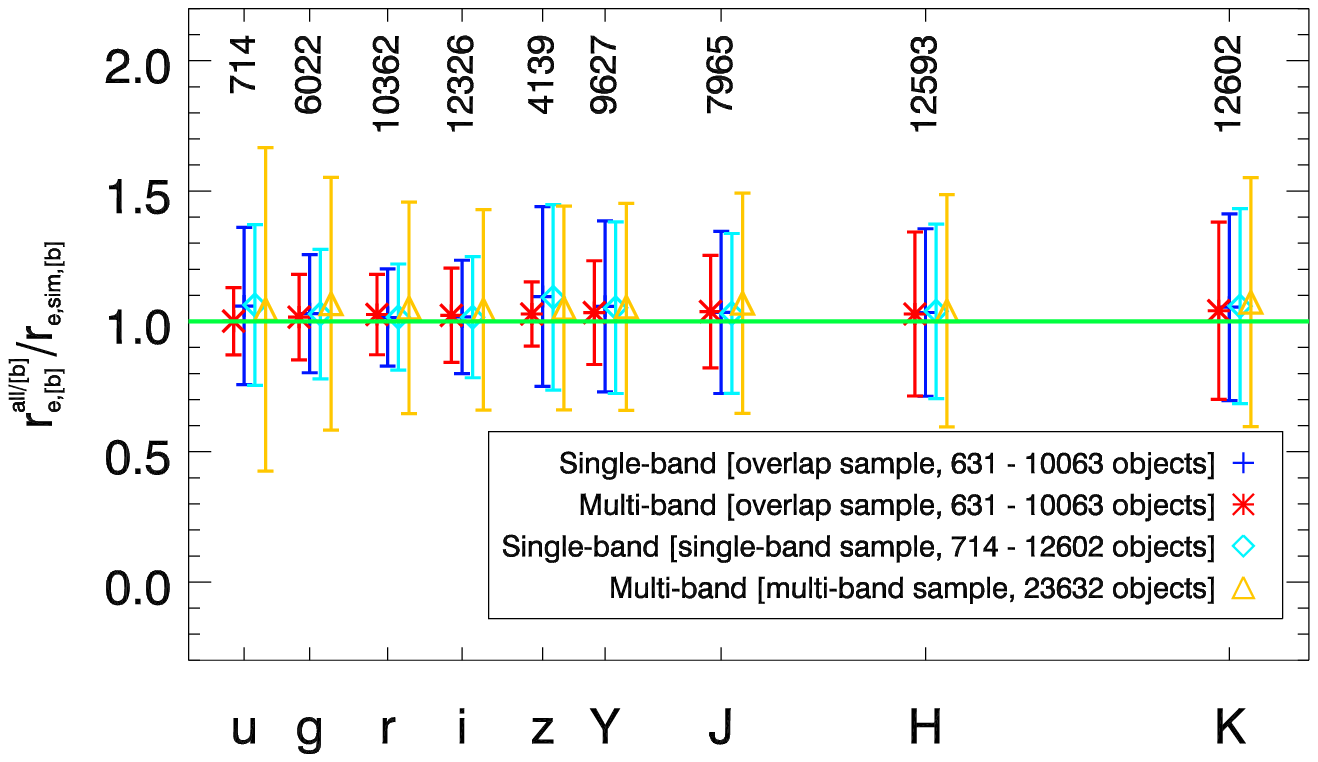}
\includegraphics[width=0.50\textwidth, trim=45 30 10 30, clip]{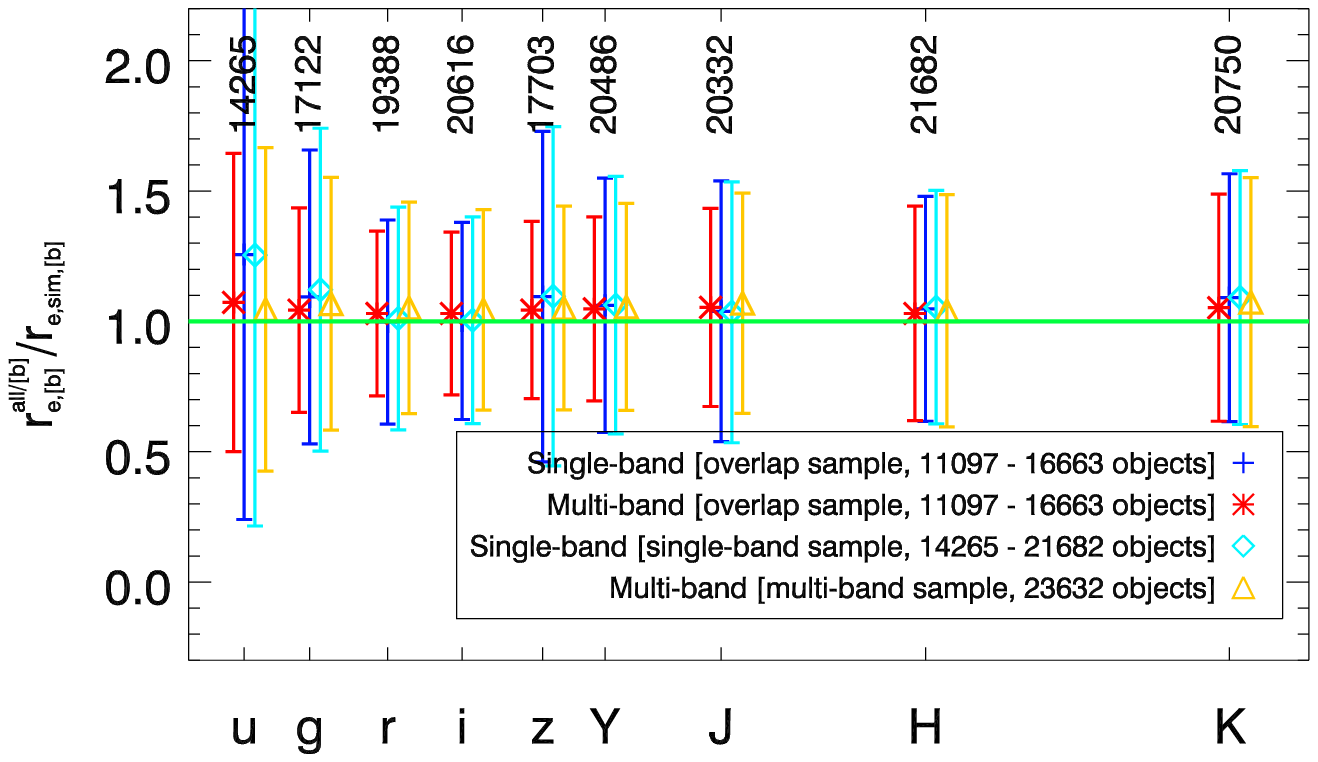}
\caption{Size recoverability for simulated data (top panel: Mode\_S1 and Mode\_M, bottom panel: Mode\_S2 and Mode\_M). Fitting results are improved when using multi-band fitting. The increased error bars in case of the entire multi-band sample (in orange), largely come from adding in much fainter galaxies, see Fig.~\ref{fig11} and discussion in the text. Only a very small systematic offset is visible, even when including the faintest galaxies.}
\label{fig9}
\end{center}
\end{figure}
\begin{table}
\centering
%\begin{minipage}{140mm}
\caption{Mean and scatter of the effective radius offsets for Mode\_S1 and Mode\_M for a sample of objects with complete Mode\_S1 measurements, as shown in the top panel of Figure~\ref{fig10}.}
\begin{tabular}{@{}lcc@{}}
\hline
Band    &   mean $\pm \sigma$ Mode\_S1 &   mean $\pm \sigma$ Mode\_M \\
\hline
\hline
u & 1.06$\pm$0.20 & 1.00$\pm$0.06\\
g & 0.99$\pm$0.05 & 1.00$\pm$0.04\\
r & 1.00$\pm$0.03 & 1.00$\pm$0.03\\
i & 1.00$\pm$0.05 & 1.00$\pm$0.03\\
z & 1.04$\pm$0.15 & 1.01$\pm$0.04\\
Y & 1.04$\pm$0.12 & 1.02$\pm$0.05\\
J & 1.04$\pm$0.17 & 1.03$\pm$0.08\\
H & 1.03$\pm$0.13 & 1.05$\pm$0.10\\
K & 1.03$\pm$0.18 & 1.02$\pm$0.14\\
\hline
\label{table_re_error}
\end{tabular}
%\end{minipage}
\end{table}

\begin{figure}
\begin{center}
\includegraphics[width=0.50\textwidth, trim=45 30 10 30, clip]{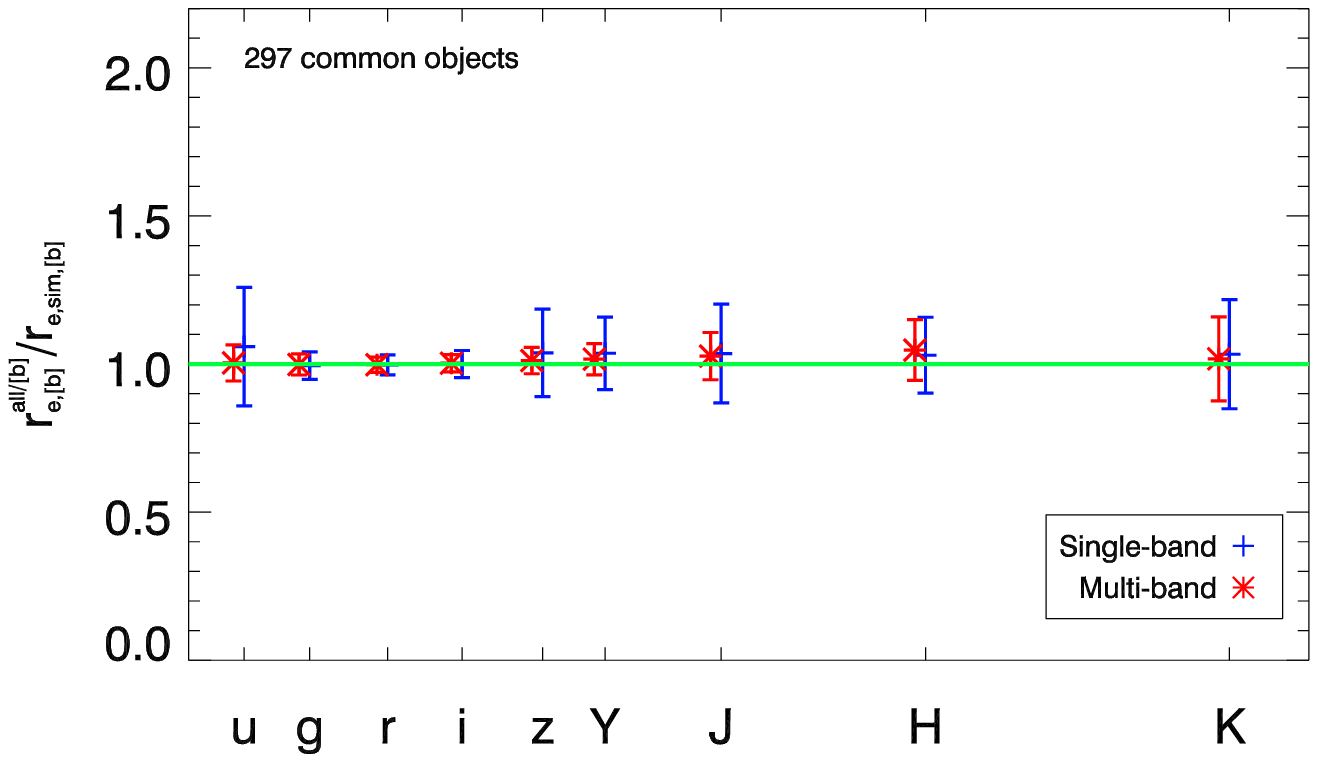}\\
\includegraphics[width=0.50\textwidth, trim=45 30 10 30, clip]{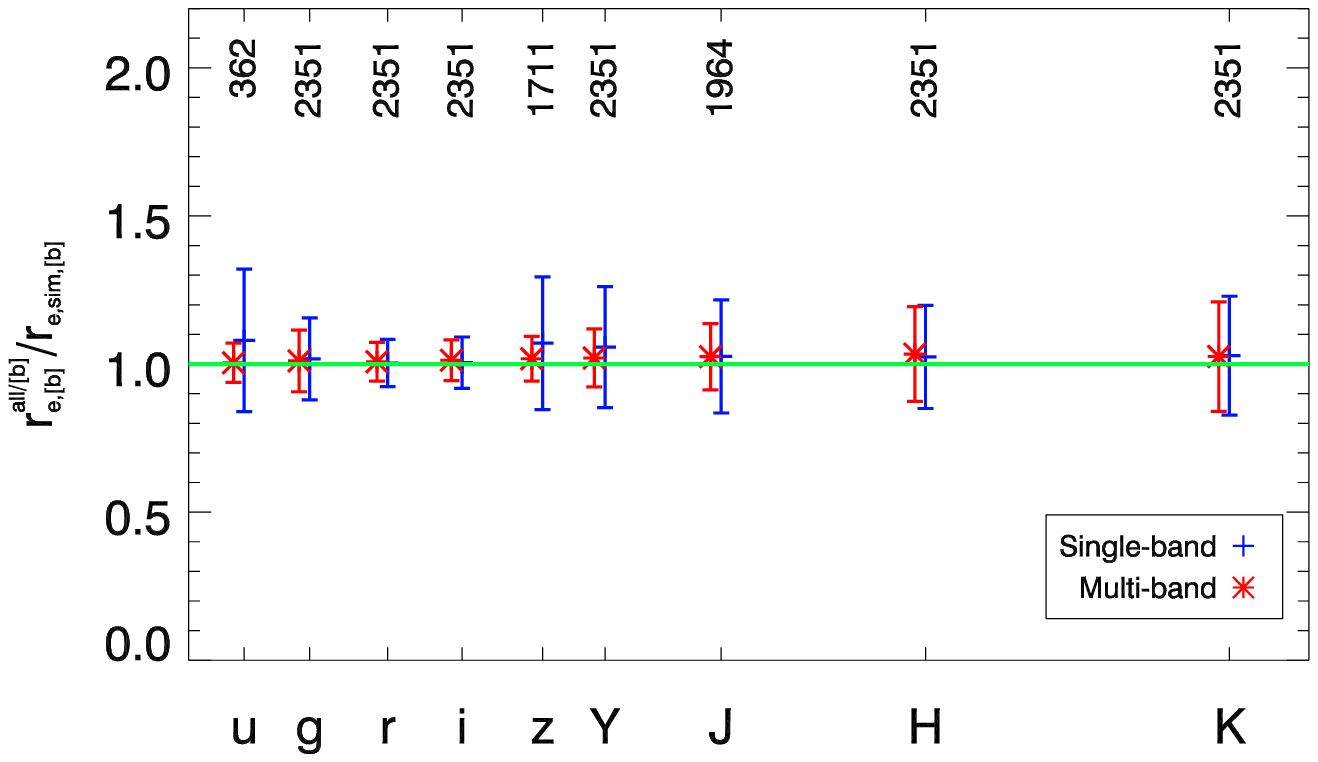}
\caption{Size recoverability for simulated data, similar to Fig.~\ref{fig9}, but only for the subset of galaxies where all Mode\_S1 fits (top) or $griYHK$ band fits (bottom) return valid results. It is evident that multi-band fitting, when comparing the same sample of galaxies, decreases the scatter and thus hugely improves the recoverability of galaxy sizes.}
\label{fig10}
\end{center}
\end{figure}
\begin{figure*}
\begin{center}
\includegraphics[width=0.88\textwidth, trim=0 0 0 0, clip]{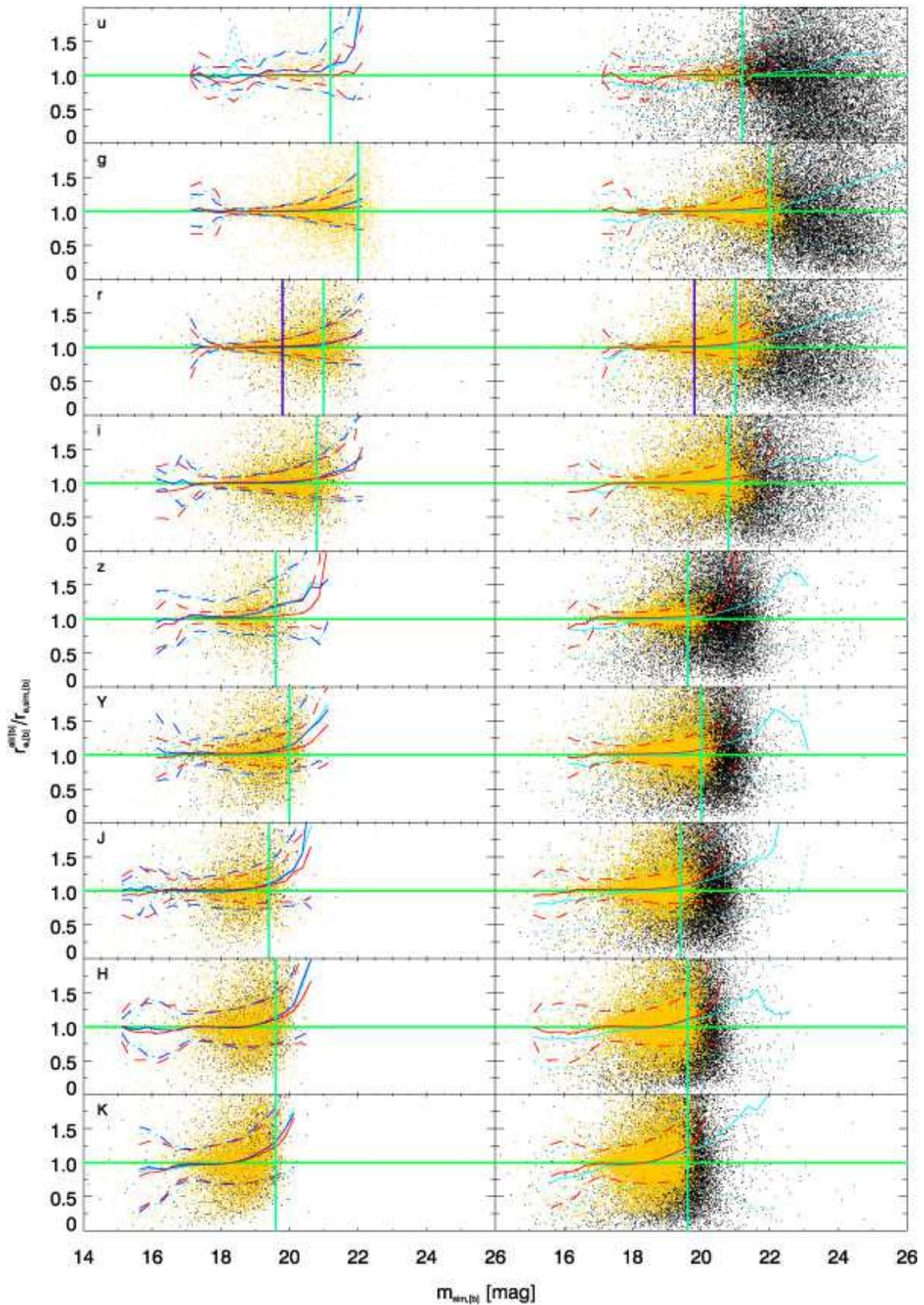}
\caption{Simulation results showing the recoverability of effective radius for each band as a function of simulated magnitude. Single-band fitting results are shown on the left, multi-band results are on the right.  Points and lines are the same as for Fig.~\ref{fig8}. The multi-band technique, while only slightly improving on magnitude fitting, greatly improves the fitting accuracy for galaxy size.}
\label{fig11}
\end{center}
\end{figure*}
In Figs.~\ref{fig9} to \ref{fig11} we show the equivalent analysis to that presented above, but now considering the recoverability of effective radii in the simulated data. This complements the similar analysis that was performed using the results of fitting real data in Section~\ref{sec_real_results}. As was seen in Section~\ref{sec_real}, our multi-band fitting technique dramatically improves the fitting accuracy for galaxy sizes.  This is true in all bands (but especially those with low \sn), and at all magnitudes. In Figs.~\ref{fig9} to \ref{fig11} we qualitatively confirm the result that we found for real galaxies.  However, using the simulations we can make more quantitative statements.  Fig.~\ref{fig10}, in particular, shows that galaxy half-light radius can be measured with much higher precision in case of Mode\_M compared to Mode\_S1.  Values from the top panel in Fig.~\ref{fig10} are given in Table~\ref{table_re_error}. Figure~\ref{fig11} shows that multi-band fitting can measure reasonably accurate sizes even for very faint objects, even in bands where the object may be below the single-band detection limit.

\begin{figure}
\begin{center}
\includegraphics[width=0.50\textwidth, trim=45 30 10 30, clip]{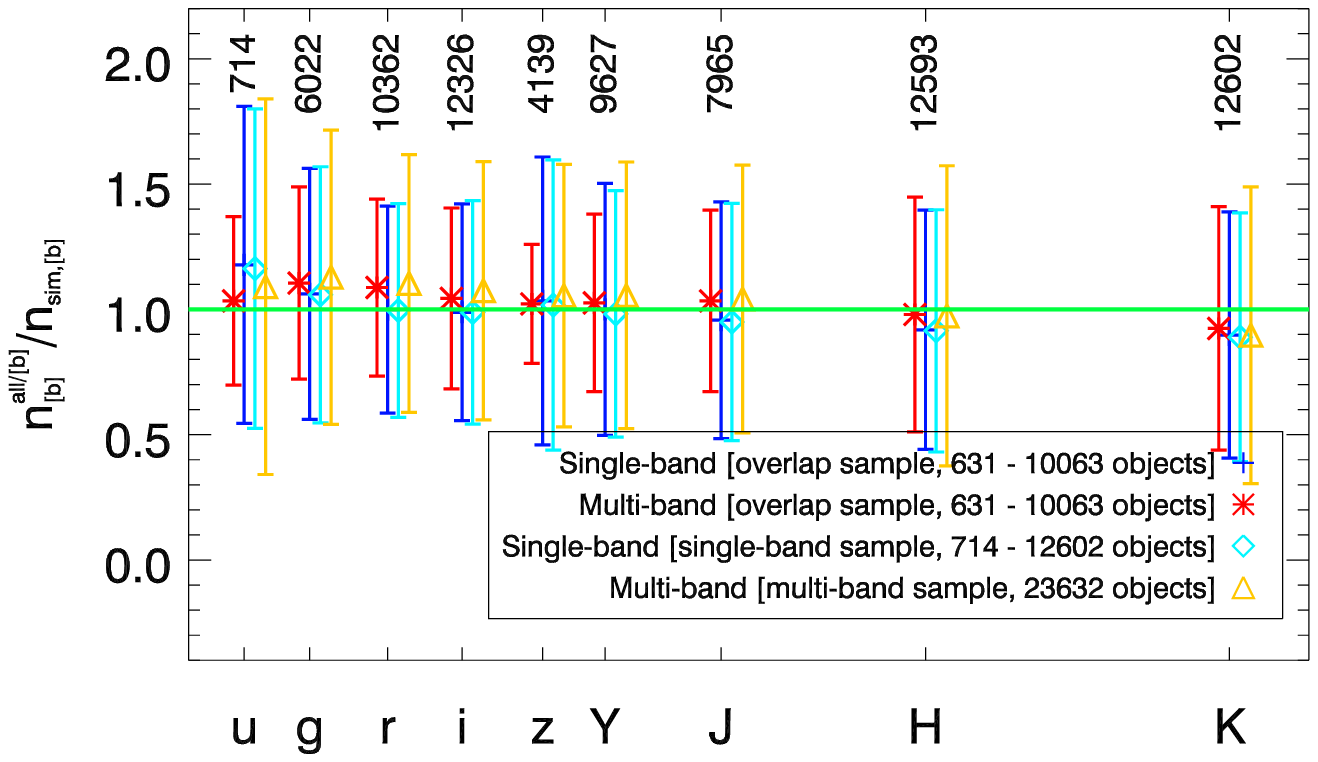}
\includegraphics[width=0.50\textwidth, trim=45 30 10 30, clip]{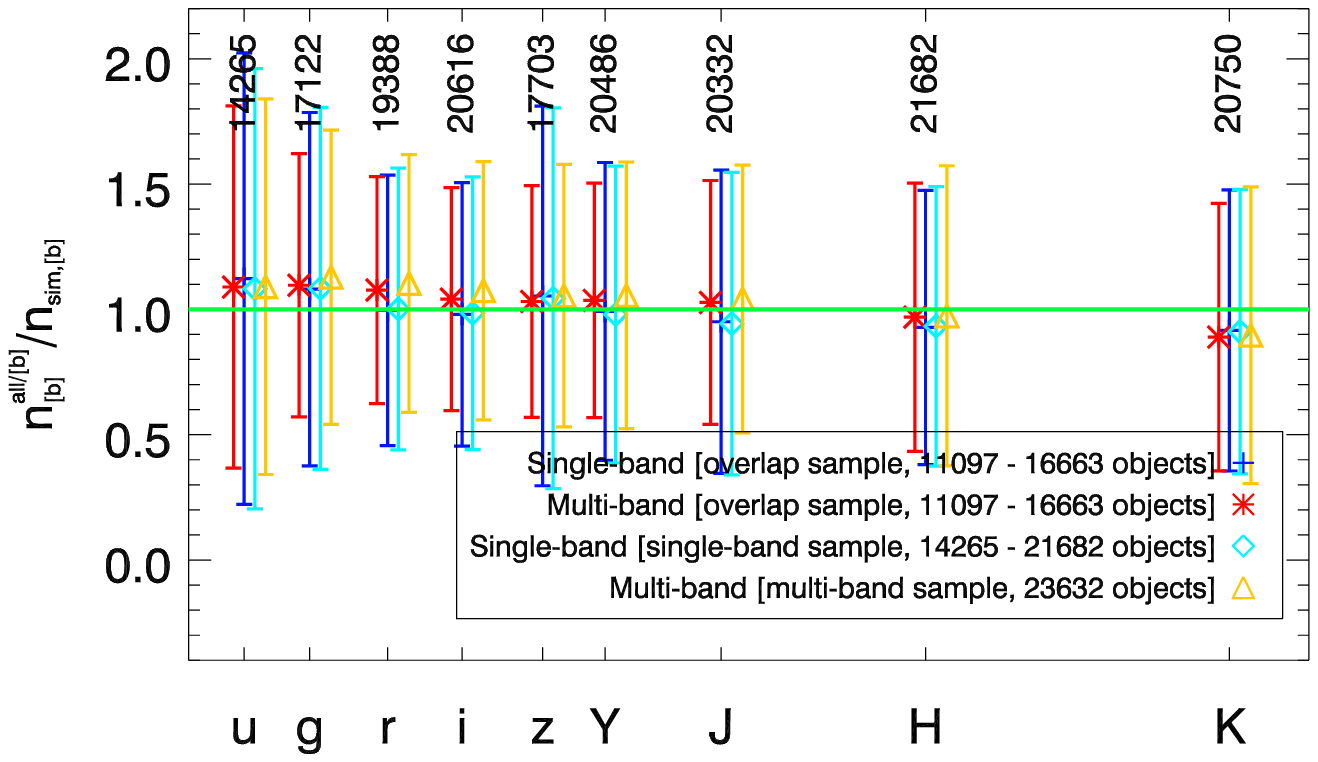}
\caption{\sersic index recoverability for the individual bands (top panel: Mode\_S1 and Mode\_M, bottom panel: Mode\_S2 and Mode\_M). As for effective radius, multi-band fitting reduces the scatter on recovered parameters substantially, without introducing significant systematic effects.}
\label{fig12}
\end{center}
\end{figure}
\begin{figure}
\begin{center}
\includegraphics[width=0.50\textwidth, trim=45 30 10 30, clip]{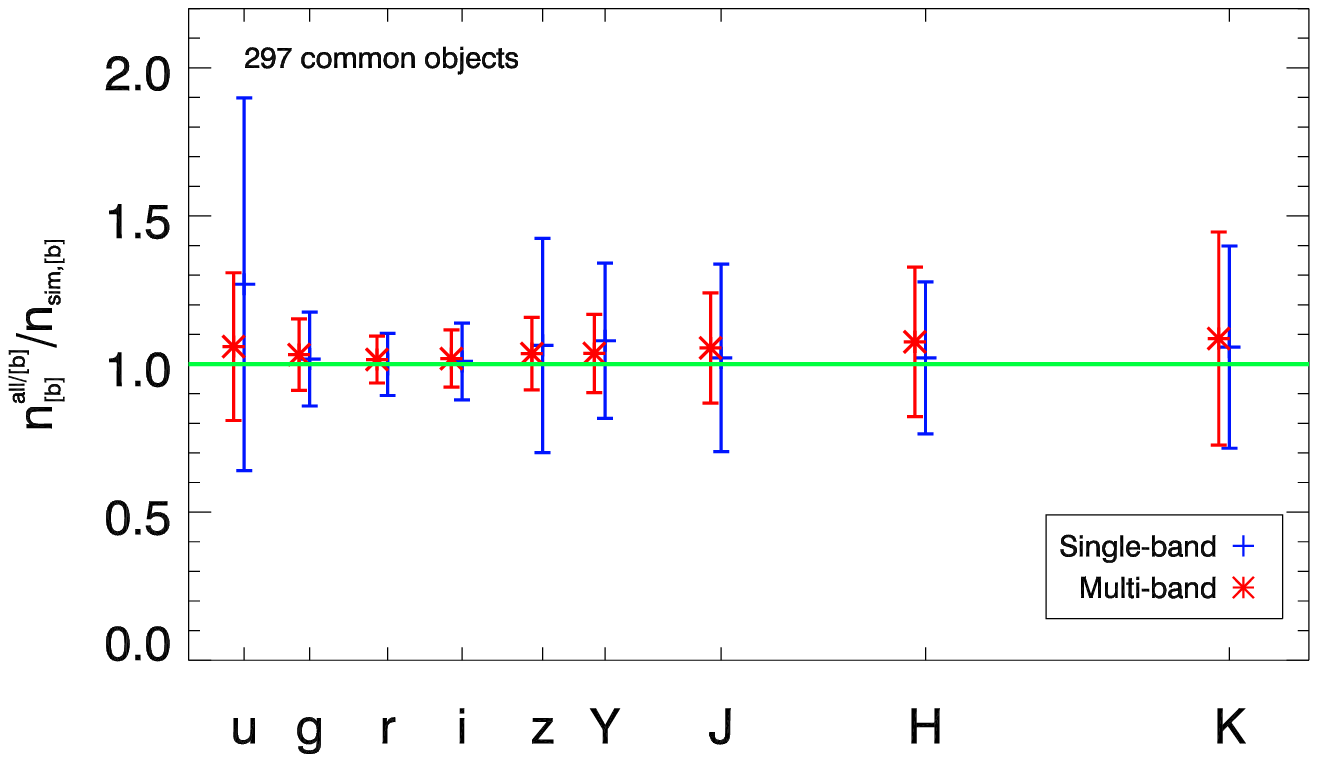}\\
\includegraphics[width=0.50\textwidth, trim=45 30 10 30, clip]{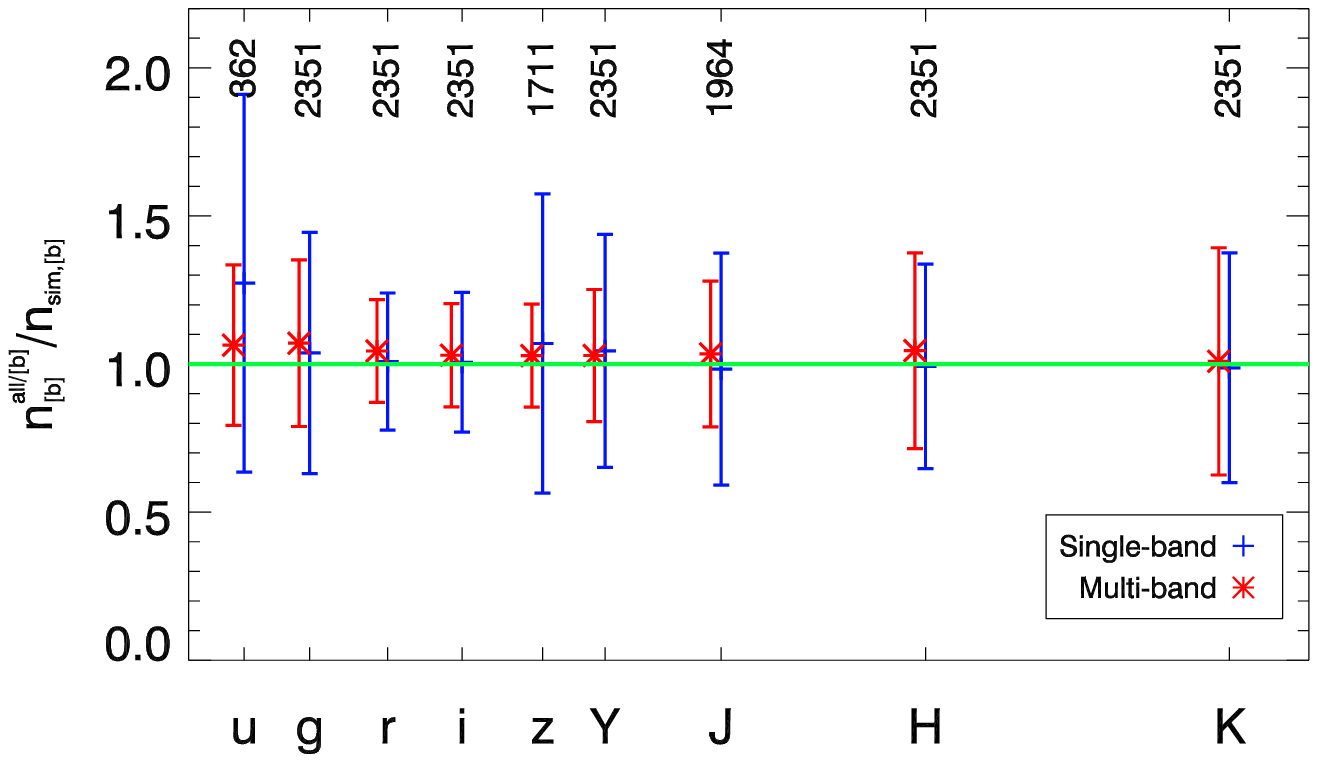}
\caption{This is the same figure as the top panel of Fig.~\ref{fig12}, but now for the subset of galaxies for which all (top) or most (bottom) single-band fits return a valid result. Again, the reduction in scatter for Mode\_M is significant.}
\label{fig13}
\end{center}
\end{figure}
\begin{table}
\centering
%\begin{minipage}{140mm}
\caption{Mean and scatter of \sersic index offsets for Mode\_S1 and Mode\_M for a sample of objects with complete Mode\_S1 measurements, as shown in the top panel of Figure~\ref{fig13}.}
\begin{tabular}{@{}lcc@{}}
\hline
Band    &   mean $\pm \sigma$ Mode\_S1 &   mean $\pm \sigma$ Mode\_M \\
\hline
\hline
u & 1.27$\pm$0.63 & 1.06$\pm$0.25\\
g & 1.02$\pm$0.16 & 1.03$\pm$0.12\\
r & 1.00$\pm$0.11 & 1.02$\pm$0.08\\
i & 1.01$\pm$0.13 & 1.02$\pm$0.10\\
z & 1.06$\pm$0.36 & 1.04$\pm$0.12\\
Y & 1.08$\pm$0.26 & 1.04$\pm$0.13\\
J & 1.02$\pm$0.32 & 1.05$\pm$0.19\\
H & 1.02$\pm$0.26 & 1.08$\pm$0.25\\
K & 1.06$\pm$0.34 & 1.09$\pm$0.36\\
\hline
\label{table_n_error}
\end{tabular}
%\end{minipage}
\end{table}

\begin{figure*}
\begin{center}
\includegraphics[width=0.88\textwidth, trim=0 0 0 0, clip]{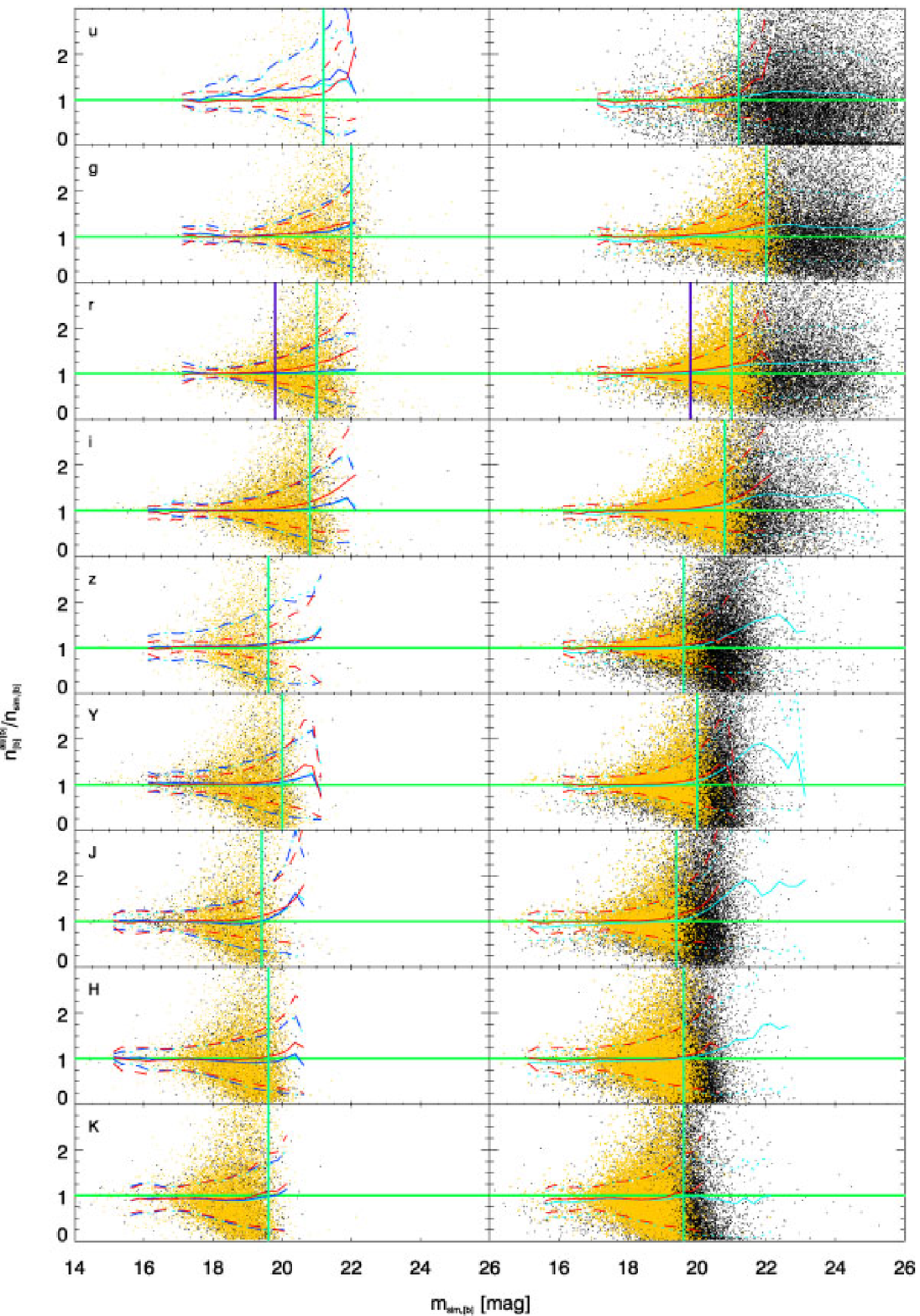}
\caption{Simulation results for each band: Mode\_S1 vs. Mode\_M. Multi-band fitting, while improving the results for simulated galaxies in most bands, multiplies the sample size without increasing the scatter more than would be expected by adding fainter galaxies.  Points and lines are the same as for Figs.~\ref{fig8} and \ref{fig11}.}
\label{fig14}
\end{center}
\end{figure*}
Figs.~\ref{fig12} to \ref{fig14} repeat the same analysis for \sersic indices. Similar to half-light radii, the improvement in fitting quality is much clearer than for magnitudes. Be aware that the error bars in all figures are relative errors, but they combine values from galaxies with different \sersic indices. A galaxy with $n=4$ is much harder to fit than a galaxy with $n=1$, mostly due to uncertainties in the sky estimation (H07), so this way of plotting reduces this effect. Values from the top panel in Fig.~\ref{fig13} are shown in Table~\ref{table_n_error}, showing that the measurements of \sersic index are significantly improved when multi-band fitting is used. In Fig.~\ref{fig14} we show the effect as a function of simulated magnitude. While small systematic effects are visible for fainter galaxies, the improvement in fitting quality when using multi-band fitting is evident, especially in the lower \sn bands.

\begin{figure}
\begin{center}
\includegraphics[width=0.50\textwidth, trim=80 20 10 30, clip]{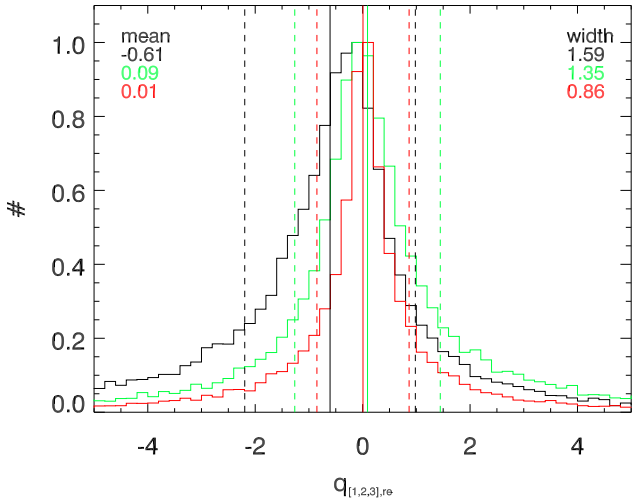}\\
\caption{Histogram of the Chebyshev coefficients describing the wavelength dependence of $r_{\rmn{e}}$: $q_1$ (black), $q_2$ (green) and $q_3$ (red), obtained from \galfitm multi-band fits to our simulated dataset.  The mean and rms scatter of each distribution are indicated by vertical solid and dashed lines, respectively, and the values printed in the corners of the figure. Histograms have been normalized to unit maximum. As expected, the distribution of $q_3$ in our simulated sample is narrower than the distribution of $q_2$.}
\label{fig14.2}
\end{center}
\end{figure}
\begin{figure}
\begin{center}
\includegraphics[width=0.50\textwidth, trim=60 20 10 30, clip]{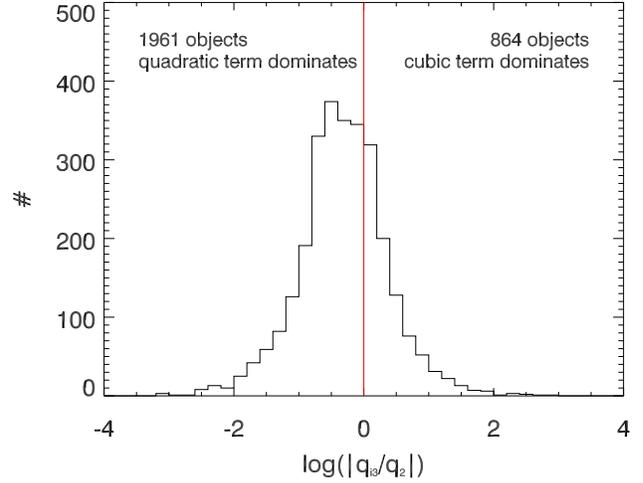}
\caption{Histogram of $\log(|q_3/q_2|)$ for galaxies with $mag_{sim}<19.8$ This figure compares the importance of the second-order and third-order terms of the Chebyshev polynomials for $r_{\rmn{e}}$. As the simulations were done to second order specifications, one would expect to see that the second order term dominates in most cases. The histogram is clearly offset to the left, confirmed by the values printed in the figure, which give the numbers either side of zero.}
\label{fig14.3}
\end{center}
\end{figure}

We have shown that using multiple band images simultaneously, with structural parameters that are required to vary smoothly as a function of wavelength, can dramatically improve both the robustness of fits to faint galaxies and the accuracy of recovered parameters for objects of all brightnesses.  The examples in Figs.~\ref{fig0} and \ref{fig6} indicate that the recovered wavelength dependencies do reflect the intrinsic variations of the structural parameters between wavebands.  We now examine the meaning we can ascribe to the measured Chebyshev coefficients more closely.

The simulated galaxies were created such that the simulated values for $r_{\rmn{e}}$ and $n$ follow second order polynomials, e.g. making a second order polynomial the perfect fit to the simulated data. We have intentionally allowed more freedom, using third-order polynomials in both $r_{\rmn{e}}$ and $n$, when fitting these data with the multi-band technique, in order to test how well we can recover the correct trends. In Fig.~\ref{fig14.2} we show histograms of the Chebyshev coefficients for $r_{\rmn{e}}$: $q_1$ (black), $q_2$ (green) and $q_3$ (red). There are several things to mention about this figure. Firstly, the histogram for $q_1$ is clearly offset from $0$. As $q_1$ describes the linear terms of the Chebyshev polynomials, this means that statistically a non-flat linear trend (in this case a negative trend) of $r_{\rmn{e}}$ with wavelength is visible. Galaxies in redder bands generally appear smaller. This effect has already been shown in Fig.~\ref{fig4} for real galaxies. As the simulations were created following this catalogue, the same trend is apparent in the simulated data, but it is pleasing that we can see this directly in the Chebyshev coefficients. Secondly, the histograms for both $q_2$ and $q_3$ are not significantly offset from $0$, so no systematic trend is visible in the population. This does not, however, mean that individual objects do not significantly show second order polynomials. Thirdly, and most importantly, the width -- and hence the importance -- of the third order term $q_3$ is significantly smaller than for $q_2$. Ideally, we would want $q_3 = 0$ for all galaxies, but given the noise properties of the image, this was not expected.

A more sensitive way to show this is via the histogram of $\log(|q_3/q_2|)$ in Fig.~\ref{fig14.3}. Plotted this way the second order term \lq dominates\rq\ the shape of the polynomial compared to the third order term in all galaxies that show $\log(|q_3/q_2|)<0$, or equivalently $|q_3|<|q_2|$.  This is important in order to justify whether the polynomial order fitted to the real data is the correct order to use -- e.g., if one fits a third order polynomial and finds that $|q_3|<|q_2|$ for most galaxies, it is a justifiable assumption that second order polynomials are appropriate for most objects. In previous tests this is what we have found, hence our decision to fit real galaxies with second order polynomials and creating simulations to this specification.

\section{Computational considerations}
\label{sec_other}
Other than fitting accuracy, there are also other issues to be considered. In this section, we will discuss some of these issues and compare the single and multi-band fitting procedures.

\subsection{Fitting time}
\label{sec_time}

\begin{figure}
\begin{center}
\includegraphics[width=0.50\textwidth, trim=60 20 0 20, clip]{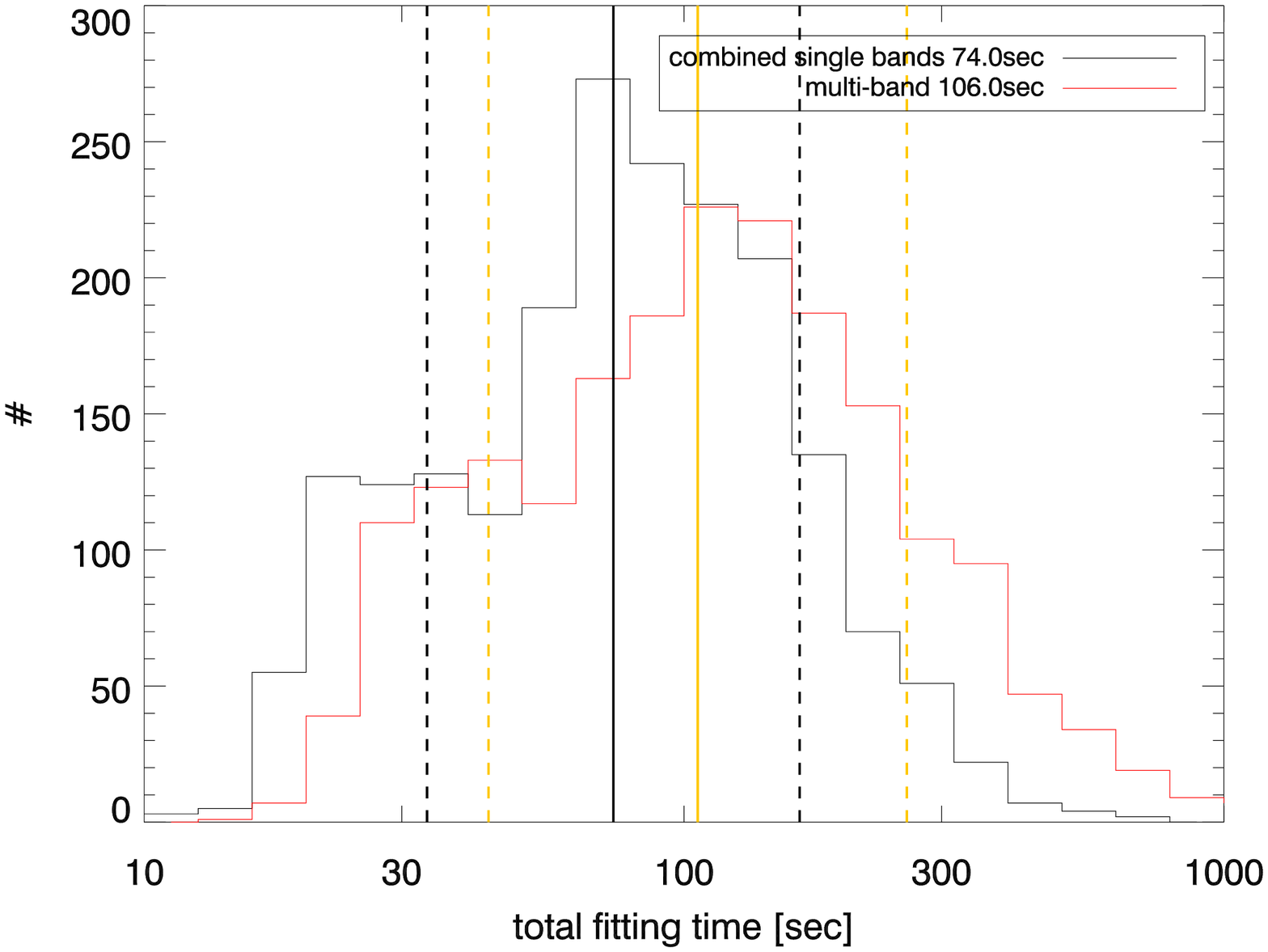}
\includegraphics[width=0.50\textwidth, trim=60 20 0 20, clip]{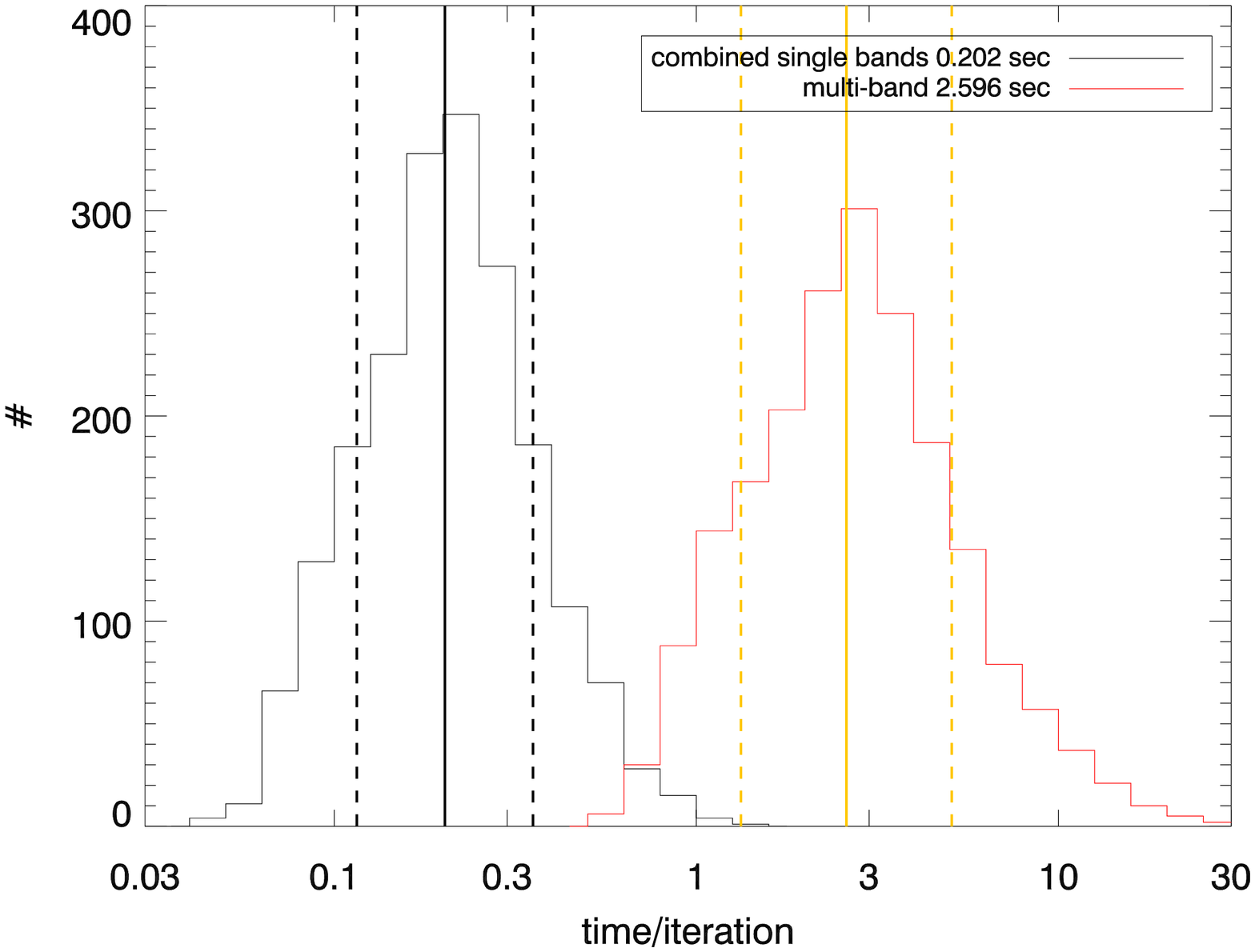}
\includegraphics[width=0.50\textwidth, trim=60 20 0 20, clip]{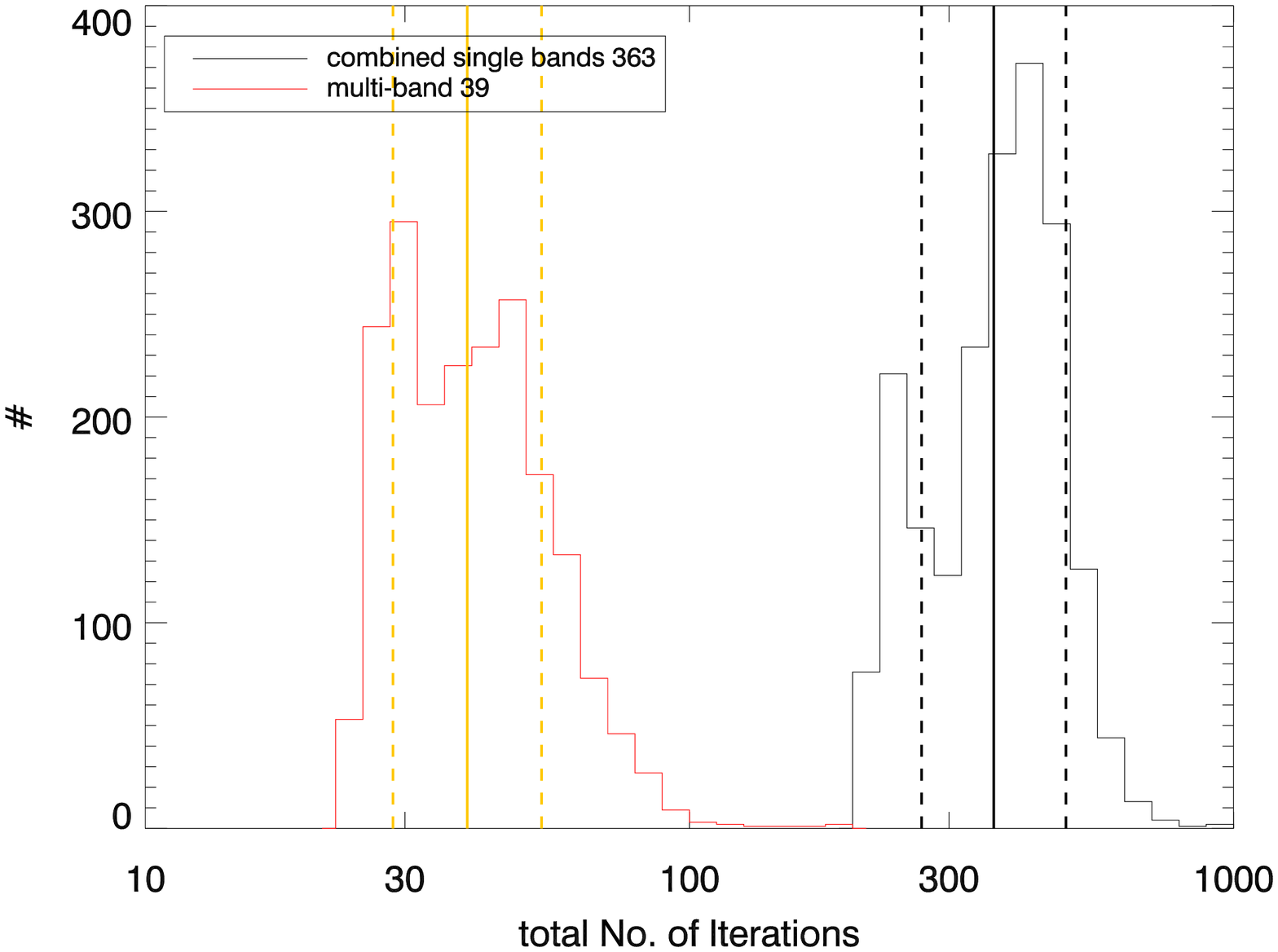}
\caption{CPU time histograms. 
Red histograms represent times for Mode\_M fitting, black histograms give the sum of the times for the nine Mode\_S1 fits. This figure only contains values from objects where all nine Mode\_S2 and Mode\_M fits exist, to allow a fair comparison between the methods. Vertical lines show the mean and rms scatter for both histograms. The top panel shows the total fitting time as returned by \galfitm. Multi-band fits take longer than single-band fits. The main reason for this is seen in the second panel, every individual iteration takes substantially longer in Mode\_M compared to Mode\_S2. This effect is nearly, but not entirely, canceled out by the smaller total number of iterations needed to derive all 9 band fits (lower panel).}
\label{fig15}
\end{center}
\end{figure}

Potentially, multi-band fitting may require less computing time to produce useful fits for large samples of galaxies, when compared with the single-band approach, as in principle we have fewer free parameters being constrained by the same dataset.  As we have fit samples using both methods, we can directly make this comparison.  Here we consider only the fitting time, i.e., the processing time required to process the galaxy with \galfitm.  We ignore the time it takes for the fit to be initially set-up, as this is similar for both methods. \galapagos does most jobs on a band-by-band basis, so the required setup time should generally be the same.  However, for the multi-band method not all steps in the code are repeated for all bands; e.g. deblending/masking decisions are common for all bands and only decided once, whereas Mode\_S1 makes these decisions on each band independently. Some further optimization of this part of \galapagos-2 is also possible in the future.

When comparing the pure fitting time for the simulated objects, we find that the multi-band fits are in fact slower than the nine individual band fits combined. The top panel of Fig.~\ref{fig15} shows a histogram of the overall fitting time for a small subset of objects for which we have all nine individual band fits, and for which such a comparison is possible. 
The fitting time will depend on how many secondary objects with free parameters, i.e. those fainter than the primary, are included in the fits. The number of secondaries depends upon the deblending, and is greater for multi-band detection.  To compare fitting times fairly, we therefore plot results for Mode\_S2 and Mode\_M.
While the combined single-band fits take around 74 seconds/object, the multi-band fits take about 106 seconds/object.  

The most time consuming part of the \galfitm algorithm is producing the model derivatives, which describe how the model images would change for small variations in each of the parameters.  The derivatives must be calculated for every pixel and convolved with the PSF, which is an expensive operation.  The Levenberg-Marquardt algorithm requires these derivative images, one set for each band being used in the fit, to be calculated with respect to each parameter, at every iteration step.  We would therefore naively assume that the fitting time per iteration should be roughly proportional to the number of free parameters. 

For instance, a single-band, single-\sersic fit has 7 free parameters (magnitude, $x$-centre, $y$-centre, half-light radius, \sersic index, axis ratio and position angle), and hence 7 derivative images must be calculated by \galfitm each iteration.  However, this is done independently for the 9 input images.  The multi-band fits carried out for this work use 19 free parameters (9 for magnitude, 1 $x$-centre, 1 $y$-centre, 3 for half-light radius, 3 for \sersic index, 1 position angle and 1 axis ratio), and derivatives must be calculated for 9 bands.  Hence, $19 \times 9 = 171$ derivatives must to be calculated for each iteration. By this estimate, each multi-band iteration should take $171/7 \approx 24$ times longer than each single-band iteration (this factor should be the same for any number of secondary objects included in the fit).  In reality, from the middle panel of Fig.~\ref{fig15}, we see that the difference is somewhat less: a factor of $\sim 13$.  This is probably because the derivatives of higher order Chebyshev parameters are less expensive to calculate than the zeroth-order parameters and, similarly, the standard parameters in single-band fitting.

When one accounts for the fact that the single-band fit must be run for each of the bands, the multi-band method gains by a factor of 9.  If the number of iterations required for each method were the same, the multi-band method would take $13/9$ times as long and hence would be $\sim 40$ per cent slower.  Indeed the number of iterations required for each fit are very similar, with the single-band approach requiring about 9 times more iterations, as the fit must be performed for each band, as shown in the lower panel of Fig.~\ref{fig15}).
For these bright galaxies, with complete sets of single-band fits, the multi-band method therefore takes roughly $40$ per cent more fitting time compare to the Mode\_S2 single-band fits. 
Note that, for single-band detection, the number of secondaries that are fit is reduced, and hence the combined time of the Mode\_S1 fits take only $\sim39$ seconds compared to $\sim81$ seconds for the same sample of galaxies in Mode\_M.  The multi-band Mode\_M method therefore takes roughly twice the time of Mode\_S1.

The number of successfully fit objects is much higher in the case of multi-band fitting (see Table~\ref{tab_2}). From the above arguments, this should result in a much larger number of galaxies with a complete set of multi-band measurements per unit processing time.  However, the above speed comparisons are limited to the cases where we obtain successful fits in all nine single-band fits and the multi-band fit.  For these, the number of iterations are similar, but for the majority of galaxies, with lower signal-to-noise, the single-band fits (many of which fail) appear to require significantly fewer iterations than the multi-band fits.  To get an overall impression of the speed of single- versus multi-band, we divide the number of galaxies with a complete set of multi-band measurements by the total time taken to fit the entire sample, to produce a \lq complete success\rq\ rate.
For  Mode\_S2 this is $\sim2.4$ \lq complete successes\rq\ per hour (if one requires all $ugrizYJHK$ band values; $23.5$ requiring just the $griYHK$ bands), while for Mode\_M it is $29.3$ \lq complete successes\rq\ per hour (full $ugrizYJHK$). In single-band fitting, much more CPU time is spent on objects that are later discarded as unreliable by the cleaning process. So in terms of useful fits per unit time, multi-band fitting compares very favourably.

\subsection{Storage requirements}
\label{sec_space}
Another practical aspect to take into account, especially for large datasets, is the disk space used. If a dataset uses less disk space it is easier to handle and less space will be needed for big surveys. For the sample of 29205 real objects, we use:
\begin{itemize}
\item 51 GB for the Mode\_M fitting,
\item 78 GB for the Mode\_S1 fitting.
The difference is largely down to having SExtractor files for each band instead of one set for the whole process. In comparison, the single-band fitting does save some disk space because not as many galaxy postage stamps have to be made for the individual bands as less objects are detected.
\item In Mode\_S2, cutting all postage stamps when running the object detection on a multi-wavelength image increases the disk space required to a total of 100GB, so a reduction of disk space $\sim50$ per cent is achieved when changing from single-band to multi-band fitting.
\end{itemize} 
Multi-wavelength fitting has the smallest space-to-result ratio, especially when taking into account that one gets usable results for a much bigger sample. We get valid results (fits that did not run into fitting constraints) for 15566 objects, compared to 209 galaxies with valid fits in $u$-band, and 3276 ($g$-band), 5802 ($r$-band), 6804 ($i$-band), 2117 ($z$-band), 3912 ($Y$-band), 2530 ($J$-band), 5432 ($H$-band) and 5190 ($K$-band), respectively. The overlap of all these samples is a mere 87 galaxies (or 992 when only $griYHK$ are taken into account).
 
For the set of simulated galaxies, we find a similar result:
For a total of 42229 simulated objects, the disk space needed is:
\begin{itemize}
\item 64 GB for Mode\_M fitting,
\item 73 GB for Mode\_S1 fitting,
\item 114GB for Mode\_S2 fitting.
\end{itemize} 
Object numbers with useful values here are: 714 ($u$-band), 6022 ($g$-band), 10362 ($r$-band), 12326 ($i$-band), 4139 ($z$-band), 9627 ($Y$-band), 7965 ($J$-band), 12593 ($H$-band) and 12602 ($K$-band) with an overlap of 305 objects (2560 objects). This compares to 23632 objects for which we get a valid result in all bands when using multi-band fitting.
Overall, our multi-band fitting procedure can help to save nearly 50 per cent of disk space required.

\section{Colour--Magnitude Diagram from multi-band fitting}
\label{sec_cmd}

The multi-wavelength Mode\_M technique that we have presented and tested, both above and in accompanying MegaMorph papers, has subsequently been applied to a larger sample of galaxies from the GAMA survey.  Here we use this sample to present preliminary colour--magnitude and size--magnitude diagrams.  This is done for two purposes: as a consistency check of the fitting results, and to demonstrate where the multi-band technique offers scope for quickly improving science results.

Single-band single-\sersic fits, in $ugrizYJHK$, have been performed by K12 for all spectroscopic targets in the GAMA sample, using their own code, \sigmakelvin, to run \galfitthree.  However, a detailed comparison between our single-band fits and those conducted in K12 is beyond the scope of this paper.  We therefore decided that comparing their single-band fitting results to our multi-band results could potentially lead to wrong conclusions regarding single vs multi-band fitting, due to the use of different wrapper scripts. However, in earlier tests, we have established that both procedures, \sigmakelvin (K12) and our own method based on \galapagos \citep{galapagos} generally return very similar results when used on single-band data, and comparable values when used on multi-band data in our setup. In order to show the strength of our code, we avoid using any additional information about the objects other than \galapagos fitting results and spectroscopic redshifts. The latter are provided by GAMA, and used only to calculate physically meaningful parameters such as absolute magnitudes and sizes, for both our single and multi-band fitting results. For these transformations, a cosmological model of $\Omega_{\Lambda}=0.7$, $\Omega_{\rm m}=0.3$ and $H_{0}=70$~$h_{70}$~km~s$^{-1}$~Mpc$^{-1}$ is assumed.

In Section~\ref{sec_real} we have carried out both single and multi-band fits on a small subarea of the GAMA survey field. Compared to the \galapagos version used and tested above, we use a minimally changed version in this section. The main change was the introduction of a feature that allows the code to only target certain objects instead of targeting all detections in a survey. 
For this chapter we applied (using Mode\_M) multi-band version of \galapagos-2 and \galfitm to all objects that have both full wavelength coverage in the GAMA 9-hour (G09) field and redshifts provided by the GAMA team \citep[][Liske, in prep]{Driver2011}. Additionally, we include all objects within 66 arcsec ($\sim200$ pixels) to one of these GAMA objects and brighter than said object, as \galapagos works on objects in brightness order from bright to faint\footnote{During the development of the code, it proved useful for fitting stability if results for brighter neighbouring objects are already known at the time they have to be used as neighbours. This issue is discussed in detail in \citet{galapagos}.}.  A good fraction of these neighbouring objects are bright stars.
Given the brightness of the objects targeted, this slightly changed code should not influence fitting results significantly but allows the code to only target 144261 objects instead of the total 1401969 detections in the field, reducing the sample size by $\sim90$ per cent and saving a huge amount of fitting time. 

Of these 144261 objects, we have spectroscopic redshifts for 43617 objects. \galapagos returned a result for 43572 ($99.9$ per cent), and 33696 ($77.3$ per cent) have useful fits and a full set of multi-wavelength parameters is provided. The fraction of galaxies with valid fits is significantly higher in this sample of galaxies compared to the samples used earlier in this paper (e.g. the left part of Table~\ref{tab_1}), mostly because for this chapter bright galaxies -- that are potentially easier to fit -- have been primarily targeted, whereas in the previous chapters, we had targeted all detected objects. Also, as stars have no redshifts assigned, they are removed when deriving these numbers, boosting the success rate.

The resulting catalogue of this procedure is used in both Vulcani et al (2012, in prep) and this section, in which we briefly examine the colour-magnitude diagram of galaxies.

\begin{figure*}
\begin{center}
\includegraphics[width=0.98\textwidth, trim=0 0 0 0, clip]{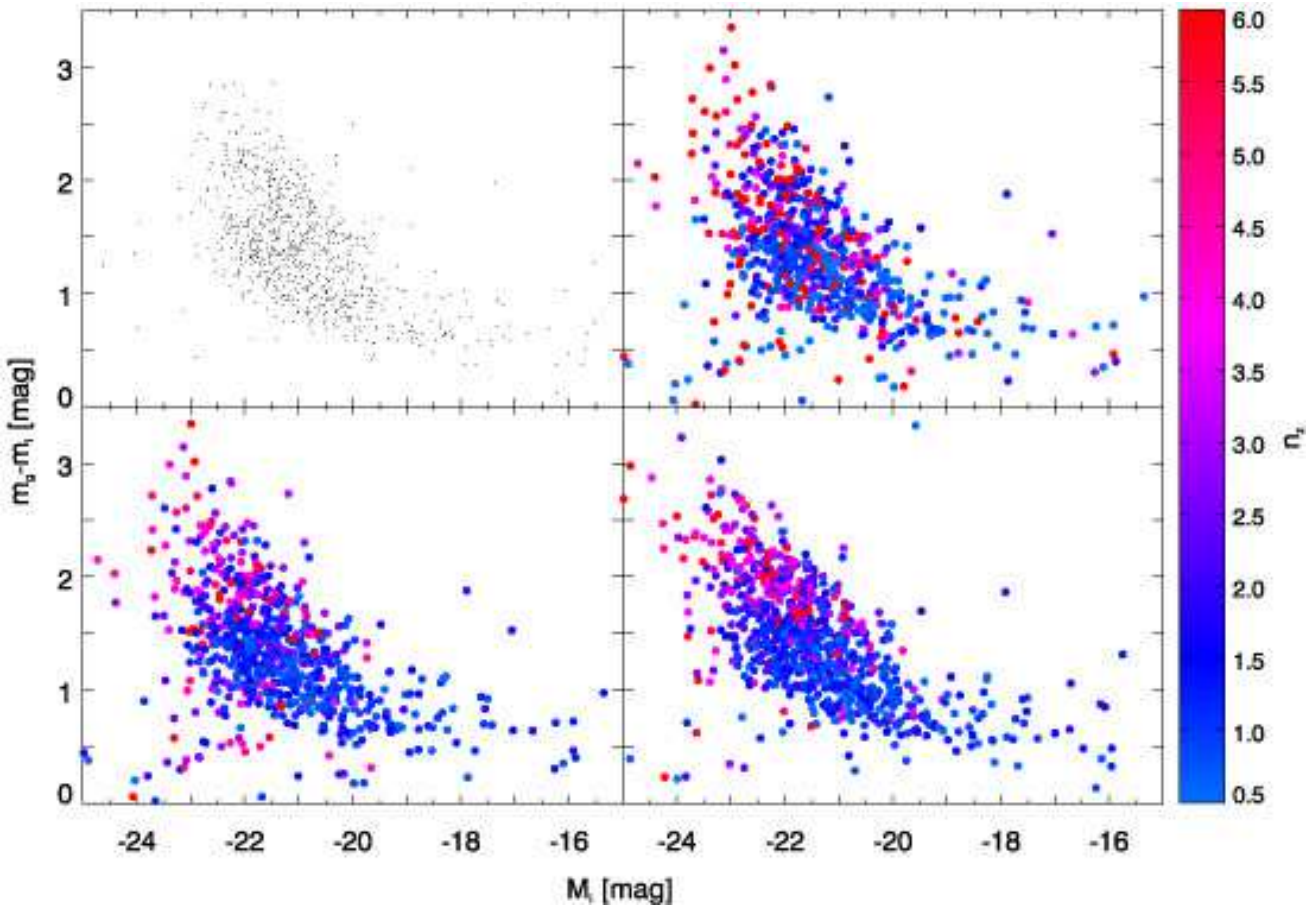}\\
\caption{Colour-magnitude diagram (y-axis shows purely observed colours). Top-left: when using aperture photometry (GAMA \sex MAG\_AUTO from catalogue ApMatchedCatv03); Top-right: when using single band fitting (Mode\_S1), colour-coding using single-band $n_z$ values; Bottom-left: Mode\_S1 fitting results, colour coding using Mode\_M $n_z$ values. Bottom-right: using only Mode\_M multi-band fitting, colour coded by $z$-band \sersic index from multi-band fitting (blue=low ($<0.5$), red=high($>6$)). $n_z$ is assumed to be the best separator and we argue in this paper for multi-band to be a more reliable value than single-band \sersic indices. A separation between \lq early-type\rq\ (high-$n$) and \lq late-type\rq\ (low-$n$) galaxies becomes much clearer when Mode\_M $n_z$ values are used for colour coding. Please be reminded that the number of points in this analysis is mostly defined by the number of redshifts available, and \emph{not} by the number of successful fits. As a consequence, single and multi-band figures contain a very similar number of points.}
\label{fig18}
\end{center}
\end{figure*}
In Fig.~\ref{fig18}, we show $g$-$i$ vs. $M_i$ Colour--Magnitude Diagrams (CMDs) as they would be recovered by the different techniques, colour-coded by \sersic index $n_z$. The top-left panel shows values from aperture photometry, for which we use matched-aperture photometry obtained in a manner similar to \citet{Hill} and to be described in Liske et al. (in prep). The points in this panel have not been colour coded as this information would have to be provided by a second code, aperture photometry itself does not give an estimate for galaxy \lq morphology\rq. The top-right figure shows the CMD as given entirely by Mode\_S1 fitting, e.g. the colour coding according to $n_z$ as provided by single-band fitting. High-$n$ and low-$n$ galaxies show mixed colours in this figure, a separation, although indicated, is weak.
The bottom-left figure shows the same single-band values, but colour-coded using multi-band $n_z$. In this figure, the separation between low-$n$ and high-$n$ galaxies is better. The reader should keep in mind that \sersic index is not a robust value to be used for classifying galaxies in a morphological sense, especially early-type galaxies are known to show a wide range of \sersic\ indices, e.g., they are following a magnitude--$n$ relation \citep[see e.g.][]{Grahamb}. Additionally, \sersic indices are sensitive to measurement errors, for example, a faint AGN or bright star cluster in the centre of a galaxy would boost $n$ to show a high value instead of the intrinsic value of the underlying host galaxy. We use $n_z$ only as a very crude estimator in this analysis and would strongly discourage users to use it for more than that. Especially, it cannot replace a visual classification of morphology. 

In the bottom-right panel of the figure we show the CMD as recovered entirely by multi-band fitting. The division between low-$n$ and high-$n$ galaxies becomes much stronger, and the scatter in the figure gets reduced, with high-$n$ galaxies showing significantly redder colours than low-$n$ galaxies. This is expected by simple morphology--stellar-population arguments. It should be stated here that galaxies are plotted in random order, not by increasing \sersic index, the lack of blue points on the red-sequence and red points in the blue cloud, even when using $n_z$ as colour index, is real and not due to plotting procedure. 

Please note that the number of points in the single-band panels (727 objects) is only slightly smaller than the one in the multi-band panel (795 objects). This is true as long as no $u$- or $z$-band data is used and only bright galaxies which were targeted for spectroscopic redshifts are considered ($m_r<19.8$). Given that this completeness effect has already been discussed in detail above, we decided to not use $u$- or $z$-band data here, although it might show a better separation of the red sequence and the blue cloud. Instead we concentrate on the fitting accuracy itself.

\begin{figure*}
\begin{center}
\includegraphics[width=0.98\textwidth, trim=0 0 0 0, clip]{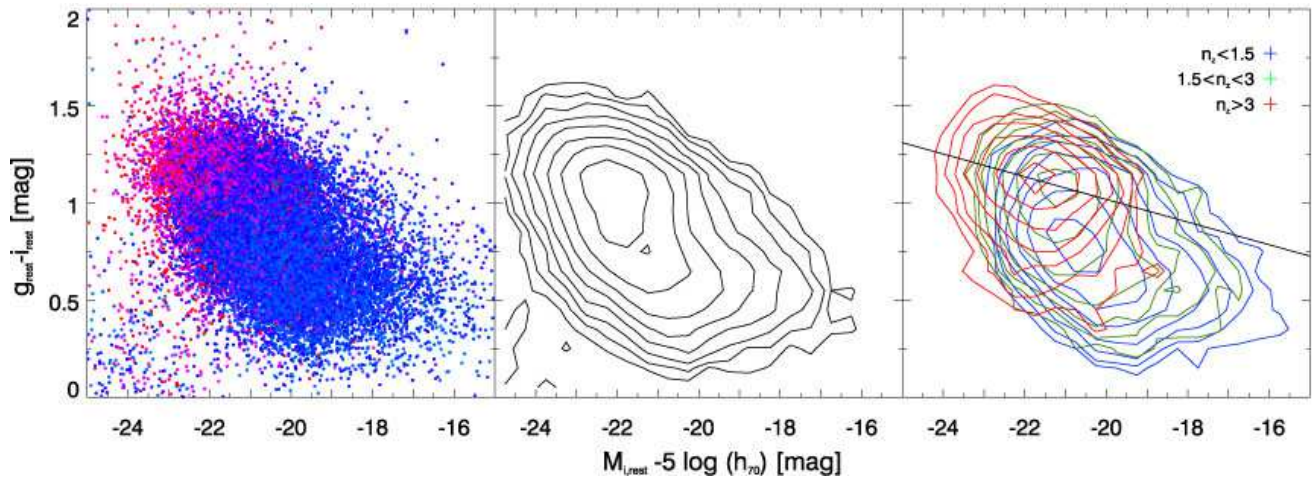}\\
\caption{Colour magnitude diagram (both axes restframe and k-corrected). Left: all GAMA G09 galaxies with redshifts; middle: contours for all galaxies in the left figure; right: contours for samples split by \sersic index. Points are colour coded by $z$-band \sersic index (blue=low ($<0.5$), red=high($>6$)). A clear separation between low-$n$ and high-$n$ galaxies is visible (please note that the points are plotted in random order, not by increasing \sersic index), especially when plotting three $n$ bins ($n<1.5$, $1.5<n<3$, $4<n$) as contours (right figure). Even in the contours for all galaxies (Center), the shape of the contours indicate 2 overlapping sequences, which is what one expects. The solid line in the right figure indicates a separator between the red sequence and the blue cloud as empirically determined by \citet{Gavazzi}}
\label{fig18.2}
\end{center}
\end{figure*}

Given the overall shape of this figure, it should be emphasized that this figure shows purely observed colours, e.g., no k-corrections or dust-corrections have been applied. The main reason for this is that we want to show the difference between the techniques, not sample sizes. Robust k-corrections would require magnitude measurements in many bands which are automatically provided by multi-band fitting. Single-band fitting, however, can produce results in some bands, not in the others, providing reliable k-correction only for a subset of galaxies. We will apply k-correction to all multi-band fits in Fig.~\ref{fig18.2}.

In Fig.~\ref{fig18.2} we show the CMD for the 28737 GAMA galaxies in the G09 field for which we were able to recover useful fitting values using multi-band fitting procedures (Mode\_M). In contrast to Fig.~\ref{fig18} values in this figure have been k-corrected to show restframe values, using \textsc{kcorrect} \citep{kcorrect}, version 4\_2. This is possible here because the multi-band fitting procedure returns magnitudes for all $ugrizYJHK$ bands, whereas any single-band routine could have some magnitude values missing in certain bands, making k-correction both harder and less accurate. The middle panel shows contours of the same distribution for all objects. An indication for a bimodality and galaxy separation can be seen. The right panel shows the same, but galaxies have been split up into broad $n_z$-value bins, each with a large number of galaxies (7000-12000 galaxies), in order to show the separation between the different \sersic index populations more clearly. The bimodality becomes very visible here, although the red sequence is not as tight as one would hope when a clean classification scheme is used instead of $n_z$ values.

\begin{figure}
\begin{center}
\includegraphics[width=0.50\textwidth, trim=0 0 0 0, clip]{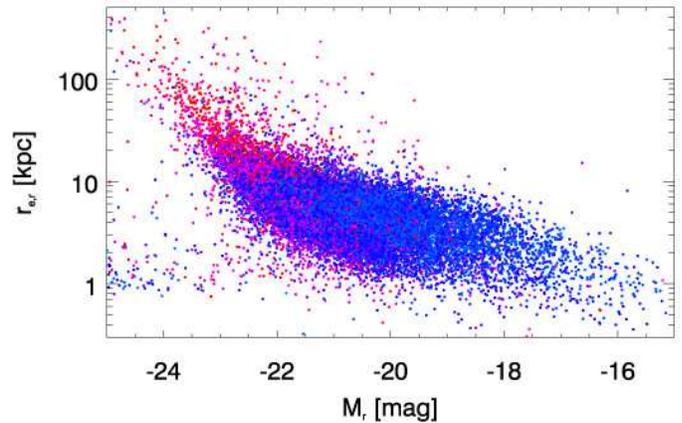}\\
\caption{The magnitude-size relation. Points are colour coded by $n_z$ matching the previous s (blue=low ($<0.5$), red=high($>6$)). As discussed in the text, there is a visible population of very bright galaxies with increased $n_z$ and very small sizes, visible as a cloud of red points at the lower-left edge of the cloud of blue points.}
\label{fig19}
\end{center}
\end{figure}

A few remarks should be made on these findings: 
\begin{itemize}
\item The solid line in the rightmost panel of Fig.~\ref{fig18.2} indicates not a fit to a red-sequence, but a separator, a fit to the \lq green valley\rq, as determined by \citet{Gavazzi} from a sample of ~4100 galaxies using visual classification. \citet{Gavazzi} use SDSS DR7 Petrosian magnitudes for their analysis. Due to the nature of these -- the $r$-band alone defines the size of the aperture in which the magnitude is measured -- we would expect a widening of our red sequence due to the ability of our method to vary galaxy size with wavelength, hence measuring a slightly different magnitude at other wavelengths than the ones provided by \citet{Gavazzi}. We argue that our magnitudes more closely resemble the true magnitudes at these wavelengths in comparison with SDSS Petrosian magnitudes which should be bulge-dominated given that the aperture size is defined in a reddish ($r$) band \citep[please also see][]{Graham}. Unfortunately for this comparison, this bulge-domination of the magnitudes measured helps to get a tight colour--magnitude relation and a better separation from the blue cloud. 
\item It is very interesting to see that the contours for galaxies with intermediate $n_z$ ($1.5<n<3$; shown in green) seem to have a peak on the red sequence and a long \lq tail\rq\ towards the blue cloud. This is exactly what would be expected for a transition population.
\item Our galaxy \lq separator\rq\ $n_z$ should only give a very rough estimation of the galaxy \lq morphology\rq\ as it is easily biased and boosted by central light sources such as faint AGN and/or bright star clusters. Hence the high-$n$ sample will include some of these objects, although they would visually be classified as disks, suggesting a place in or closer to the blue cloud. A way to test this hypothesis is to look at the magnitude-size relation of all objects. When a single \sersic profile is used to fit an object which contains both a disk and a point source, one would expect to fit both a high \sersic index and a small size, as well as a bright magnitude as this would include the flux of the nuclear source. When plotting the magnitude--size relation of all the objects in our sample (see Fig.~\ref{fig19}), we can see that indeed many of these high-$n$ galaxies are the brightest and smallest galaxies, seen as a purple/red cloud of points, in what would traditionally be called the \lq blue cloud\rq\ of disk-dominated galaxies.
\item Even using perfect classifications, one would always find both blue ellipticals in the blue cloud and disks that show red colours given their dust properties and possibly seen edge-on \citep[e.g.][]{Rowlands, Wolf2009}. An entirely clean separation is impossible to achieve.
\item Please again bear in mind that no other correction (e.g., for dust content) than k-correction has been applied to any of the values in Fig.~\ref{fig18.2}.   This preliminary analysis is intended only as a demonstration of the advantages of multi-band fitting.
\end{itemize}

Our simple analysis, using no other information than redshifts and colours as directly measured by our multi-band fitting code, clearly shows the known separation between the red sequence and the blue cloud of galaxies, even when using a crude and inaccurate proxy for \lq galaxy morphology\rq\ such as \sersic index $n_z$. Although a separation of the two galaxy populations by using \sersic index as a \lq classifier\rq\ will never be ideal, it proves a powerful statistical tool in this example. 

We assume that most of the outliers in Fig.~\ref{fig18.2} are due to an unflagged error/uncertainty in our fitting pipeline and our quick and rough analysis. Especially, we are not (yet) running bulge-disk decomposition on our data, and fitting single \sersic profiles to B/D composite galaxies is known to be non-ideal. We assume that the fitting process went wrong in these examples in a way that we were unable to detect in a simple automated fashion. Statistically, these $\sim$100--200 outliers become insignificant in a sample of 28737 galaxies.

\section{Summary}
\label{sec_summary}
In this paper, we have presented new versions of both \galfit (which we call \galfitm) and \galapagos (\galapagos-2) to automate the process of fitting two-dimensional, single-Sersic models to large samples of galaxies.  Our novel approach uses multiple images taken at different wavelengths simultaneously, instead of single-band data, as used by the current published versions of these codes. We have tested our multi-band modifications extensively, both on real and simulated data and compared their performance to the single-band versions of the codes. For real data, we have selected 9-band ($ugrizYJHK$) data from the GAMA survey \citep{Driver2011}. Simulations were created following H07 to largely reproduce the parameter distributions of the real data. This creates a dataset as realistic as necessary, while taking out the issues of image confusion by nearby stars and internal galaxy structure, as well as providing a truth value for comparison against those recovered recovered by the fitting process.

From both real and simulated data, we can draw very similar conclusions:
\begin{itemize}
\item Multi-band fitting significantly improves the recovery of galaxy magnitudes, at least in the lower \sn bands (particularly $u$ and $z$-band in this analysis, see e.g. Fig.~\ref{fig6.2} and Table~\ref{table_mag_error}).  The improvement for both sizes (see e.g. Fig.~\ref{fig9} and Table~\ref{table_re_error}) and \sersic indices (see e.g. Fig.~\ref{fig12} and Table~\ref{table_n_error}) is significant over the entire wavelength range tested in this paper. The multi-band approach returns more accurate results than treating the bands  independently, and is able to make useful measurements for very faint galaxies. In our analysis, multi-band fitting is able to derive reliable values for $\sim75$ times more galaxies in SDSS $u$-band data than the single-band approach.
\item The multi-band version of our code is more stable, in terms of providing fitting results for objects where single-band fits fail to return a result.  This is because the single-band fits occasionally return results which are affected by the constraints that are provided by \galapagos and used by \galfitm during the fitting process. \galfitm more frequently returns a valid results when multi-band data is used. This is always the case for real galaxy images, while in our simulations this is true when fits in several bands are required, e.g. for SED modelling. 
\item Physically meaningful variations of galaxy parameters with observed wavelength are used in the fitting method presented in this paper. They enable accurate wavelength interpolation, useful to derive restframe galaxy parameters, and provide the user with a full set of parameters at all wavelengths. Single-band fits on the other hand are more likely to fail in at least individual bands, while returning a result in other wavelengths, thus leaving a less complete sample from which to interpolate values.
\item While running single-band fits returns more or less equivalent results for bright galaxies, multi-band fitting provides a more homogenous dataset as decisions about object detection, deblending and masking are necessarily the same. Single-band fits can vary these decisions between different bands.
\item In addition to providing results for a higher fraction of galaxies, the multi-band code can be run on fainter objects, as the object detection can be sensibly carried out on a co-added image. The number of galaxies that can be used for scientific purposes is increased by a factor of $\sim2$ if only single-band values are needed up to $\sim16$ if multi-band data ($griYHK$) values are required -- a factor of $\sim180$ is achieved when a full set of parameters in $ugrizYJHK$ data are needed.
\end{itemize}

Over all, we conclude that using multi-band fitting instead of single-band fitting improves both quality and quantity of galaxy fits on survey data. It therefore provides an excellent tool for modern surveys and enables a step forward in many areas of astronomy, especially where sample size and magnitude constraints limit current work.

\section{Prospects}
\label{sec_prospect}
In this paper, we have exclusively tested \galfitm and \galapagos-2 on single-\sersic profiles. Further plans most importantly include implementation of bulge-disk decomposition into \galapagos-2. As some of the conclusions in this paper might change due to both the additional flexibility and the more challenging fits even on bright objects, we will discuss these changes in detail in the follow-up paper. The intention is to separate galaxy bulges and disks more accurately than previously possible by using the information provided by the different colours of the respective components.

The colour gradients (variations of $n$ and $r_{\rmn{e}}$ with wavelength) seen in this paper come from a combination of both mixing the bulge and disk stellar populations when fitting them with a single-\sersic profile and the effects of dust (e.g., \citealt{Moellenhoff,2012IAUS..284..306P}).  
Fully understanding the observed trends will require careful consideration of both multi-wavelength bulge-disk decompositions and spatially-resolved models of dust attenuation (e.g., \citealt{Popescu}),  constrained by spaceborne UV \& FIR/submm photometry."

Another issue, which we have tried to encompass by testing with both real and simulated data in this work, but which has otherwise been neglected, is the performance of multi-band \galfitm (and profile fitting in general) in the presence of morphological features, such as bars and spiral arms.  We aim to investigate this in future by drawing on the data and resources of Galaxy Zoo \citep{2011MNRAS.410..166L,2011MNRAS.411.2026M}

Although \galfitm already supports multiple components, an adapted version of \galapagos-2 will be needed for this work.  This version will be presented in a different paper (\Haeussler et al., in prep). It is currently under development and testing and we aim for it to be ready for public release in $\sim$6 months.  We will publish our codes and tools in a follow-up paper when bulge-disk decomposition is successfully incorporated into the code and encourage astronomers to use these tools on their data.

\section*{Acknowledgements}
We would like to thank the editor and an anonymous referee for very useful comments that improved the paper.
We thank Chien Peng for productive meetings and helpful discussions during the development of the code and Eric Bell for useful comments and discussion on a first draft of this paper. 

This publication was made possible by NPRP grant \# 08-643-1-112 from the Qatar National Research Fund (a member of Qatar Foundation). The statements made herein are solely the responsibility of the authors.  BH and MV are supported by this NPRP grant.
SPB is supported by an STFC Advanced Fellowship. We thank Carnegie Mellon University in Qatar and The University of Nottingham for their hospitality.

GAMA is a joint European-Australasian project based around a spectroscopic campaign using the Anglo-Australian Telescope. The GAMA input catalogue is based on data taken from the Sloan Digital Sky Survey and the UKIRT Infrared Deep Sky Survey. Complementary imaging of the GAMA regions is being obtained by a number of independent survey programs including GALEX MIS, VST KIDS, VISTA VIKING, WISE, Herschel-ATLAS, GMRT and ASKAP providing UV to radio coverage. GAMA is funded by the STFC (UK), the ARC (Australia), the AAO, and the participating institutions. The GAMA website is http://www.gama-survey.org/ .

\bsp

\bibliography{references}

\begin{thebibliography}{52}
\expandafter\ifx\csname natexlab\endcsname\relax\def\natexlab#1{#1}\fi

\bibitem[{Abramowitz \& Stegun(1965)}]{cheb_poly}
Abramowitz M., Stegun I., 1965, Handbook of Mathematical Functions: With
  Formulas, Graphs, and Mathematical Tables, Applied mathematics series. Dover
  Publications

\bibitem[{{Allen} {et~al}\mbox{.}(2006){Allen}, {Driver}, {Graham}, {Cameron},
  {Liske}, \& {de Propris}}]{Allen2006}
{Allen} P.~D., {Driver} S.~P., {Graham} A.~W., {Cameron} E., {Liske} J., {de
  Propris} R., 2006, \mnras, 371, 2

\bibitem[{{Barden} {et~al}\mbox{.}(2012){Barden}, {H{\"a}u{\ss}ler}, {Peng},
  {McIntosh}, \& {Guo}}]{galapagos}
{Barden} M., {H{\"a}u{\ss}ler} B., {Peng} C.~Y., {McIntosh} D.~H., {Guo} Y.,
  2012, \mnras, 422, 449

\bibitem[{{Bell} \& {de Jong}(2000)}]{2000MNRAS.312..497B}
{Bell} E.~F., {de Jong} R.~S., 2000, \mnras, 312, 497

\bibitem[{{Bertin} \& {Arnouts}(1996)}]{Bertin}
{Bertin} E., {Arnouts} S., 1996, A\&AS, 117, 393

\bibitem[{{Bertin} {et~al}\mbox{.}(2002){Bertin}, {Mellier}, {Radovich},
  {Missonnier}, {Didelon}, \& {Morin}}]{terapix}
{Bertin} E., {Mellier} Y., {Radovich} M., {Missonnier} G., {Didelon} P.,
  {Morin} B., 2002, in Astronomical Society of the Pacific Conference Series,
  Vol. 281, Astronomical Data Analysis Software and Systems XI, {Bohlender}
  D.~A., {Durand} D., {Handley} T.~H., eds., p. 228

\bibitem[{{Blanton} \& {Roweis}(2007)}]{kcorrect}
{Blanton} M.~R., {Roweis} S., 2007, \aj, 133, 734

\bibitem[{{Conselice}(2003)}]{Conselice2003}
{Conselice} C.~J., 2003, ApJS, 147, 1

\bibitem[{{de Souza} {et~al}\mbox{.}(2004){de Souza}, {Gadotti}, \& {dos
  Anjos}}]{budda}
{de Souza} R.~E., {Gadotti} D.~A., {dos Anjos} S., 2004, \apjs, 153, 411

\bibitem[{{de Vaucouleurs}(1948)}]{deVauc}
{de Vaucouleurs} G., 1948, Annales d'Astrophysique, 11, 247+

\bibitem[{{Driver} {et~al}\mbox{.}(2011){Driver}, {Hill}, {Kelvin}, {Robotham},
  {Liske}, {Norberg}, {Baldry}, {Bamford}, {Hopkins}, {Loveday}, {Peacock},
  {Andrae}, {Bland-Hawthorn}, {Brough}, {Brown}, {Cameron}, {Ching}, {Colless},
  {Conselice}, {Croom}, {Cross}, {de Propris}, {Dye}, {Drinkwater}, {Ellis},
  {Graham}, {Grootes}, {Gunawardhana}, {Jones}, {van Kampen}, {Maraston},
  {Nichol}, {Parkinson}, {Phillipps}, {Pimbblet}, {Popescu}, {Prescott},
  {Roseboom}, {Sadler}, {Sansom}, {Sharp}, {Smith}, {Taylor}, {Thomas},
  {Tuffs}, {Wijesinghe}, {Dunne}, {Frenk}, {Jarvis}, {Madore}, {Meyer},
  {Seibert}, {Staveley-Smith}, {Sutherland}, \& {Warren}}]{Driver2011}
{Driver} S.~P. {et~al.}, 2011, \mnras, 413, 971

\bibitem[{{Driver} {et~al}\mbox{.}(2007){Driver}, {Popescu}, {Tuffs}, {Liske},
  {Graham}, {Allen}, \& {de Propris}}]{2007MNRAS.379.1022D}
{Driver} S.~P., {Popescu} C.~C., {Tuffs} R.~J., {Liske} J., {Graham} A.~W.,
  {Allen} P.~D., {de Propris} R., 2007, \mnras, 379, 1022

\bibitem[{{Gavazzi} {et~al}\mbox{.}(2010){Gavazzi}, {Fumagalli}, {Cucciati}, \&
  {Boselli}}]{Gavazzi}
{Gavazzi} G., {Fumagalli} M., {Cucciati} O., {Boselli} A., 2010, \aap, 517, A73

\bibitem[{{Gonzalez-Perez} {et~al}\mbox{.}(2011){Gonzalez-Perez}, {Castander},
  \& {Kauffmann}}]{2011MNRAS.411.1151G}
{Gonzalez-Perez} V., {Castander} F.~J., {Kauffmann} G., 2011, \mnras, 411, 1151

\bibitem[{{Graham}(2011)}]{Grahamb}
{Graham} A.~W., 2011, arXiv/1108.0997

\bibitem[{{Graham} \& {Driver}(2005)}]{GrahamDriver}
{Graham} A.~W., {Driver} S.~P., 2005, PASA, 22, 118

\bibitem[{{Graham} {et~al}\mbox{.}(2005){Graham}, {Driver}, {Petrosian},
  {Conselice}, {Bershady}, {Crawford}, \& {Goto}}]{Graham}
{Graham} A.~W., {Driver} S.~P., {Petrosian} V., {Conselice} C.~J., {Bershady}
  M.~A., {Crawford} S.~M., {Goto} T., 2005, AJ, 130, 1535

\bibitem[{{H{\"a}ussler} {et~al}\mbox{.}(2007){H{\"a}ussler}, {McIntosh},
  {Barden}, {Bell}, {Rix}, {Borch}, {Beckwith}, {Caldwell}, {Heymans},
  {Jahnke}, {Jogee}, {Koposov}, {Meisenheimer}, {S{\'a}nchez}, {Somerville},
  {Wisotzki}, \& {Wolf}}]{Haeussler2007}
{H{\"a}ussler} B. {et~al.}, 2007, ApJS, 172, 615

\bibitem[{{Hill} {et~al}\mbox{.}(2011){Hill}, {Kelvin}, {Driver}, {Robotham},
  {Cameron}, {Cross}, {Andrae}, {Baldry}, {Bamford}, {Bland-Hawthorn},
  {Brough}, {Conselice}, {Dye}, {Hopkins}, {Liske}, {Loveday}, {Norberg},
  {Peacock}, {Croom}, {Frenk}, {Graham}, {Jones}, {Kuijken}, {Madore},
  {Nichol}, {Parkinson}, {Phillipps}, {Pimbblet}, {Popescu}, {Prescott},
  {Seibert}, {Sharp}, {Sutherland}, {Thomas}, {Tuffs}, \& {van Kampen}}]{Hill}
{Hill} D.~T. {et~al.}, 2011, \mnras, 412, 765

\bibitem[{{Kelvin} {et~al}\mbox{.}(2012){Kelvin}, {Driver}, {Robotham}, {Hill},
  {Alpaslan}, {Baldry}, {Bamford}, {Bland-Hawthorn}, {Brough}, {Graham},
  {H{\"a}ussler}, {Hopkins}, {Liske}, {Loveday}, {Norberg}, {Phillipps},
  {Popescu}, {Prescott}, {Taylor}, \& {Tuffs}}]{Kelvin2012}
{Kelvin} L.~S. {et~al.}, 2012, \mnras, 421, 1007

\bibitem[{{La Barbera} \& {de Carvalho}(2009)}]{2009ApJ...699L..76L}
{La Barbera} F., {de Carvalho} R.~R., 2009, \apjl, 699, L76

\bibitem[{{La Barbera} {et~al}\mbox{.}(2010){La Barbera}, {de Carvalho}, {de La
  Rosa}, {Lopes}, {Kohl-Moreira}, \& {Capelato}}]{2010MNRAS.408.1313L}
{La Barbera} F., {de Carvalho} R.~R., {de La Rosa} I.~G., {Lopes} P.~A.~A.,
  {Kohl-Moreira} J.~L., {Capelato} H.~V., 2010, \mnras, 408, 1313

\bibitem[{{La Barbera} {et~al}\mbox{.}(2008){La Barbera}, {de Carvalho},
  {Kohl-Moreira}, {Gal}, {Soares-Santos}, {Capaccioli}, {Santos}, \&
  {Sant'anna}}]{2008PASP..120..681L}
{La Barbera} F., {de Carvalho} R.~R., {Kohl-Moreira} J.~L., {Gal} R.~R.,
  {Soares-Santos} M., {Capaccioli} M., {Santos} R., {Sant'anna} N., 2008,
  \pasp, 120, 681

\bibitem[{{Lackner} \& {Gunn}(2012)}]{LacknerGunn}
{Lackner} C.~N., {Gunn} J.~E., 2012, \mnras, 421, 2277

\bibitem[{{Lawrence} {et~al}\mbox{.}(2007){Lawrence}, {Warren}, {Almaini},
  {Edge}, {Hambly}, {Jameson}, {Lucas}, {Casali}, {Adamson}, {Dye}, {Emerson},
  {Foucaud}, {Hewett}, {Hirst}, {Hodgkin}, {Irwin}, {Lodieu}, {McMahon},
  {Simpson}, {Smail}, {Mortlock}, \& {Folger}}]{ukidss}
{Lawrence} A. {et~al.}, 2007, \mnras, 379, 1599

\bibitem[{{Lilly} {et~al}\mbox{.}(1998){Lilly}, {Schade}, {Ellis}, {Le Fevre},
  {Brinchmann}, {Tresse}, {Abraham}, {Hammer}, {Crampton}, {Colless},
  {Glazebrook}, {Mallen-Ornelas}, \& {Broadhurst}}]{Lilly1998}
{Lilly} S. {et~al.}, 1998, ApJ, 500, 75

\bibitem[{{Lintott} {et~al}\mbox{.}(2011){Lintott}, {Schawinski}, {Bamford},
  {Slosar}, {Land}, {Thomas}, {Edmondson}, {Masters}, {Nichol}, {Raddick},
  {Szalay}, {Andreescu}, {Murray}, \& {Vandenberg}}]{2011MNRAS.410..166L}
{Lintott} C. {et~al.}, 2011, \mnras, 410, 166

\bibitem[{{Lintott} {et~al}\mbox{.}(2008){Lintott}, {Schawinski}, {Slosar},
  {Land}, {Bamford}, {Thomas}, {Raddick}, {Nichol}, {Szalay}, {Andreescu},
  {Murray}, \& {Vandenberg}}]{GZ1}
{Lintott} C.~J. {et~al.}, 2008, \mnras, 389, 1179

\bibitem[{{MacArthur} {et~al}\mbox{.}(2004){MacArthur}, {Courteau}, {Bell}, \&
  {Holtzman}}]{2004ApJS..152..175M}
{MacArthur} L.~A., {Courteau} S., {Bell} E., {Holtzman} J.~A., 2004, \apjs,
  152, 175

\bibitem[{{Masters} {et~al}\mbox{.}(2010){Masters}, {Nichol}, {Bamford},
  {Mosleh}, {Lintott}, {Andreescu}, {Edmondson}, {Keel}, {Murray}, {Raddick},
  {Schawinski}, {Slosar}, {Szalay}, {Thomas}, \&
  {Vandenberg}}]{2010MNRAS.404..792M}
{Masters} K.~L. {et~al.}, 2010, \mnras, 404, 792

\bibitem[{{Masters} {et~al}\mbox{.}(2011){Masters}, {Nichol}, {Hoyle},
  {Lintott}, {Bamford}, {Edmondson}, {Fortson}, {Keel}, {Schawinski}, {Smith},
  \& {Thomas}}]{2011MNRAS.411.2026M}
{Masters} K.~L. {et~al.}, 2011, \mnras, 411, 2026

\bibitem[{{M{\"o}llenhoff} {et~al}\mbox{.}(2006){M{\"o}llenhoff}, {Popescu}, \&
  {Tuffs}}]{Moellenhoff}
{M{\"o}llenhoff} C., {Popescu} C.~C., {Tuffs} R.~J., 2006, \aap, 456, 941

\bibitem[{{Pastrav} {et~al}\mbox{.}(2012){Pastrav}, {Popescu}, {Tuffs}, \&
  {Sansom}}]{2012IAUS..284..306P}
{Pastrav} B.~A., {Popescu} C.~C., {Tuffs} R.~J., {Sansom} A.~E., 2012, in IAU
  Symposium, Vol. 284, IAU Symposium, pp. 306--308

\bibitem[{{Peng} {et~al}\mbox{.}(2002){Peng}, {Ho}, {Impey}, \&
  {Rix}}]{Peng2002}
{Peng} C.~Y., {Ho} L.~C., {Impey} C.~D., {Rix} H., 2002, Astronomical J., 124,
  266

\bibitem[{{Peng} {et~al}\mbox{.}(2010){Peng}, {Ho}, {Impey}, \&
  {Rix}}]{Peng2010}
{Peng} C.~Y., {Ho} L.~C., {Impey} C.~D., {Rix} H.-W., 2010, \aj, 139, 2097

\bibitem[{{Pignatelli} {et~al}\mbox{.}(2006){Pignatelli}, {Fasano}, \&
  {Cassata}}]{gasphot}
{Pignatelli} E., {Fasano} G., {Cassata} P., 2006, \aap, 446, 373

\bibitem[{{Popescu} {et~al}\mbox{.}(2011){Popescu}, {Tuffs}, {Dopita},
  {Fischera}, {Kylafis}, \& {Madore}}]{Popescu}
{Popescu} C.~C., {Tuffs} R.~J., {Dopita} M.~A., {Fischera} J., {Kylafis} N.~D.,
  {Madore} B.~F., 2011, \aap, 527, A109

\bibitem[{{Rowlands} {et~al}\mbox{.}(2012){Rowlands}, {Dunne}, {Maddox},
  {Bourne}, {Gomez}, {Kaviraj}, {Bamford}, {Brough}, {Charlot}, {da Cunha},
  {Driver}, {Eales}, {Hopkins}, {Kelvin}, {Nichol}, {Sansom}, {Sharp}, {Smith},
  {Temi}, {van der Werf}, {Baes}, {Cava}, {Cooray}, {Croom}, {Dariush}, {de
  Zotti}, {Dye}, {Fritz}, {Hopwood}, {Ibar}, {Ivison}, {Liske}, {Loveday},
  {Madore}, {Norberg}, {Popescu}, {Rigby}, {Robotham}, {Rodighiero}, {Seibert},
  \& {Tuffs}}]{Rowlands}
{Rowlands} K. {et~al.}, 2012, \mnras, 419, 2545

\bibitem[{{Schade} {et~al}\mbox{.}(1997){Schade}, {Barrientos}, \&
  {Lopez-Cruz}}]{schade97}
{Schade} D., {Barrientos} L.~F., {Lopez-Cruz} O., 1997, ApJ, 477, L17+

\bibitem[{{S\'ersic}(1968)}]{Sersic}
{S\'ersic} J.~L., 1968, {Atlas de galaxias australes}. Cordoba, Argentina:
  Observatorio Astronomico, 1968

\bibitem[{{Simard}(1998)}]{Simard98}
{Simard} L., 1998, in ASP Conf. Ser. 145: Astronomical Data Analysis Software
  and Systems VII, {Albrecht} R., {Hook} R.~N., {Bushouse} H.~A., eds., pp.
  108--+

\bibitem[{{Simard} {et~al}\mbox{.}(2011){Simard}, {Mendel}, {Patton},
  {Ellison}, \& {McConnachie}}]{Simard2011}
{Simard} L., {Mendel} J.~T., {Patton} D.~R., {Ellison} S.~L., {McConnachie}
  A.~W., 2011, \apjs, 196, 11

\bibitem[{{Simard} {et~al}\mbox{.}(2002){Simard}, {Willmer}, {Vogt},
  {Sarajedini}, {Phillips}, {Weiner}, {Koo}, {Im}, {Illingworth}, \&
  {Faber}}]{Simard2002}
{Simard} L. {et~al.}, 2002, ApJS, 142, 1

\bibitem[{{Strateva} {et~al}\mbox{.}(2001){Strateva}, {Ivezi{\'c}}, {Knapp},
  {Narayanan}, {Strauss}, {Gunn}, {Lupton}, {Schlegel}, {Bahcall}, {Brinkmann},
  {Brunner}, {Budav{\'a}ri}, {Csabai}, {Castander}, {Doi}, {Fukugita}, {Gy{\H
  o}ry}, {Hamabe}, {Hennessy}, {Ichikawa}, {Kunszt}, {Lamb}, {McKay},
  {Okamura}, {Racusin}, {Sekiguchi}, {Schneider}, {Shimasaku}, \&
  {York}}]{Strateva}
{Strateva} I. {et~al.}, 2001, AJ, 122, 1861

\bibitem[{{Suh} {et~al}\mbox{.}(2010){Suh}, {Jeong}, {Oh}, {Yi}, {Ferreras}, \&
  {Schawinski}}]{2010ApJS..187..374S}
{Suh} H., {Jeong} H., {Oh} K., {Yi} S.~K., {Ferreras} I., {Schawinski} K.,
  2010, \apjs, 187, 374

\bibitem[{{Szalay} {et~al}\mbox{.}(1999){Szalay}, {Connolly}, \&
  {Szokoly}}]{1999AJ....117...68S}
{Szalay} A.~S., {Connolly} A.~J., {Szokoly} G.~P., 1999, \aj, 117, 68

\bibitem[{{Tasca} \& {White}(2011)}]{TascaWhite}
{Tasca} L.~A.~M., {White} S.~D.~M., 2011, \aap, 530, A106

\bibitem[{{Tortora} {et~al}\mbox{.}(2010){Tortora}, {Napolitano}, {Cardone},
  {Capaccioli}, {Jetzer}, \& {Molinaro}}]{2010MNRAS.407..144T}
{Tortora} C., {Napolitano} N.~R., {Cardone} V.~F., {Capaccioli} M., {Jetzer}
  P., {Molinaro} R., 2010, \mnras, 407, 144

\bibitem[{{Vika} {et~al}\mbox{.}(2012){Vika}, {Driver}, {Cameron}, {Kelvin}, \&
  {Robotham}}]{Vika}
{Vika} M., {Driver} S.~P., {Cameron} E., {Kelvin} L., {Robotham} A., 2012,
  \mnras, 419, 2264

\bibitem[{{White} {et~al}\mbox{.}(2005){White}, {Clowe}, {Simard}, {Rudnick},
  {De Lucia}, {Arag{\'o}n-Salamanca}, {Bender}, {Best}, {Bremer}, {Charlot},
  {Dalcanton}, {Dantel}, {Desai}, {Fort}, {Halliday}, {Jablonka}, {Kauffmann},
  {Mellier}, {Milvang-Jensen}, {Pell{\'o}}, {Poggianti}, {Poirier},
  {Rottgering}, {Saglia}, {Schneider}, \& {Zaritsky}}]{EDisCS}
{White} S.~D.~M. {et~al.}, 2005, \aap, 444, 365

\bibitem[{{Wolf} {et~al}\mbox{.}(2009){Wolf}, {Arag{\'o}n-Salamanca}, {Balogh},
  {Barden}, {Bell}, {Gray}, {Peng}, {Bacon}, {Barazza}, {B{\"o}hm}, {Caldwell},
  {Gallazzi}, {H{\"a}u{\ss}ler}, {Heymans}, {Jahnke}, {Jogee}, {van Kampen},
  {Lane}, {McIntosh}, {Meisenheimer}, {Papovich}, {S{\'a}nchez}, {Taylor},
  {Wisotzki}, \& {Zheng}}]{Wolf2009}
{Wolf} C. {et~al.}, 2009, MNRAS, 393, 1302

\bibitem[{{York} {et~al}\mbox{.}(2000){York}, {Adelman}, {Anderson},
  {Anderson}, {Annis}, {Bahcall}, {Bakken}, {Barkhouser}, {Bastian}, {Berman},
  {Boroski}, {Bracker}, {Briegel}, {Briggs}, {Brinkmann}, {Brunner}, {Burles},
  {Carey}, {Carr}, {Castander}, {Chen}, {Colestock}, {Connolly}, {Crocker},
  {Csabai}, {Czarapata}, {Davis}, {Doi}, {Dombeck}, {Eisenstein}, {Ellman},
  {Elms}, {Evans}, {Fan}, {Federwitz}, {Fiscelli}, {Friedman}, {Frieman},
  {Fukugita}, {Gillespie}, {Gunn}, {Gurbani}, {de Haas}, {Haldeman}, {Harris},
  {Hayes}, {Heckman}, {Hennessy}, {Hindsley}, {Holm}, {Holmgren}, {Huang},
  {Hull}, {Husby}, {Ichikawa}, {Ichikawa}, {Ivezi{\'c}}, {Kent}, {Kim},
  {Kinney}, {Klaene}, {Kleinman}, {Kleinman}, {Knapp}, {Korienek}, {Kron},
  {Kunszt}, {Lamb}, {Lee}, {Leger}, {Limmongkol}, {Lindenmeyer}, {Long},
  {Loomis}, {Loveday}, {Lucinio}, {Lupton}, {MacKinnon}, {Mannery}, {Mantsch},
  {Margon}, {McGehee}, {McKay}, {Meiksin}, {Merelli}, {Monet}, {Munn},
  {Narayanan}, {Nash}, {Neilsen}, {Neswold}, {Newberg}, {Nichol}, {Nicinski},
  {Nonino}, {Okada}, {Okamura}, {Ostriker}, {Owen}, {Pauls}, {Peoples},
  {Peterson}, {Petravick}, {Pier}, {Pope}, {Pordes}, {Prosapio},
  {Rechenmacher}, {Quinn}, {Richards}, {Richmond}, {Rivetta}, {Rockosi},
  {Ruthmansdorfer}, {Sandford}, {Schlegel}, {Schneider}, {Sekiguchi}, {Sergey},
  {Shimasaku}, {Siegmund}, {Smee}, {Smith}, {Snedden}, {Stone}, {Stoughton},
  {Strauss}, {Stubbs}, {SubbaRao}, {Szalay}, {Szapudi}, {Szokoly}, {Thakar},
  {Tremonti}, {Tucker}, {Uomoto}, {Vanden Berk}, {Vogeley}, {Waddell}, {Wang},
  {Watanabe}, {Weinberg}, {Yanny}, {Yasuda}, \& {SDSS
  Collaboration}}]{York2000}
{York} D.~G. {et~al.}, 2000, \aj, 120, 1579

\end{thebibliography}

\appendix
\section{Example images and fit results}
In this appendix, we show some sample images of simulated galaxies, their fitting models (if applicable) and their fitting residuals in all 9 bands for both multi-band and single-band fitting. These galaxies are not randomly chosen, but rather chosen to show typical difficulties and effects that separate single-band and multi-band fitting. In particular, the objects shown are the same objects shown in Fig.~\ref{fig1} and Fig.~\ref{fig6}.

\begin{figure*}
\begin{center}
\includegraphics[width=0.80\textwidth, trim=0 0 0 0, clip]{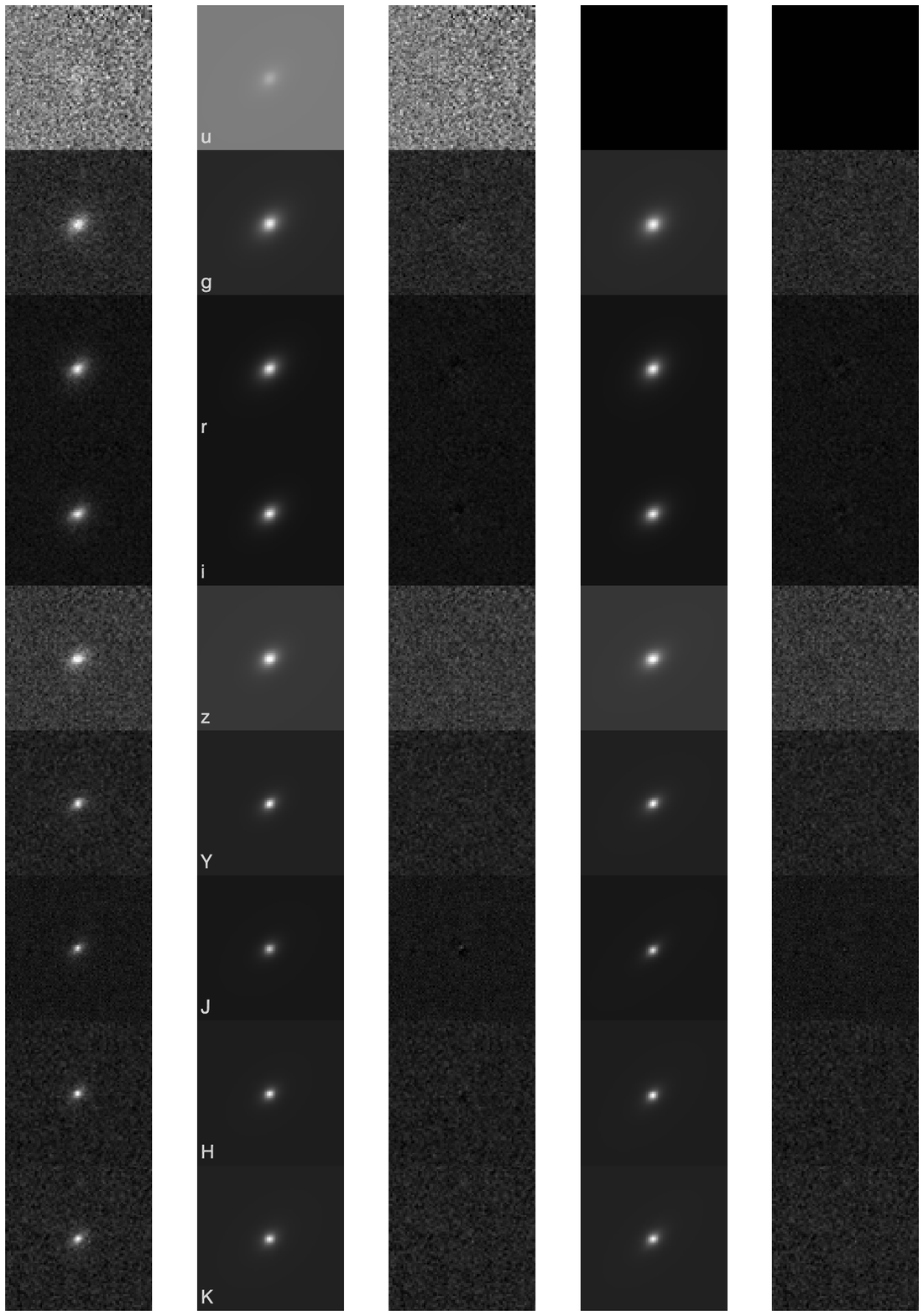}
\caption{Example real object. From left to right: input image, model \& residual multi-band fit, model \& residual single-band fit. Black images in case of single-band fitting indicate where no \galfitm results exist (while ignoring constraints and cleaning of the catalogue).}
\label{figA1}
\end{center}
\end{figure*}
\begin{figure*}
\begin{center}
\includegraphics[width=0.80\textwidth, trim=0 0 0 0, clip]{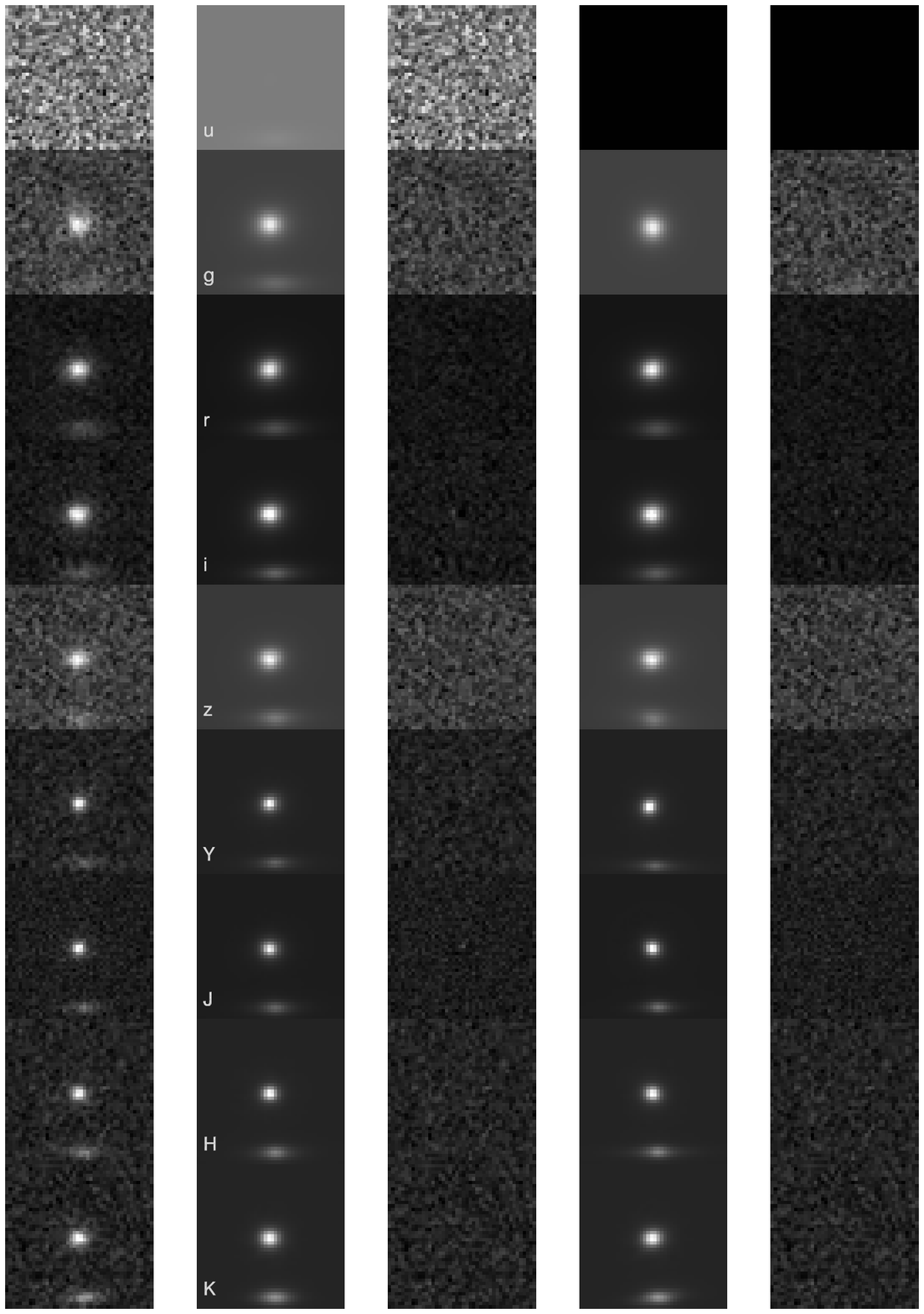}
\caption{Example real object, similar to Fig. \ref{figA1}: This example shows a galaxy that has not been detected in single-band $u$-band. The neighbouring galaxy has also not been detected in the single-band $g$-band, potentially biasing the fit.}
\label{figA2}
\end{center}
\end{figure*}
\begin{figure*}
\begin{center}
\includegraphics[width=0.80\textwidth, trim=0 0 0 0, clip]{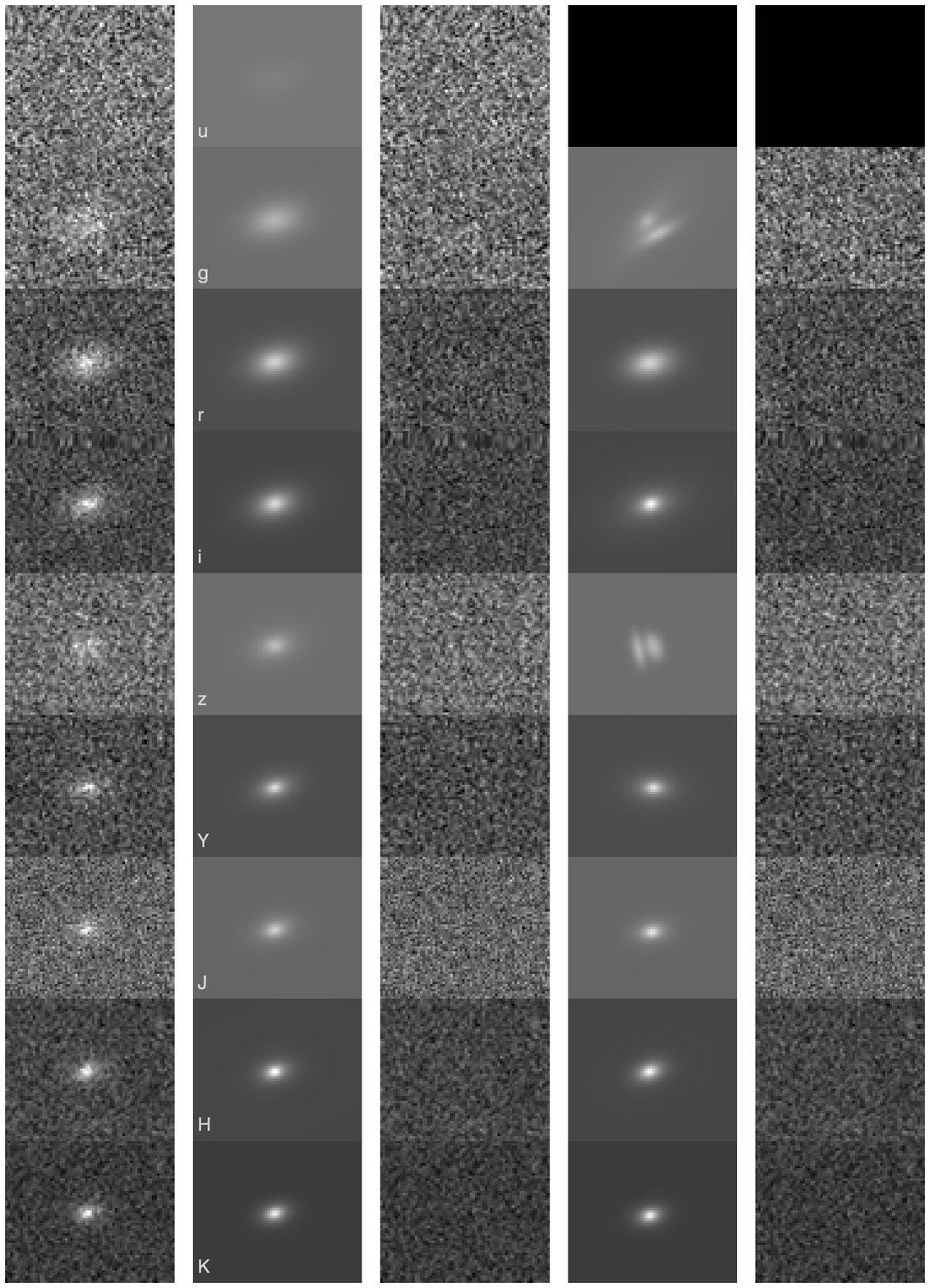}
\caption{Example real object, similar to Fig. \ref{figA1}: This is a very interesting object, because it becomes immediately apparent why some of the single-band fits failed. In both $g$-band and $i$-band, the object is split up into 2 objects that are fit.}
\label{figA3}
\end{center}
\end{figure*}
\begin{figure*}
\begin{center}
\includegraphics[width=0.80\textwidth, trim=0 0 0 0, clip]{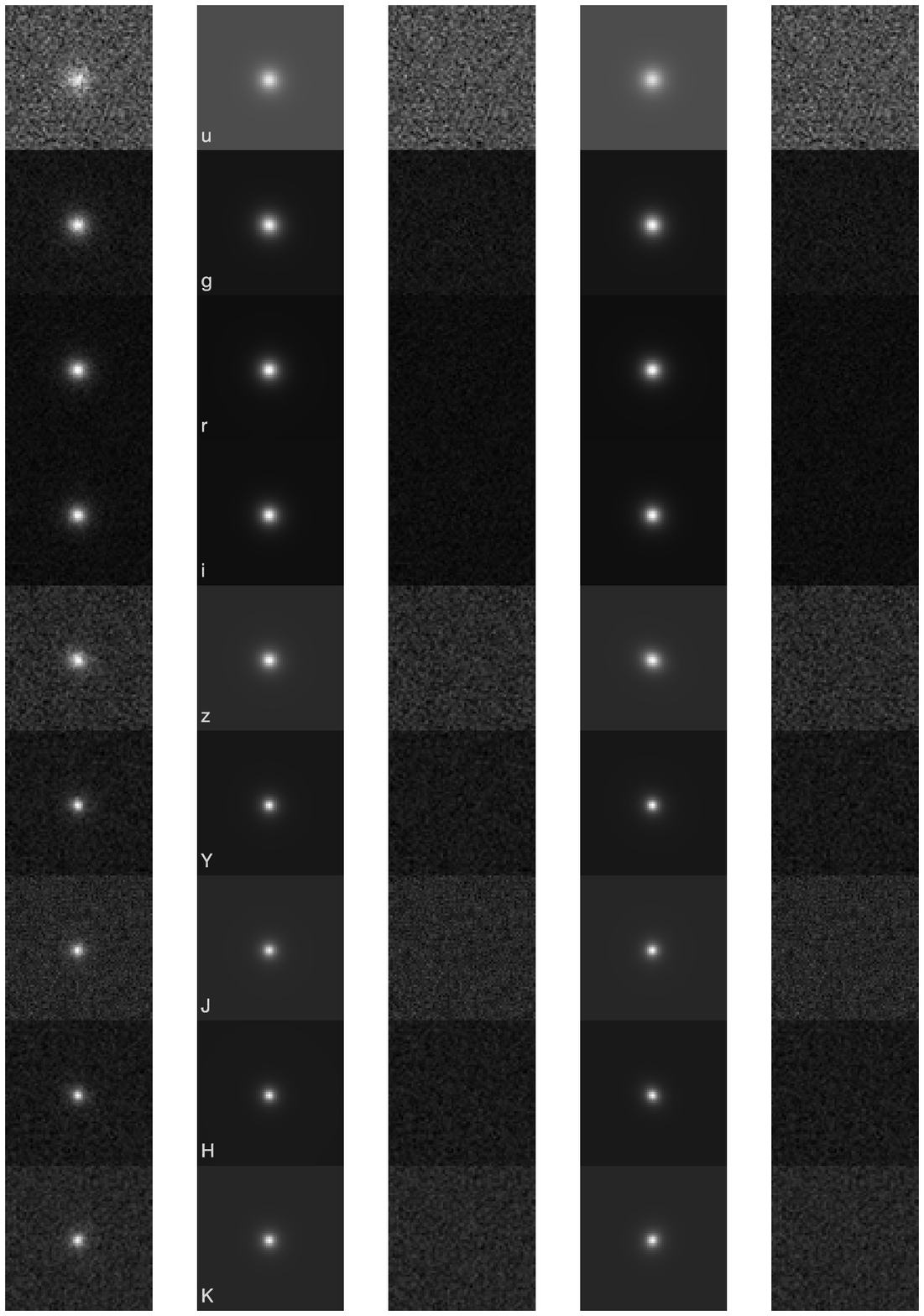}
\caption{Example simulated object. This object failed to fit in some of the single-band fits. Whereas in the $u$-band data, it was missed in the detection procedure, in both $z$- and $J$-band, the fit crashed, returning no valid result. Thanks to the multi-band detection that is used for the multi-band fitting, we can recover parameters for all bands, e.g. creating a complete broad-band SED.}
\label{figA4}
\end{center}
\end{figure*}
\begin{figure*}
\begin{center}
\includegraphics[width=0.80\textwidth, trim=0 0 0 0, clip]{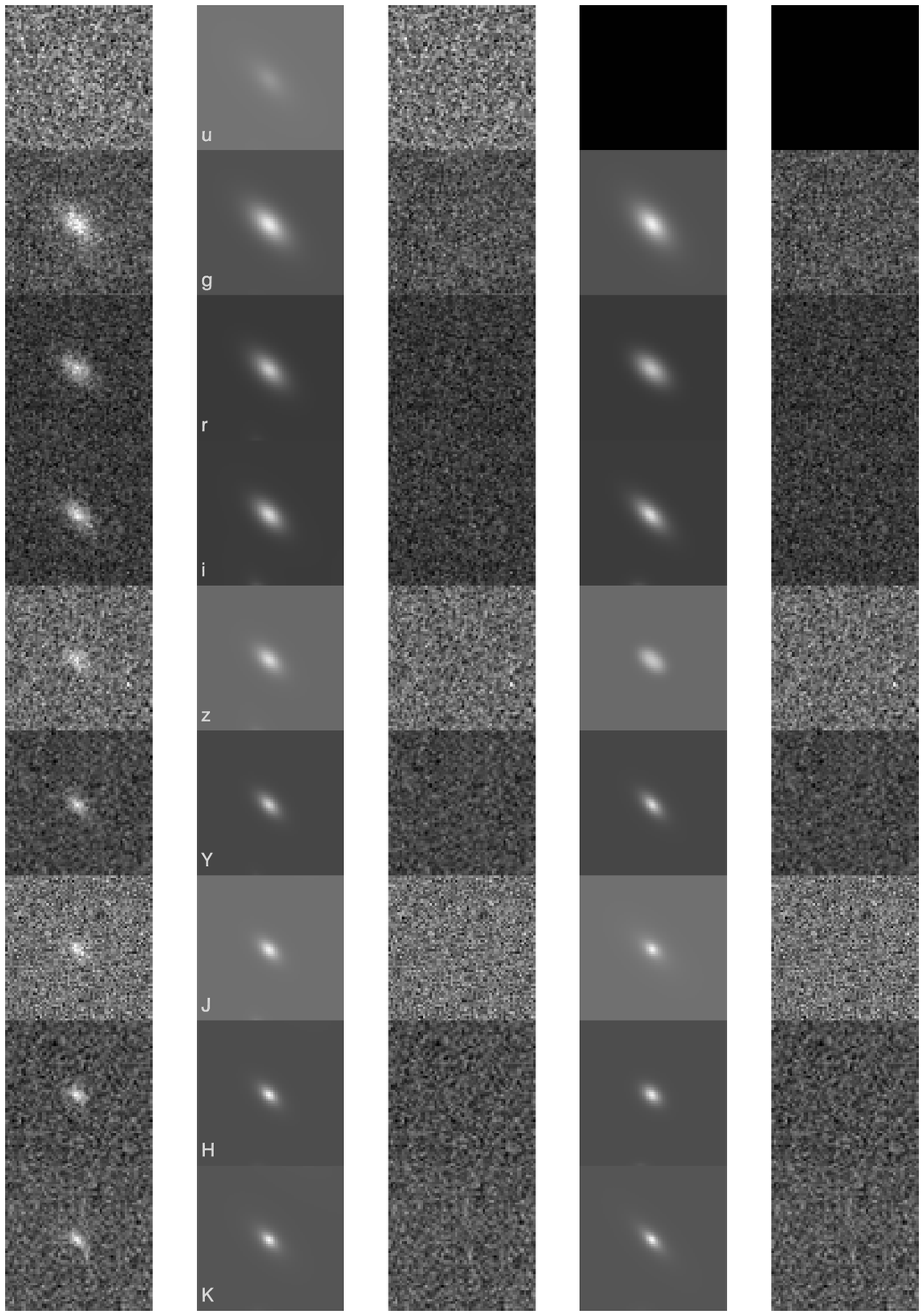}
\caption{Example simulated object. This object is generally well fit by both single and multi-band fitting. Only the single-band fit on the $u$-band image did not return a result, most likely because the object was not detected in the $u$-band image.}
\label{figA5}
\end{center}
\end{figure*}
\begin{figure*}
\begin{center}
\includegraphics[width=0.80\textwidth, trim=0 0 0 0, clip]{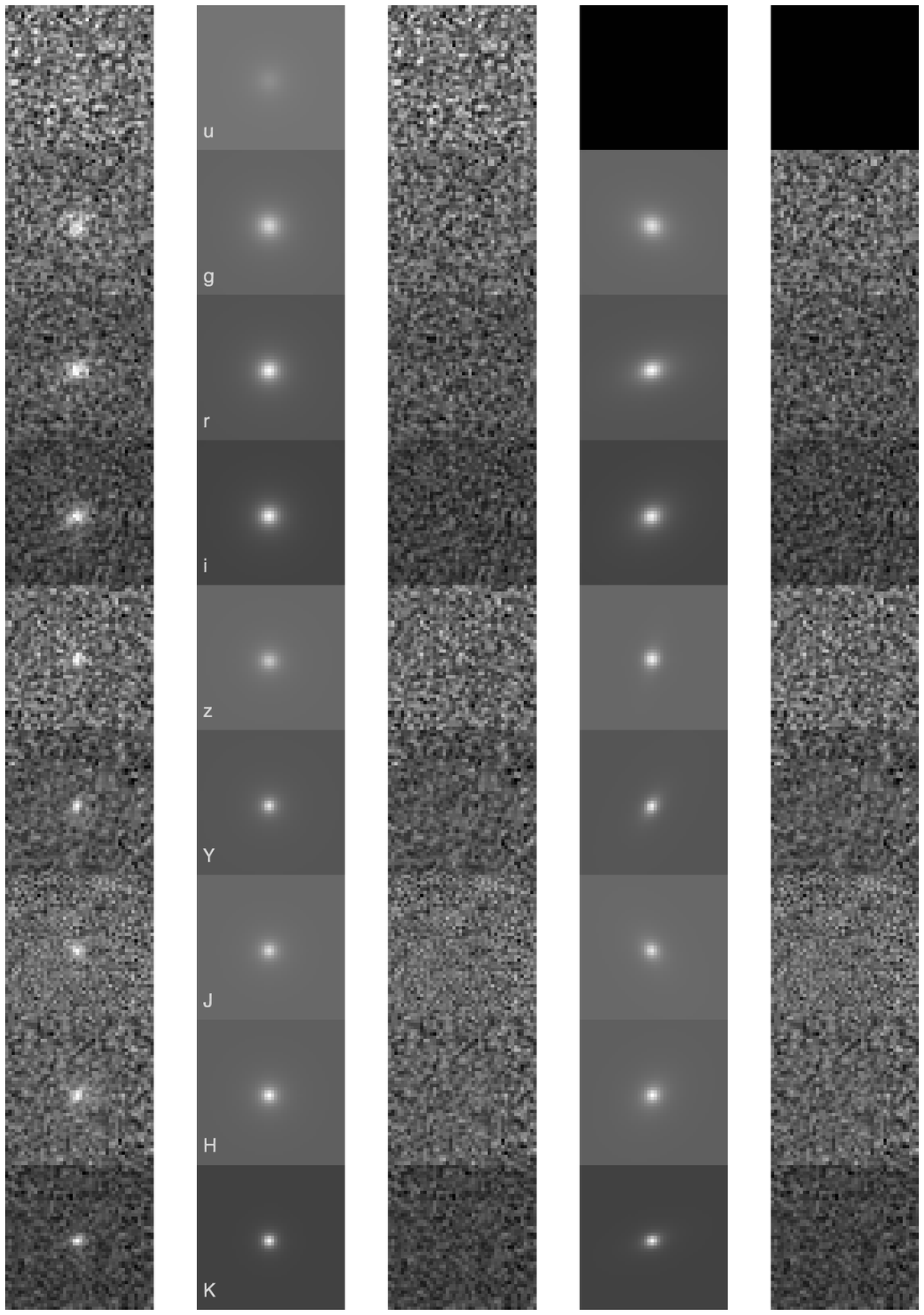}
\caption{Example simulated object. From left to right: input image, model \& residual multi-band fit, image, model \& residual single-band fit. Black images in case of single-band fitting indicate where no \galfitm results exist (while ignoring constraints and cleaning of the catalogue).}
\label{figA6}
\end{center}
\end{figure*}
\begin{figure*}
\begin{center}
\includegraphics[width=0.80\textwidth, trim=0 0 0 0, clip]{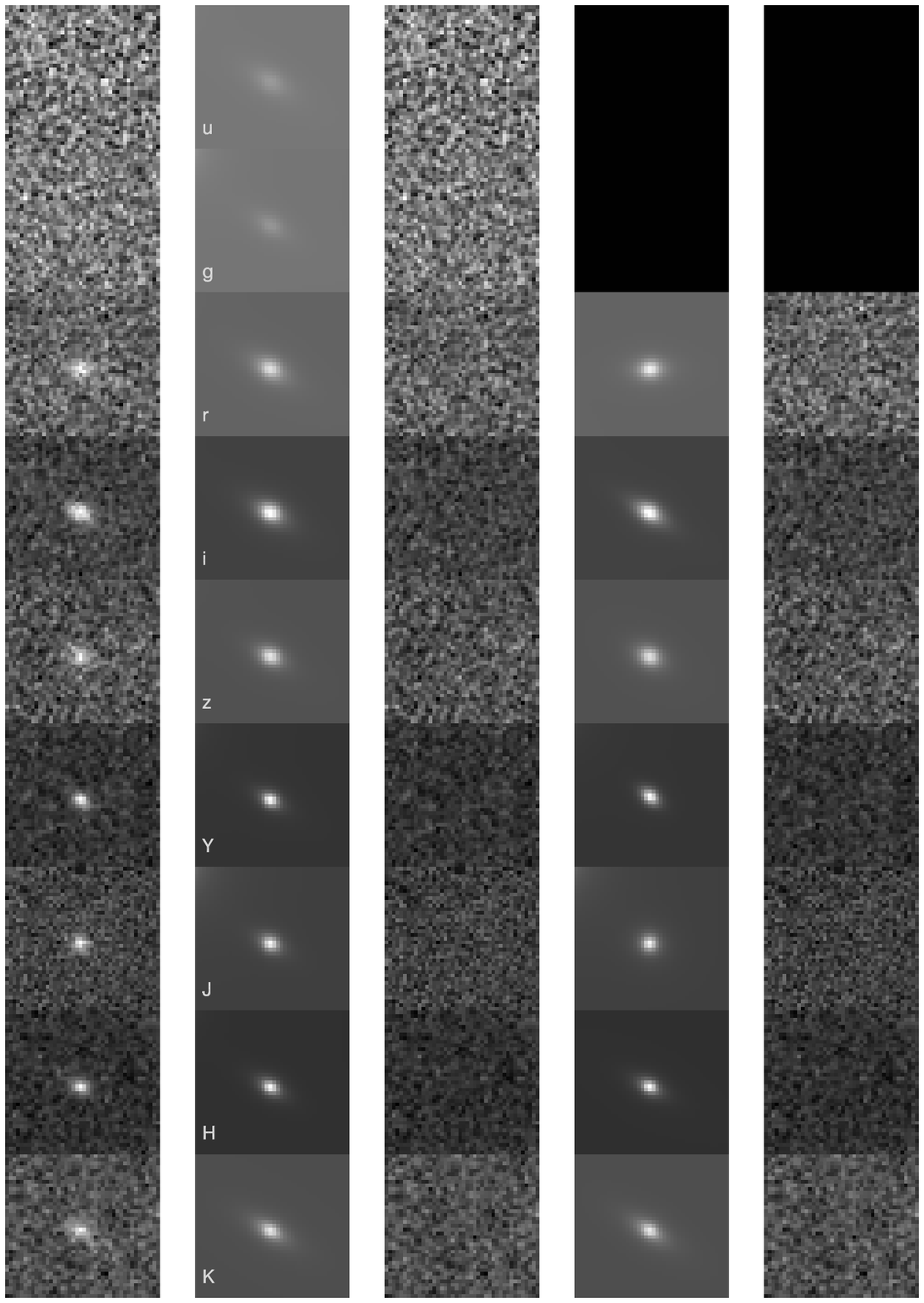}
\caption{This is an extreme simulated object. It can really only be seen in $Y$- and $J$-band data, but thanks to the multi-band capabilities, we can recover a full SED and parameter in all bands on this galaxy. In comparison, single-band fitting only returns one value in the $J$-band fit. Please keep in mind that all examples in this Appendix ignore the cleaning of the catalogue, e.g. showing this example does not mean that this objects would end up in the science catalogue.}
\label{figA7}
\label{lastpage}
\end{center}
\end{figure*}
\clearpage

\end{document}